\documentclass[submission, PhysLectNotes]{SciPost}

\hypersetup{
    colorlinks,
    linkcolor={red!50!black},
    citecolor={blue!50!black},
    urlcolor={blue!80!black}
}
 
\usepackage[bitstream-charter]{mathdesign}
\DeclareSymbolFont{usualmathcal}{OMS}{cmsy}{m}{n}
\DeclareSymbolFontAlphabet{\mathcal}{usualmathcal}

\usepackage{pzccal} 
\DeclareFontFamily{OT1}{pzc}{}
\DeclareFontShape{OT1}{pzc}{m}{it}{<-> s * [1.10] pzcmi7t}{}
\DeclareMathAlphabet{\mathpzc}{OT1}{pzc}{m}{it}

\usepackage{bm}
\usepackage{amsthm}

\newcommand{\mz}{\mathpzc{z}}
\newcommand{\be}{\begin{equation}}
\newcommand{\ee}{\end{equation}}
\newcommand{\bd}{\begin{displaymath}}
\newcommand{\ed}{\end{displaymath}}
\newcommand{\BE}{\begin{eqnarray}}
\newcommand{\EE}{\end{eqnarray}}

\newcommand{\erf}{{\rm erf}}

\newcommand{\bb}{\ensuremath{\mathbf{b}}}
\newcommand{\bh}{\ensuremath{\mathbf{h}}}

\newcommand{\bs}{\ensuremath{\mathbf{s}}}

\newcommand{\bv}{\ensuremath{\mathbf{v}}}

\newcommand{\bx}{\ensuremath{\mathbf{x}}}

\newcommand{\bz}{\ensuremath{\mathbf{z}}}

\newcommand{\bnull}{\ensuremath{\mathbf{0}}}

\newcommand{\bA}{\ensuremath{\mathbf{A}}}

\newcommand{\bC}{\ensuremath{\mathbf{C}}}

\newcommand{\D}{{\cal D}}

\newcommand{\bpsi}{{\mbox{\boldmath $\psi$}}}

\newcommand{\boldpsi}{{\mbox{\boldmath $\psi$}}}
\newcommand{\bulpsi}{{\mbox{\boldmath $\underline{\psi}$}}}

\newcommand{\bxi}{\bm{\xi}}

\newcommand{\avg}[1]{\left\langle{#1}\right\rangle}
\newcommand{\olavg}[1]{\overline{\left\langle{#1}\right\rangle}}

\newcommand{\ulh}{\underline{h}}
\newcommand{\ulM}{\underline{M}}
\newcommand{\ulm}{\underline{m}}
\newcommand{\ulP}{\underline{P}}
\newcommand{\ulp}{\underline{p}}
\newcommand{\ulx}{\underline{x}}
\newcommand{\ulf}{\underline{f}}
\newcommand{\ulb}{\underline{b}}
\newcommand{\ulbx}{\underline{\mathbf{x}}}
\newcommand{\ulbxi}{\underline{\bxi}}
\newcommand{\uleta}{\underline{\eta}}
\newcommand{\ulz}{\underline{z}}
\newcommand{\ulv}{\underline{v}}
\newcommand{\uly}{\underline{y}}
\newcommand{\ulpsi}{\underline{\psi}}
\newcommand{\ulbf}{\underline{\mathbf{f}}}
\newcommand{\ulbpsi}{\underline{\bpsi}}

\newcommand{\ulbh}{\underline{\bh}}

\newcommand{\ulxi}{\underline{\xi}}
\newcommand{\ulnull}{\underline{0}}
\newcommand{\ullnull}{\underline{\underline{0}}}
\newcommand{\ulbnull}{\underline{\mathbf{0}}}
\newcommand{\ullA}{\underline{\underline{A}}}
\newcommand{\ullC}{\underline{\underline{C}}}
\newcommand{\ullc}{\underline{\underline{c}}}
\newcommand{\ullL}{\underline{\underline{L}}}
\newcommand{\ullell}{\underline{\underline{\ell}}}
\newcommand{\ullG}{\underline{\underline{G}}}
\newcommand{\ullK}{\underline{\underline{K}}}
\newcommand{\ullk}{\underline{\underline{k}}}
\begin{document}

\begin{center}{\Large \textbf{
Generating-functional analysis of random Lotka--Volterra systems: A step-by-step guide\\
}}\end{center}

\begin{center}
Tobias Galla
\end{center}

\begin{center}
Institute for Cross-Disciplinary Physics and Complex Systems IFISC (CSIC-UIB),\\
Campus Universitat de les Illes Balears, E-07122 Palma de Mallorca, Spain
\\
{\small \sf tobias.galla@ifisc.uib-csic.es}
\end{center}

\begin{center}
\today
\end{center}


\section*{Abstract}
{\bf
This paper provides what is hopefully a self-contained set of notes describing the detailed steps of a generating-functional analysis of systems of generalised Lotka--Volterra equations with random interaction coefficients. Nothing in these notes is original, instead the generating-functional method (also known as the Martin-Siggia-Rose-DeDominic-Janssen formalism) and the resulting dynamic mean field theories have been used for the study of disordered systems and spin glasses for decades. But it is hard to find unifying sources which would allow a beginner to learn step-by-step how these methods can be used. My aim is to provide such a source. Most of the calculations are specific to generalised Lotka--Volterra systems, but much can be transferred to disordered systems in more general. 
}

\vspace{10pt}
\noindent\rule{\textwidth}{1pt}
\tableofcontents\thispagestyle{fancy}
\noindent\rule{\textwidth}{1pt}
\vspace{10pt}

\section{Introduction: What this is about, who might be interested in this, and how to use these notes}\label{sec:intro}

\subsection{What this is about}
In these notes I describe how so-called `generating functionals' can be used to study generalised Lotka-Volterra dynamics with random interaction matrices. These equations describe the time-evolution of non-negative abundances $x_1,\dots,x_N$ of $N$ species, and are of the form \cite{bunin2016interaction,bunin2017,Galla_2018}
\be\label{eq:lv}
\dot x_i=x_i\left(1-x_i+\sum_{j}\alpha_{ij}x_j\right),
\ee
where $i=1,\dots,N$, and where the $\alpha_{ij}$ are `quenched' random variables. This means that the $\alpha_{ij}$ are drawn at the beginning from a joint probability distribution (to be specified). Then they are kept fixed, and we are interested in the behaviour of the solution of the coupled differential equations in (\ref{eq:lv}), i.e., of the resulting $x_i(t)$. The initial condition for the $x_i(t=0)>0$ can also be random, i.e. , the $x_1(0),\dots,x_N(0)$ are (or can be) drawn from some joint probability distribution. Crucially however, the generalised random Lotka-Volterra dynamics contains no further stochasticity during the time-evolution of the system.

\medskip

The initial Lotka--Volterra model only described two species (see e.g. \cite{hofbauersigmund}). The term `generalised' is commonly used to describe generalisations to systems with more than two species. Throughout this document we will be looking at systems with a  large number $N$ of species (technically, we will be considering the thermodynamic limit $N\to\infty$), but we will often omit the word `generalised'. We also have random interaction coefficients, and we will simply call this the random Lotka--Volterra model, or the disordered Lotka--Volterra equations. Sometimes we will use combinations of these terms.

\medskip

Given that Eqs.~(\ref{eq:lv}) are nonlinear, it is hopeless to expect analytical progress for any particular realisation of the $\alpha_{ij}$. However, and as we will see, the {\em typical} behaviour of the dynamics can be characterised, at least in the thermodynamic limit. To do this one derives what is called a `dynamical mean field theory', capturing what a typical `representative' species experiences. The purpose of these notes is to describe how to do this, and how to analyse the resulting mean field dynamics.

The methods we will use, generating functionals and dynamic mean-field theory, are well established in the theory of disordered systems. There are plenty of sources, see for example \cite{mezard1987, dedominicis1978dynamics, sompolinsky1982relaxational, kirkpatrick1987p,Coolen_MG, coolen_kuehn_sollich,altlandsimons,hertz2016path}. The application of these methods to random Lotka--Volterra equations or related replicator equations is not new either, see \cite{opper1992phase, galla2006,Galla_2018,sidhomgalla,froy}, and in a sense these notes are an expanded variant of the Supplementary Material of \cite{Galla_2018}.   More recent work includes \cite{baron_et_al_prl, park, poley2, aguirre, azaele}.

\medskip

{\bf Note: This version is not the final version. Please send me comments and corrections.} \\
\noindent The document you are looking at on your screen (or on paper) is not the final version of these notes. I have done a fair amount of checking, and a number of my students have pointed out a number of typos (and I have corrected those). But inevitably there will still be factors of $i$ or $2\pi$ missing on occasion. This is the version I decided to put on the ArXiv for the time being. If anyone really reads this and finds mistakes, please email me with corrections. If you think the text could be made clearer somewhere, let me know.  

\subsection{Who might be interested in this and why?}
An earlier version of these notes was initially written as an internal document for masters and PhD students who were trying get into the topic. This remains the natural primary audience, along with other non-expert researchers. I hope that what I have written is sufficiently didactic and self-contained. There might also be readers who are already familiar with tools for disordered systems, and who'd like to apply these methods to models of complex ecosystems. And then there are perhaps readers who are already experts, but who'd like to look up particular technical details.
\medskip

\subsection{How to use these notes}
Readers who are experienced with generating-functional methods for disordered systems will not need any advice on how to use the notes. They know what they are doing. They will probably not read through this paper line-by-line. Instead they'll just skim through the first pages, recognise most equations, mumble something to the effect of `ahh, yes, this here is this, and this here is that', turning the pages quite quickly. Then they might look at a few technical points. Perhaps they'll identify some of sort of subtlety that I didn't see. If you are such a reader and find anything that is wrong, incomplete, or requires further discussion, please let me know. If you think the presentation is unnecessarily complicated and that things could be explained more efficiently, please also get in touch.

\medskip

Things could look a little more daunting for someone who is new to all of this. There are so many quantities and symbols here, individual steps can be quite technical, and the whole thing is just very long. It is hard to see the forest for the trees. There are many little things to get hung up on. A beginner might read the notes line-by-line, diligently going from one equation to the next, and they will then (most likely) get stuck at some point. It'll then probably take a bit of courage to proceed.

If these methods are all new to you, then my advice is that you initially try to go through the notes line by line. But if you get stuck somewhere, don't spend hours and hours thinking about one little detail. Proceed. As a rule of thumb, if it takes you more than  5-10 minutes to understand a step, just park it, and keep going. You can return to it later. There are so many subtle points here, and you can't worry about everything from the beginning. Adapting from ACC Coolen {\em et al}'s draft book manuscript \cite{coolen_survival}, if you build a house you first construct the foundations and decide how many rooms you want, and what room is next to what other room etc. Then you build the main walls, and perhaps the roof. You worry about the colour of the tiles in the bathroom and the shape of the door handles at the end. It is the same here, first try to get an idea of the general structure of the calculation, the main steps. That's figuring out how many rooms the house has, and where the doors are and the staircase. You worry about the many small and subtle issues later.

\medskip

I tried to write these notes to make this as easy as possible. Occasionally there are steps in the calculation where too much worry about overly technical points would be distracting. On the other hand, I wanted to be as precise as I could. Occasionally, I had to be slightly sloppy to keep the story straight. But I'll tell you when I do this (or more accurately, when I am aware that I am doing this). I then discuss the finer points later, for example in an appendix. Having said this, there will inevitably be subtleties that I missed. If you find any, please let me know.

\medskip

\subsection{The notes are not a review}
I am not aware of a review of the large (and growing) body of work by statistical physicists on random Lotka--Volterra dynamics or related systems.  Perhaps someone will write one at some point. These notes are definitely not such a review. This is not the intention. Instead, I am giving a relatively detailed account of the necessary calculations. I only focus on the most basic random Lotka--Volterra model. A number of variations and extensions have been studied with generating-functional methods or other techniques from the theory of disordered systems, see for example \cite{bunin2017,birolibunin,sidhomgalla,froy,altieri2021properties, barbier2021fingerprints, baron_et_al_prl}. I hope that the interested reader will be able to adapt the methods once they have worked through these notes. The notes end at the point where the phase diagram has been established, but of course much more is known about the model \cite{bunin2017,birolibunin, froy, altieri2021properties, barbier2021fingerprints,baron_et_al_prl,akjouj,Pirey_Bunin}. Again, this is not included here, but I hope that studying the present notes will make it easier to read the more advanced literature.

I am also conscious that I am not really motivating {\em why} it is interesting to study Lotka--Volterra equations with random interactions. I assume that anyone who is reading this is already motivated. If you are not, then you can find some more background in the references above. More widely, the study of random ecosystems is central to the so-called `stability-complexity' debate in theoretical ecology. There is plenty of material on this, see for example \cite{maybook,allesinatang1, may, may71, allesinatang2, gravel, namba, grilli, gibbs, stone, fyodorov2016, barabas2017,gross2009, landi,song2018}. I do not include details here, to keep the notes focused on the main purpose, to introduce the reader to the mathematical techniques.
\subsection{Structure of this document}
These notes can be divided into three parts:

\medskip

{\bf \underline{Part I:}}
\begin{itemize}
\item {\bf Section~\ref{sec:intro}:} General introduction;
\item {\bf Section~\ref{sec:mft}:} Brief background on `mean-field theory';  
\item {\bf Section~\ref{sec:maths}:} Collection of the main mathematical tools we will be using in the generating-functional analysis.
\end{itemize}
Sections~\ref{sec:mft} and \ref{sec:maths} are included for completeness, and can be skipped by readers who are familiar with these concepts.

\bigskip

{\bf \underline{Part II:}}\\
This part intended as an introduction for readers who are not familiar with the idea of a generating function for probability distributions, and/or generating functionals for stochastic processes.
\begin{itemize}
\item {\bf Section~\ref{sec:gf}:} The idea of generating functions and functionals for probability distributions, and simple stochastic processes;
\item {\bf Section~\ref{sec:toy_model}:} Application of generating functionals to a high-dimensional toy model, with no disorder. 
\end{itemize}

\bigskip

{\bf \underline{Part III:}}\\
This part contains the actual generating-functional analysis for the Lotka--Volterra model.
\begin{itemize}
\item {\bf Section~\ref{sec:gf_lv_1}:} We set up the generating functional for the disordered Lotka--Volterra model, carry out the disorder average, introduce the relevant macroscopic dynamic order parameters, and derive the saddle-point equations for these order parameters in the limit of an infinite number of species;
 \item {\bf Section~\ref{sec:gf_lv_2}:} We first present a simplified (`vanilla') derivation of the corresponding dynamic mean field theory from the saddle-point equations.  This is based on certain assertions, which can only be justified retrospectively.
 \item {\bf Section~\ref{sec:gf_lv_3}:} This section contains are more complete bottom-up derivation of the same effective dynamics as in Sec.~\ref{sec:gf_lv_2}. We also relate the order parameters at the saddle point to those in the original Lotka--Volterra problem.
 \item {\bf Section~\ref{sec:gf_lv_4}:} We discuss the mathematical structure and the interpretation of the dynamic mean field theory for the Lotka--Volterra model. This theory consists of the effective single-species process, and the self-consistency relations for the order parameters that complement the representative-species dynamics.
 \item {\bf Section~\ref{sec:fixed_point}:} Here, we make a fixed-point ansatz for the effective process, and derive the self-consistency relations for the corresponding static order parameters. We describe how these can be solved parametrically in closed form.
  \item {\bf Section~\ref{sec:stability_pg}:}   We study the stability of the fixed point from Sec.~\ref{sec:fixed_point}, and establish a phase (or stability) diagram for the disordered generalised Lotka--Volterra system.
  \item{\bf Section~\ref{sec:cavity}:} In this supplementary section we briefly describe an alternative derivation of the effective dynamics using the so-called `cavity' approach.  The section is not required for the analysis in Secs.~\ref{sec:gf_lv_1}-\ref{sec:stability_pg}, and is included mostly for completeness.
\end{itemize}
Parts I-III are complemented by a summary and discussion in Sec.~\ref{sec:summary}, and by two appendices in which we provide supplementary details for some parts of the calculation in Part III.
\medskip

 \section{Mean-field theory}\label{sec:mft}
\subsection{Equilibrium mean field theory for the Ising model}
In most textbooks or lecture notes on statistical physics mean-field theory is introduced in the context of the Ising model (Refs.~\cite{yeomans,evans} are examples of good sources). We'll briefly summarise this here. 

A typical model describes $N$ spins, $s_i$, sitting on some sort of network (usually a regular lattice). For simplicity, we assume a regular network, i.e., each spin has the same number of nearest neighbours. We write $k$ for this coordination number.

The spins can each be up or down, $s_i=\pm 1$. We write $\bs=(s_1,\dots,s_N)$ for a spin configuration. The total energy of the system is given by
\be
H(\bs)= -h\sum_i s_i - J \sum_{i}s_i\sum_{j\in i} s_j.
\ee
The notation $j\in i$ describes the nearest neighbours of spin $i$. The quantity $J$ is a coupling strength, and $h$ is an external field. Without loss of generality we set $J=1$.

The idea of mean field theory is to replace the sum $\sum_{j\in i} s_j$ by a `mean field' acting on any one spin. More precisely, we replace
\be
\sum_{j\in i} s_j \to k m,
\ee
where $m$ is the mean magnetisation of the spins neighbouring $i$, i.e., 
\be\label{eq:m}
m=\frac{1}{k}\sum_{j\in i}\avg{s_j}.
\ee
This quantity is taken to be the same for all spins $i$. Just to be clear, the word `mean' describes an average over the Boltzmann distribution for the model\footnote{When you see (or use) angle brackets $\avg{\dots}$ it is always a good idea to ask yourself what type of average exactly these brackets represent.}. We will come to this in a minute. 

We can then write
\be
H=-h_{mf}\sum_i s_i,
\ee
with the `mean field' 
\be
h_{mf}= h + k m.
\ee
Within this approximation and using the usual Boltzmann distribution at inverse temperature $\beta$, the probability for spin $i$ to be in state $s_i=\pm 1$ is proportional to $e^{-\beta h_{mf} s_i}$. Therefore we have
\be\label{eq:p_updown}
P(\mbox{spin $i$ is in state $s_i=\pm 1$})= \frac{e^{-\beta h_{mf} s_i}}{2\cosh(\beta h_{mf})}.
\ee
It remains to determine the mean field self-consistently. Eq.~(\ref{eq:p_updown}) implies that $\avg{s_i}$ is the same for all $i$, and this value must then be the same as the magnetisation $m$ in Eq.~(\ref{eq:m}). Therefore, we have
\BE
m&=&P(\mbox{spin $i$ is up}) - P(\mbox{spin $i$ is down}) \nonumber \\
&=& \tanh(\beta h_{mf}) \nonumber \\
&=& \tanh[\beta (h+km)].
\EE
So, the result from this analysis is that the magnetisation of the model as a function of temperature, the mean coordination number $k$, and the external field $h$, is obtained as the solution of the equation
\be\label{eq:scm}
m=\tanh[\beta (h+km)].
\ee
This can then be analysed to identify paramagnetic and ferromagnetic phases, the transition between the two and so on. We won't go into detail here. Our main point is that we have obtained the self-consistent equation (\ref{eq:scm}) for the {\em macroscopic order parameter} $m$. The word `macroscopic' indicates that $m$ is a property of the system as a whole, not a spin-specific quantity ($m$ carries no site index $i$).

We note that this was a static equilibrium analysis. We have not said anything about time-evolution, nothing in our analysis was `dynamic'. Instead we assumed a Boltzmann distribution at equilibrium. The order parameter $m$ is not time dependent, it is a {\em static} macroscopic order parameter.

\subsection{Dynamic mean-field analysis for a kinetic Ising model}
We'll now look at the Ising model again, but from a dynamic point of view. This means that we start from an actual dynamic process governing the evolution of the spins in time. Of course, this will be closely related to the static picture in the previous section, but I hope you see the difference.

Say the spins at time $t$ are in the configuration $\bs(t)=[s_1(t),\dots,s_N(t)]$. We then assume that, in the next time step, each spin is updated according to the following rule,
\be\label{eq:update}
P[s_i(t+1)=\pm 1]=\frac{e^{\mp\beta h_i[\bs(t)]}}{2\cosh[\beta h_i[\bs(t)]},
\ee
where $h_i[\bs(t)]$ is the total field acting on spin $i$ at that time,
\be
h_i[\bs(t)]=h+\sum_{j\in i} s_j(t).
\ee
We now make a similar approximation as in the static case, and introduce
\be
h_{mf}[m(t)]=h+km(t),
\ee
where
\be
m(t)=\frac{1}{k}\sum_{j\in i}\avg{s_j(t)}.
\ee
We again assume that the quantity on the right does not depend on the site index $i$. Inserting into Eq.~(\ref{eq:update}), and averaging, we then arrive at
\be
\avg{s_i(t+1)}=\tanh[\beta(h+km(t))],
\ee
for all $i$. The average on the left-hand side is $m(t+1)$, and we have therefore arrived an the following equation for the {\em dynamical} macroscopic order parameter $m(t)$,
\be\label{eq:dyn_m}
m(t+1)=\tanh[\beta(h+km(t))].
\ee
The fixed points of this dynamics fulfill Eq.~(\ref{eq:scm}), but Eq.~(\ref{eq:dyn_m}) delivers more than that, it describes the dynamical evolution of the magnetisation in time (within the mean-field approximation).
\subsection{A word of caution}
Our main focus is on the generalised Lotka-Volterra equations, 
\be\label{eq:lv2}
\dot x_i=x_i\left(1-x_i+\sum_{j}\alpha_{ij}x_j\right),
\ee
with quenched random coefficients $\alpha_{ij}$. Say the distribution of the $\alpha_{ij}$ is such that the mean of each $\alpha_{ij}$ (for given $i$ and $j$) is $\mu/N$. One might then be tempted to think that a mean-field approach could look something like this:
\be
\dot m(t) =m(t) \left[1-m(t)+\mu m(t)\right],
\ee
where $m(t)=\overline{x_i(t)}$ (we generall use an overbar $\overline{\cdots}$ for the average over the disorder, i.e., the random couplings $\alpha_{ij}$). Unfortunately, things are not as simple as this -- if they were, the variance of the $\alpha_{ij}$ would be irrelevant, and a model in which all $\alpha_{ij}$ are the same would lead to the same mean-field dynamics as a truly disordered model. That's not the case. For example, if $\mu=0$, the random Lotka--Volterra system will not just converge to $m(t)\to 1$ if started from positive $x_i(t=0)$. Instead, as we will see later, there is a transition between a stable regime for small variance of the interaction coefficients and an unstable regime for larger variances.  This means that we'll need to work a little harder to derive a dynamical mean-field theory for the random Lotka--Volterra model.

\section{Mathematical tools: delta functions, Gaussian integrals and saddle-point method (skip if you are familiar with this)}\label{sec:maths}
Before we can start with our mean-field analysis of the random Lotka--Volterra equations, we have to introduce some notation, and a few general concepts.

\subsection{Notation}
Throughout the paper, bold-face letters indicate column vectors, e.g. $\bx=(x_1,\dots,x_N)^T$. The superscript $T$ stands for `transposed'. Vector components are indicated by latin indices, for example $x_i$ is the $i$-th component of $\bx$.

\medskip

Underlined symbols represent scalar functions of time. If time is discrete, we write for example $\ulx=(x(t=0),x(t=1),\dots)^T$. Occasionally these objects are also treated as vectors (for example to be multiplied by a matrix). We understand these to be column vectors, hence the transpose sign. We will also use this notation when time is continuous. Time is indicated as an argument in brackets, for example $x(t)$.

\medskip

Multi-component functions of time are written as underlined bold-face symbols. For example, $\ulbx$ stands for $\ulx_1, \ulx_2,\dots,\ulx_N$, where each of the $\ulx_i$ is a function of time.

\medskip

Matrices in `time space' are indicated by double underline, e.g. $\ullC$, with entries $C(t,t')$.
\medskip

Matrices in component space are denoted by bold face capital letters, e.g. $\bA$ has elements $a_{ij}$, where $i,j=1,\dots,N$.
\medskip

We use the following convention for Fourier transforms and the inverse transform:
\BE
\widetilde y(\omega)&=&\frac{1}{\sqrt{2\pi}}\int dt\,e^{-i\omega t} y(t),\nonumber \\
y(t)&=&\frac{1}{\sqrt{2\pi}}\int dt\,e^{i\omega t} \widetilde y(t).
\EE

\subsection{Useful identities for delta functions}

First, we recall the definition of the Dirac delta function,
\be\label{eq:delta}
\int_a^b dx~f(x)\delta(x-x_0)=f(x_0),
\ee
if $x_0\in[a,b]$. Otherwise the integral is zero.
The delta function can be written in its so-called `exponential representation'. This can be derived by carrying out a Fourier transform of $\delta(\cdot)$:
\be\label{eq:deltaf}
{\cal F}[\delta](\widehat x)=\frac{1}{\sqrt{2\pi}}\int_{-\infty}^\infty dx ~\delta(x)~ e^{-i\widehat x x}=\frac{1}{\sqrt{2\pi}}.
\ee
In the last step we have used Eq.~(\ref{eq:delta}). We have written ${\cal F}$ for the Fourier transform, and we have used $\widehat x$ as the variable conjugate to $x$. Eq. (\ref{eq:deltaf}) means that the Fourier transform of the $\delta$-function is flat (independent of $\widehat x$). 

Applying the reverse Fourier transform we then have
\be\label{eq:deltaexp}
\delta(x) = \frac{1}{2\pi}\int_{-\infty}^\infty d\widehat x ~ e^{i\widehat x x}.
\ee

Using $1=\int_{-\infty}^\infty dx~ \delta(x)$, we therefore have
\be
1=\int_{-\infty}^\infty \frac{dx
d\widehat x}{2\pi} ~ e^{i\widehat x x}
\ee
We also have $1=\int_{-\infty}^\infty dx\, \delta(x-a)=1$, for any constant $a$ (i.e., an $a$ which does not depend on $x$). Therefore
\be\label{eq:deltaexp2}
1=\int_{-\infty}^\infty \frac{dx
d\widehat x}{2\pi} ~ e^{i\widehat x (x-a)}.
\ee
\subsection{Fourier transforms of correlation functions}\label{sec:ft_corr_fct}
Consider a real-valued stochastic process $y(t)$ in continuous time, with zero average, i.e., $\avg{y(t)}=0$ for all $t$. We define the correlation function 
\be
C_y(t,t')=\avg{y(t)y(t')},
\ee
where the angle bracket stands for an average over realisations of the process. Assuming that the process is such that it reaches a stationary state eventually, we have time-translation invariance at long times, i.e., $\lim_{t \to \infty} C_y(t+\tau,t)$ is a function of time difference $\tau$ only. We write
\be
C(\tau)=\lim_{t\to\infty} C(t+\tau,t).
\ee
With the definition
\be
\widetilde y(\omega)=\frac{1}{\sqrt{2\pi}}\int dt\,e^{-i\omega t} y(t)
\ee
for the Fourier transform of $y(\cdot)$. Using the fact that $y(t')$ is real valued, and writing $t'=t-\tau$, we then have
\BE\label{eq:FT_aux1}
\avg{\widetilde y(\omega')^* \widetilde y(\omega)}&=&\frac{1}{2\pi}\int dt dt'\, e^{i(\omega' t'-\omega t)} \avg{y(t) y(t')}\nonumber \\
&=&\frac{1}{2\pi}\int dt d\tau\, e^{-i\omega \tau} e^{i(\omega'-\omega) t} \avg{y(t)y(t-\tau)} \nonumber \\
&=& \delta(\omega-\omega')\int d\tau C(\tau) e^{-i\omega' \tau}\nonumber \\
&=& \sqrt{2\pi} \delta(\omega-\omega')\widetilde C(\omega).
\EE
We have used the fact that $\avg{y(t)y(t+\tau)}=\avg{y(t)y(t-\tau)}$ in the stationary regime.

\medskip

This result is often summarised as follows: $\avg{|\widetilde y(\omega)|^2}$ is the Fourier transform of $C(\tau)$. There are two issues here. One is the pre-factor $\sqrt{2\pi}$, that's not a problem `is the Fourier transform' simply means 'is the Fourier transform up to trivial pre-factors'. 

The other issue is more subtle, namely, the fact that the left-hand side of Eq.~(\ref{eq:FT_aux1}) contains two frequencies, $\omega$ and $\omega'$, and the right-hand side the delta function $\delta(\omega-\omega')$. Nonetheless it is common to use abbreviated statements of the form $\widetilde C(\omega)=(2\pi)^{-1/2}\avg{|\widetilde y(\omega)|^2}$.
\medskip

If $y(t)$ is white noise of unit variance, i.e., $\avg{y(t)y(t')}=\delta(t-t')$, we have $C(\tau)=\delta(\tau)$, and therefore $\widetilde C(\omega)\equiv \frac{1}{\sqrt{2\pi}}$ for all $\omega$ (a flat spectrum is why this is called white noise after all). It then follows that
\be
\avg{\widetilde y(\omega')^* \widetilde y(\omega)}=\delta(\omega-\omega').
\ee
Often this is simply written as $\avg{|\widetilde y(\omega)|^2}=1$.

\medskip

We mention a third subtlety. In the third step in Eq.~(\ref{eq:FT_aux1}) we have made the replacement $\avg{y(t)y(t-\tau)}\to C(\tau)$. This is valid only in the stationary state, i.e., for large times $t$. But $t$ is an integration variable in Eq.~(\ref{eq:FT_aux1}). There are two ways of interpreting this. One can implicitly assume that the starting point of the dynamics is at $t\to -\infty$. Any finite time $t$ is then automatically in the stationary state. Alternatively one can think of starting the dynamics at $t=0$ but from initial conditions drawn from the stationary distribution of the process.

\subsection{Gaussian integrals}\label{sec:gauss}
We will need various multivariate Gaussian integrals, in particular ones of the form
\be
\int_{-\infty}^\infty dx_1\cdots dx_N \exp\left(-\frac{1}{2}\sum_{i,j=1}^N
x_i A_{ij} x_j+\sum_{i=1}^N b_i x_i\right) 
\equiv \int d\bx ~\exp\left(-\frac{1}{2}\bx^T \bA \bx+ \bb^T\bx\right).
\ee
The $N\times N$ matrix $\bA$ is assumed to be symmetric and positive definite.

The integral can be carried out (for example by transforming to the space spanned by the eigenvectors of $\bA$), and one has
\be\label{eq:Gauss}
\int d\bx ~\exp\left(-\frac{1}{2}\bx^T \bA \bx+ \bb^T\bx\right)=\sqrt{\frac{(2\pi)^N}{\det \bA}}\exp\left(\frac{1}{2}\bb^T\bA^{-1}\bb\right).
\ee
In the case $N=1$, and setting $\bA=\frac{1}{\sigma^2}$ this reduces to the well-known identity

\be\label{eq:gausslin1}
\frac{1}{\sqrt{2\pi\sigma^2}}\int dx \exp\left(-\frac{x^2}{2\sigma^2}+bx\right)=\exp\left(\frac{1}{2}\sigma^2 b^2\right).
\ee

\medskip

{\bf Multivariate Gaussian distribution}:\\
We also note that 
\be
p_{\bC}(\bx)=(2\pi)^{-N/2}(\det \bC)^{-1/2} \exp\left(-\frac{1}{2}\bx^T \bC^{-1} \bx\right)
\ee
is a normalised probability distribution, for a symmetric and positive definite $N\times N$ matrix, $\bC$. In other words $\int d\bx ~p_{\bC}(\bx)=1$. This follows from Eq.~(\ref{eq:Gauss}), with the replacement $\bA\to \bC^{-1}$, and $\bb=\mathbf{0}$.

We also have
\be\label{eq:gausscorr}
\int d\bx ~p_{\bC}(\bx) x_i x_j = C_{ij}.
\ee
This can be derived from Eq.~(\ref{eq:Gauss}) by taking derivatives with respect to $b_i$ and $b_j$, evaluated at $\bb=\bnull$.

Equation (\ref{eq:gausscorr}) indicates that the matrix $\bC$ is the correlation matrix of the Gaussian random variables $x_i$, $i=1,\dots,N$. This also justifies why we assume that the matrix $\bA=\bC^{-1}$ was symmetric and positive definite. Eq.~(\ref{eq:gausscorr}) shows that $\bC$ is symmetric, and this in turn means that $\bC^{-1}$ is symmetric. Further we have for any (non-zero) vector $\bv$ of length $N$,
\BE
\bv^T\bC \bv&=&\sum_{ij} v_i v_j \int d\bx ~p_{\bC}(\bx) x_i x_j \nonumber \\
&=&\int d\bx~ p_{\bC}(\bx) \left(\sum_k v_k x_k\right)^2>0.
\EE
Hence $\bC$ is positive definite.
\medskip

{\bf Continuous Gaussian distribution:}\\
Equation~(\ref{eq:Gauss}) can directly be transferred to the time domain:
\be\label{eq:Gauss_time}
\int d\ulx ~\exp\left(-\frac{1}{2}\ulx^T \ullA \,\ulx+ \ulb^T\ulx\right)=\sqrt{\frac{(2\pi)^N}{\det \ullA}}\exp\left(\frac{1}{2}\ulb^T\ullA^{-1}\ulb\right).
\ee
If there is no endpoint in time ($t=0,1,2,\dots$ unbounded), then the vectors and matrices are infinite dimensional. The space of $\ulx$ is then an infinite-dimensional vector space.

The formula in Eq.~(\ref{eq:Gauss_time}) can also be used when the time index is continuous. We then have $\ulb^T \ulx=\int dt ~b(t) x(t)$ for the scalar product, and the object $\ulx^T \ullA\,\ulx$ is given by
\be
\ulx^T \ullA\, \ulx = \int dt ~dt'~ x(t) A(t,t') x(t').
\ee
The object $\ulx$ is now an element of a space of functions with one continuous argument $t$, and $\ullA$ is a linear operator, mapping one such function onto another.

\subsection{Saddle-point integration (Laplace's method)}\label{sec:saddlepoint}
In this section we briefly discuss Laplace's method to carry out integrals of the form $\int dx g(x) \exp[Nf(x)]$ in the limit $N\to\infty$. Physicists sometimes also refer to this as `saddle-point integration'. In mathematics, saddle-point integration or `the method of steepest descent' refers to wider set of integration techniques in the complex plane \cite{wong,debruijn}. A concise summary can also be found in \cite{evans_mp}.

\subsubsection{Exponential function only}
Suppose we are faced with an integral of the type
\be
I\equiv\int dx  e^{Nf(x)},
\ee
and we are interested in the behaviour for $N\gg 1$. We also assume that $f(\cdot)$ has a unique global maximum at $x_0$. For large $N$ the integral $I$ will therefore be dominated by the region around $x_0$. We can then expand
\be
f(x)=f(x_0)+\frac{1}{2} f''(x_0) (x-x_0)^2 +\cdots.
\ee
We note that $f'(x_0)=0$ and $f''(x_0)<0$ (because $f$ has a local maximum at $x_0$). We then have
\BE
I&=&\int dx  e^{N f(x_0)+N\frac{1}{2} f''(x_0) (x-x_0)^2 +\cdots} \nonumber \\
&=&e^{N f(x_0)} \int dx  e^{N\frac{1}{2} f''(x_0) (x-x_0)^2 +\cdots},
\EE
and we can proceed by Gaussian integration. We find
\be
I \approx \sqrt{\frac{2\pi}{N |f''(x_0)|}}e^{Nf(x_0)}.
\ee

\subsubsection{Product of exponentially dominated density and general function}

Suppose now, we know that $p(x)$ is normalised, $\int dx ~p(x)=1$, and of the form $p(x)=Ae^{Nf(x)}$, with $f$ as above. Then we must have $A=\sqrt{\frac{|f''(x_0)|N}{2\pi}}e^{-Nf(x_0)}$. 

\medskip

We now follow for example \cite{kaplunovsky} and write $x=x_0+\frac{y}{\sqrt{N}}$, and think of things as an expansion in $N^{-1/2}$ (as opposed to in $x-x_0$). We then have
\be
f(x)=f(x_0)+\frac{1}{2N} y^2 f''(x_0)+{\cal O}(N^{-3/2}),
\ee
i.e.
\be
Nf(x)=Nf(x_0)+\frac{1}{2} y^2 f''(x_0)+{\cal O}(N^{-1/2}).
\ee
This means
\be
e^{Nf(x)}=e^{Nf(x_0)}e^{\frac{1}{2} y^2 f''(x_0)}\left(1+{\cal O}(N^{-1/2})\right).
\ee
For a function $g(x)$ we also have
\be
g(x)=g(x_0)\left(1+\frac{y g'(x_0)}{g(x_0) \sqrt{N}}+{\cal O}(N^{-1})\right).
\ee
Using $p(x)=Ae^{Nf(x)}$, with $A=\sqrt{\frac{|f''(x_0)|N}{2\pi}}e^{-Nf(x_0)}$as given above, we then find to leading order in powers of $N^{-1/2}$,
\be
\int dx\, p(x) g(x) \approx \sqrt{\frac{|f''(x_0)|}{2\pi}} g(x_0)\int dy \,e^{\frac{1}{2} y^2 f''(x_0)},
\ee
 where we have shifted the integration variable from $x$ to $y$ (using $dx=dy/\sqrt{N}$). We then carry out the remaining Gaussian integral over $y$ (using the fact that $f''(x_0)<0$), and find
\be\label{eq:sp_avg_g}
\int dx\, p(x) g(x) \approx  g(x_0).
\ee

\medskip

\underline{Bottom line:} If we are doing averages against a normalised density of the form $p(x)=A e^{Nf(x)}$, with a global maximum at $x_0$, then these averages reduce to $g(x_0)$. I.e., we can just evaluate the function to be averaged at the `saddle point', $x_0$.

\medskip

We remind the reader again that we have here only considered real integrands, and that the technique known as Laplace's method, steepest descent, or saddle-point integration extends much further. Information be found in \cite{wong,debruijn,evans_mp,kaplunovsky}, and/or many textbooks or sets of lecture notes on asymptotic analysis and methods.

\section{The idea of generating functions and generating functionals}\label{sec:gf}
We now introduce the idea of generating functions (for static random variables), and then generating functionals (for stochastic processes in time).
\subsection{Generating functions for random variables}
\subsubsection{Scalar random variables}
Suppose $X$ is a real-valued random variable, with probability density $p(\cdot)$. The generating function is then defined as
\be\label{eq:gfrv}
Z(\psi)=\int dx ~p(x) e^{ix\psi}.
\ee
This is a function of the real variable $\psi$, and the function $Z(\cdot)$ is (up to a pre-factor) nothing else than the Fourier transform of $p(\cdot)$.

The generating function has the following property
\be
Z(\psi=0)=1, 
\ee
this follows from the normalisation $\int dx~ p(x)=1$. 

We also have
\be\label{eq:moments}
(-i)^n\left.\frac{d^n}{d\psi^n}Z(\psi)\right|_{\psi=0}=\avg{x^n},
\ee
where $\avg{f(x)}\equiv \int dx ~p(x) f(x)$ denotes averages of over $x$. This is why $Z(\cdot)$ is called the {\em (moment) generating} function. If the function $Z(\psi)$ is known, the moments of the random variable $X$ can be obtained from the derivatives of $Z(\psi)$ with respect to $\psi$, evaluated at $\psi=0$. Conversely, $Z$ can be expressed as its Taylor series,
\be
Z(\psi)=\sum_{n} \frac{(i\psi)^n}{n!} \avg{x^n}.
\ee
This can be seen from writing the exponential in Eq.~(\ref{eq:gfrv}) as its series expansion. Therefore, if all moments $\avg{x^n}$ are known, this fully determines $Z(\cdot)$, and hence the probability density $p(\cdot)$.

\subsubsection{Vector-valued random variables}
The definition of a generating function generalises to vector-valued random variables, say ${\mathbf X}=(x_1,\dots, x_N)$. We then define

\be\label{eq:gfv}
Z(\boldpsi)=\int dx_1 \cdots dx_N ~p(x_1,\dots,x_N) \exp\left(i\sum_{i=1}^N x_i\psi_i\right),
\ee
where $\boldpsi=(\psi_1,\dots,\psi_N)$. 

Again we have $Z(\boldpsi={\mathbf 0})=1$ from the normalisation of the distribution $p(x_1,\dots,x_N)$. Moments can now be obtained as follows

\be\label{eq:momentsvec}
\avg{x_1^{k_1}\cdots x_N^{k_N}}=(-i)^{k_1+\dots+k_N}
\left.\frac{\partial^{k_1+\dots +k_N}}{\partial \psi_1^{k_1}\cdots \partial \psi_N^{k_N}}\right|_{\boldpsi=0} Z(\boldpsi).
\ee
\subsection{Generating functionals for stochastic processes}\label{sec:stoch_proc}
\subsubsection{Discrete-time processes}
Suppose now that we have a discrete-time process $x(0), x(1), x(2),\dots $, i.e., the $x(t)$, $t=0,1,2,\dots$ are a sequence of random numbers.
\medskip

The details of the process are determined by the exact rules by which $x(t)$ evolves in time, and this results in an overall \underline{joint} distribution for the $x(t)$ ($t=0,1,\dots$). 
\medskip

The generating function for a scalar process in discrete time is very similar to the one in Eq.~(\ref{eq:gfv}) for a vector-valued random variable,

\be\label{eq:gfdp}
Z(\ulpsi)=\int d\ulx ~p(\ulx) \exp\left(i\sum_{t=0}^\infty x(t)\psi(t)\right).
\ee
We note that the source-field $\ulpsi$ is now a function (of the discrete time variable $t$). The notation $d\ulx$ is to be understood as
\be
d\ulx = \prod_{t=0,1, 2, \dots} dx(t).
\ee

\medskip

Suppose now that the process $x(\cdot)$ is Markovian. This means that the distribution of the random variable $x(t+1)$ is determined by the value $x(t)$, and that no information about $x(t')$ at earlier times $t'<t$ is required. We write this in the following form
\be\label{eq:discrete_time_process}
x(t+1)=g[x(t),\xi(t)].
\ee
This is to be read as follows. Assume, we have run the stochastic process up to time $t$, i.e., we have a particular realisation for $x(0), x(1),\dots, x(t)$. The next point $x(t+1)$ of this realisation is then obtained from the value of $x(t)$ and a random element $\xi(t)$. The nature of the function $g$ determines the details of the process.

\medskip

\underline{Example (unbiased random walk):}\\
Let us consider the unbiased random walk on the set of (positive and negative) integers. At each step the walker hops to the left with probability $1/2$, or to the right, also with probability $1/2$. This can be represented by a binary random variable $\xi(t)$, taking values $\xi(t)=\pm 1$, each with probability $1/2$, and by choosing $g[x(t),\xi(t)]=x(t)+\xi(t)$. The $\xi(t)$ are not correlated in time.
\medskip

Alternatively, $x(t)$ could be continuous (but we keep time discrete for the time being). For example, the walker could make steps with a size drawn from a Gaussian distribution. I.e., we'd again have $g[x(t),\xi(t)]=x(t)+\xi(t)$, but the $\xi(t)$ are now Gaussian random variables.

\medskip

For discrete-time processes Markovian processes of the form $x(t+1)=g[x(t),\xi(t)]$ the generating function turns into

\BE\label{eq:gfdp2}
Z(\ulpsi)&=&\int d\ulx ~d\ulxi ~p_0[x(t=0)] ~p(\ulxi) \nonumber \\
&&\times \left[\prod_{t=0}^\infty \delta[x(t+1)-g(x(t),\xi(t)]\right] \exp\left(i\sum_{t=0}^\infty x(t)\psi(t)\right).
\EE

The integral is over (i) the potentially random initial condition $x(0)$, (ii) all subsequent states at later times [$x(1),x(2),\dots$], and all realisations of the noise process $\ulxi$, i.e, over $\xi(0), \xi(1), \xi(2),\dots$. The notation $d\ulx$ stands for $dx(0) dx(1) dx(2) \dots$, and similarly for $d\ulxi$.
\medskip

The $\{x(t),\xi(t)\}$ are of course not independent from each other, instead we need to impose $x(t+1)=g[x(t),\xi(t)]$. That's what the product of delta functions in the first term on the second line in Eq.~(\ref{eq:gfdp2}) does. The last term finally is the standard `source term'. 

\medskip

We can now use Eq.~(\ref{eq:deltaexp}) to write 

\BE\label{eq:gfdp2b}
Z(\ulpsi)&=&\int d\ulxi \D\ulx ~\D\widehat\ulx ~p_0[x(0)] p(\ulxi) \nonumber \\
&&\times  \exp\left(i\sum_{t=0}^\infty \widehat x(t)\left\{x(t+1)-g[x(t),\xi(t)]\right\}\right)\nonumber \\
&&\times \exp\left(i\sum_{t=0}^\infty x(t)\psi(t)\right),
\EE
where we have introduced the notation 
\be
\D\ulx=\prod_{t=0,1,2,\dots} \frac{dx(t)}{\sqrt{2\pi}},
\ee
and similarly $\D\widehat\ulx=\prod_{t=0,1,2,\dots} \frac{d\widehat x(t)}{\sqrt{2\pi}}$.

Depending on the properties of the noise and on the nature of the function $g(\cdot,\cdot)$ this can be evaluated further. For example, let us consider the Gaussian random walk
\be
g[x(t),\xi(t)]=x(t)+\xi(t),
\ee
with independent Gaussian random variables $\xi(t)$, with mean $\mu$ and variance $\sigma^2$. We can then write $\xi(t)=\mu+\sigma z(t)$, with $z(t)$ independent standard Gaussian random variables with mean zero and unit variance.

Then
\BE\label{eq:gfdp3}
Z(\ulpsi)&=&\int \D\ulz \D\ulx \D\widehat\ulx ~p_0[x(0)] \exp\left(-\frac{1}{2}\sum_{t=0}^\infty z(t)^2\right) \exp\left(i\sum_{t=0}^\infty \widehat x(t)\left\{x(t+1)-x(t)-\mu-\sigma z(t)\right\}\right)\nonumber \\
&& ~~~~~~~~~~~~~~~~~~~~~~~~~~~~~\times \exp\left(i\sum_{t=0}^\infty x(t)\psi(t)\right),
\EE
with $\D\ulz = \prod_{t=0}^\infty \frac{dz(t)}{\sqrt{2\pi}}$.

Given that the $z(t)$ are all independent from one another, we can now carry out the average over each of the $z(t)$ separately. We have factors of the form $\int \frac{dz(t)}{\sqrt{2\pi}} \exp\left[-\frac{1}{2}z(t)^2-i\sigma\widehat x(t) z(t)\right]$. These can be evaluated using the identity in Eq.~(\ref{eq:gausslin1}), and we obtain
\be
\int \frac{dz(t)}{\sqrt{2\pi}} \exp\left(-\frac{1}{2}z(t)^2-i\sigma\widehat x(t) z(t)\right)=\exp\left(-\frac{1}{2}\sigma^2 \widehat x(t)^2\right).
\ee
The generating function in Eq.~(\ref{eq:gfdp3}) then `simplifies' to 
\BE\label{eq:gfdp4}
Z(\ulpsi)&=&\int  \D\ulx \D\widehat\ulx ~p_0[x(0)]\nonumber \\
&&\hspace{-3em}\times  \exp\left(i\sum_{t=0}^\infty \widehat x(t)\left\{x(t+1)-x(t)-\mu\right\}-\frac{1}{2}\sigma^2\sum_{t=0}^\infty\widehat x(t)^2+i\sum_{t=0}^\infty x(t)\psi(t)\right).
\EE
Now, this expression is admittedly not particularly enlightening, and not immediately useful. But at least we have a chance to practice the basic manipulation of generating functionals, and our skills in Gaussian integration.

\medskip

\underline{Exercise:}\\
By successively performing the integral over the $\{\widehat x(t)\}$, and then the $\{x(t)\}$, show that $Z(\ulpsi=\ulnull)=1$ for $Z(\ulpsi)$ in Eq.~(\ref{eq:gfdp4}).

\bigskip

\subsubsection{Continuous-time stochastic processes}
Let us now look at continuous-time processes. More specifically, we assume that $x(t)$ follows a stochastic differential equation (SDE) of the form
\be\label{eq:sde}
\dot x = g(x)+\sigma \xi(t),
\ee
where $\xi(t)$ is Gaussian white noise of mean zero, and unit variance, i.e., 
\be
\avg{\xi(t)}=0, ~~~\avg{\xi(t)\xi(t')}=\delta(t-t').
\ee

In order to set up the generating functional, we will discretise time into steps of size $\Delta t$. We then have
\be
x(t+\Delta)=x(t)+\Delta\times g[x(t)]+\sqrt{\Delta}\sigma z(t),
\ee
where the $z(t)$ are independent standard Gaussian random variables (mean zero, variance one). (If you are not familiar with the discretisation of stochastic differential equations I recommend the book by Jacobs `Stochastic Processes for Physicists' \cite{jacobs}).

We can read off the generating functional for this discretised version from Eq.~(\ref{eq:gfdp4}). All we need to do is to replace $\sigma^2\to \Delta \sigma^2$, and to adapt the term multiplying $\widehat x(t)$ in the curly bracket (this term comes from the deterministic part of the process). We find

\BE\label{eq:gfcp1}
Z(\ulpsi)&=&\int  \D\ulx \D\widehat\ulx ~p_0[x(0)]  \exp\left(i\sum_{t=0}^\infty \widehat x(t)\left\{x(t+\Delta)-x(t)-\Delta g[x(t)]\right\}\right. \nonumber \\
&&\hspace{5em}\left.-\frac{1}{2}\sigma^2\Delta\sum_{t=0}^\infty\widehat x(t)^2+i\sum_{t=0}^\infty x(t)\psi(t)\right),
\EE
After minor re-arrangements we then have
\BE\label{eq:gfcp2}
Z(\ulpsi)&=&\int  \D\ulx \D\widehat\ulx ~p_0[x(0)]  \exp\left(i\Delta\sum_{t=0}^\infty \widehat x(t)\left\{\frac{x(t+\Delta)-x(t)}{\Delta}- g[x(t)]\right\}\right.\nonumber \\
&&\hspace{5em}\left.-\frac{1}{2}\sigma^2\Delta\sum_{t=0}^\infty\widehat x(t)^2+i\sum_{t=0}^\infty x(t)\psi(t)\right),
\EE
As a next step, we re-name $\psi(t)\to \Delta \psi(t)$ (this is merely a question of choosing the `scale' for $\psi$). The expression for the generating functional then becomes
\BE\label{eq:gfcp3}
Z(\ulpsi)&=&\int  \D\ulx \D\widehat\ulx ~p_0[x(0)]  \exp\left(i\Delta\sum_{t=0}^\infty \widehat x(t)\left\{\frac{x(t+\Delta)-x(t)}{\Delta}- g[x(t)]\right\}\right.\nonumber \\
&&\hspace{5em}\left.-\frac{1}{2}\sigma^2\Delta\sum_{t=0}^\infty\widehat x(t)^2+i\Delta\sum_{t=0}^\infty x(t)\psi(t)\right),
\EE
We can now take the continuum limit $\Delta \to 0$ (we won't worry too much about what exactly happens to the measure $\D\ulx \D\widehat\ulx$ in this limit). Taking $\Delta\to 0$ turns the sums into integrals. We find
\BE\label{eq:gfcp3}
Z(\ulpsi)&=&\int  \D\ulx \D\widehat\ulx ~p_0[x(0)]  \exp\left(i\int_{0}^\infty dt~\widehat x(t)\left\{\dot x(t)- g[x(t)]\right\}\right.\nonumber \\
&&\hspace{5em}\left.-\frac{1}{2}\sigma^2\int_0^\infty dt~\widehat x(t)^2+i\int_{0}^\infty dt~x(t)\psi(t)\right).
\EE
\subsection{Correlation and response functions for discrete-time processes}\label{sec:corresp}
\subsubsection{Averages over realisations of the underlying stochastic process}
In the following we will write $\avg{\cdots}$ for averages over the  stochastic process we are studying. In this short subsection we'll briefly discuss how this average relates to the generating functional for the process. 

Take for example the discrete-time process in Eq.~(\ref{eq:discrete_time_process}), $x(t+1)=x(t)+g[x(t),\xi(t)]$. The average of a function(al) $f(\ulx)$ is then given by
\BE\label{eq:avg_f}
\avg{f(\ulx)}&=&\int d\ulxi \D\ulx \D\widehat\ulx ~p_0[x(0)] p(\ulxi)~f(\ulx) \nonumber \\
&&\times  \exp\left(i\sum_{t=0}^\infty \widehat x(t)\left\{x(t+1)-g[x(t),\xi(t)]\right\}\right).
\EE

The generating functional on the other hand is
\BE\label{eq:avg_gf}
Z[\ulpsi]&=&\int d\ulxi \D\ulx \D\widehat\ulx ~p_0[x(0)] p(\ulxi)\nonumber \\
&&\times   \exp\left(i\sum_{t=0}^\infty \widehat x(t)\left\{x(t+1)-g[x(t),\xi(t)]\right\}\right)\exp\left(i\int dt \psi(t) x(t)\right)
\EE
Suppose now that $f$ is a simple function(al), for example $f(\ulx)=x(t_1)x(t_2)$ for two fixed times $t_1$ and $t_2$. It is then clear that the average of $f$ can be generated from the generating functional as derivatives with respect to $\psi(t_1)$ and $\psi(t_2)$, and by subsequently setting $\ulpsi=0$. We'll make this a little more formal in the next subsections.

\medskip

To do this, we extend the definition of an `average' to include objects involving $\widehat \ulx$, and write for example
\BE\label{eq:avg_f_hat}
\avg{f(\ulx,\widehat\ulx)}&=&\int d\ulxi \D\ulx \D\widehat\ulx ~p_0[x(0)] p(\ulxi)~f(\ulx,\widehat\ulx) \nonumber \\
&&\times  \exp\left(i\int dt  \widehat x(t)\left\{\dot x(t)-g(x(t))-\sigma\xi(t)\right\}\right).
\EE

\subsubsection{Derivatives of the generating functional and correlation functions}

With $\avg{\dots}$ the average over realisations of the process, the generating function (or functional) is of the general form [see e.g. Eqs.~(\ref{eq:gfrv}) and (\ref{eq:gfdp})]
\be
Z[\ulpsi]=\avg{e^{i\ulx\cdot\ulpsi}},
\ee
where $\ulx\cdot\ulpsi$ represents $\int dt \psi(t)x(t)$ in the continuous case, and $\sum_t \psi(t) x(t)$ if time is discrete.
\medskip

We first focus on the discrete-time case, $Z(\ulpsi)=\avg{e^{i\sum_t \psi(t)x(t)}}$. It is then easy to see that moments of the $\{x(t)\}$ can be obtained by taking derivatives with respect to the $\{\psi(t)\}$, evaluated at $\ulpsi=0$. This is similar to what we have seen in Eq.~(\ref{eq:moments}) for scalar and in Eq.~(\ref{eq:momentsvec}) for vector-valued random variables. 

For example, we have
\be
\avg{x(t)}=-i\left.\frac{\partial}{\partial \psi(t)}Z[\ulpsi]\right|_{\ulpsi=\ulnull}
\ee
To see this, use the identity \be
\frac{\partial}{\partial \psi(t)} \sum_{t'} \psi(t')x(t')=x(t).
\ee

Similarly, we also have
\be
\avg{x(t)x(t')}=-\left.\frac{\partial^2}{\partial \psi(t)\partial \psi(t')}Z[\ulpsi]\right|_{\ulpsi=\ulnull}
\ee
This object is the correlation function of the process $x(\cdot)$.

\subsubsection{Response functions}
We now look at the discrete-time process in Eq.~(\ref{eq:discrete_time_process}), again, but we add a `perturbation field', $h(t)$, i.e., we start from
\be\label{eq:spperturb}
x(t+1)=g[x(t),\xi(t)]+h(t).
\ee
We would now like to know things such as `How does $x(t_2)$ react to a perturbation $h(t_1)$ at an earlier time?'

\medskip

The generating functional for the perturbed process can be obtained as a minor modification of Eq.~(\ref{eq:gfdp}):

\BE\label{eq:gfdp_h}
Z(\ulpsi,\ulh)&=&\int d\ulxi \D\ulx \D\widehat\ulx ~p_0[x(0)] p(\ulxi) \exp\left(i\sum_{t=0}^\infty \widehat x(t)\left\{x(t+1)-g[x(t),\xi(t)]-h(t)\right\}\right)\nonumber \\
&& \hspace{5em}\times\exp\left(i\sum_{t=0}^\infty x(t)\psi(t)\right),
\EE
Response functions are of the type
\be\label{eq:g_def}
G(t,t')=\frac{\partial }{\partial h(t')}\avg{x(t)}.
\ee
We recall that $\avg{\cdots}$ is an average over realisations of the stochastic process. The quantity $G(t,t')$ therefore describes how the field $x(t)$ `typically' reacts to a perturbation at time $t'$. More precisely, the matrix $\ullG$ captures linear response to small perturbations. By causality we must have $G(t,t')=0$ for $t'>t$.
\medskip

Using 
\BE\label{eq:avg_x}
\avg{x(t)}&=&\int d\ulxi \D\ulx \D\widehat\ulx ~p_0[x(0)] p(\ulxi) ~x(t)\nonumber \\
&& \times \exp\left(i\sum_{t''=0}^\infty \widehat x(t'')\left\{x(t''+1)-g[x(t''),\xi(t'')]-h(t'')\right\}\right)
\EE
and Eq.~(\ref{eq:g_def}) one sees that
\be
G(t,t')=-i\avg{x(t)\widehat x(t')}.
\ee

\subsubsection{Moments of conjugate variables vanish}\label{sec:avg_hat_eq_0}
We'll now derive an important fact, namely, averages of objects involving only conjugate fields vanish. We will use this at various points later on.

\medskip

We start by noting that
\be\label{eq:z_eq_1}
Z(\ulpsi=\ulnull,\ulh)=1
\ee
for any fixed choice of the perturbation field $\ulh$. This can either be seen by carrying out the $\widehat x$-integration in Eq.~(\ref{eq:gfdp_h}) (restoring the original $\delta$-functions), or perhaps easier directly from the definition $Z(\ulpsi)=\avg{e^{i\ulpsi\cdot\ulx}}$. This definition is valid for {\em any} stochasic process, so also the one in Eq.~(\ref{eq:spperturb}), including the field $\ulh$. (For any fixed choice of the function $h(t)$, we could absorb the perturbation field into $g$.) It then follows that $Z(\ulpsi=\ulnull,h)=\avg{1|\ulh}$, where the average $\avg{\cdots|\ulh}$ is over realisations of the stochastic process in Eq.~(\ref{eq:spperturb}) with the (fixed) perturbation field $h(t)$ included. Thus,  $Z(\ulpsi=\ulnull,h)$ is the average of one, and therefore equal to one.

\medskip

Eq.~(\ref{eq:z_eq_1}) in turn implies
\be
\frac{\partial}{\partial h(t)} Z(\ulpsi=\ulnull,\ulh)=0.
\ee
Using Eq.~(\ref{eq:gfdp_h}), the derivative on the right-hand side is recognised as $-i\avg{\widehat x(t)|\ulh}$, and we therefore conclude that $\avg{\widehat x(t)|\ulh}=0$. This is true for all choices of the perturbation field $h(\cdot)$, so in particular also if $h(t)\equiv 0$. We conclude
\be
\avg{\widehat x(t)}=0,~~\forall t.
\ee
Higher-order derivatives of $ Z(\ulpsi=\ulnull,\ulh)$ with respect to the components of $\ulh$ are also zero. Following similar steps as above, one then sees that {\em any} moment involving only conjugate variables $\widehat x (\cdot)$ vanishes, e.g.
\be
\avg{\widehat x(t)\widehat x(t')}=0.
\ee
Given that any (sensible) function $f$ of the $\{\widehat x(t)\}$ can be expressed in a Taylor series, we can therefore conclude that
\be
\avg{f(\{\widehat x(t)\}}=0
\ee
for all such functions.
\subsection{Continuous-time processes and functional derivatives}
The conventional partial derivative of a function of the form $f(a_1x_1+\dots+a_N x_N)$ is as follows,
\be\frac{\partial}{\partial x_i}f(a_1x_1+\dots+a_N x_N)=f'(a_1x_1+\dots+a_N x_N) a_i,
\ee
where $f'(y)=\frac{df(y)}{dy}$.

Imagine now that the index $i$ in $\sum_i a_i x_i$ is continuous (we'll call it $t$). Then we'd be dealing with an object of the form $f\left(\int dt a(t) x(t)\right)$. The functional derivative then works just as in the discrete case, it picks out the right `term' in the integral. We write this as
\be
\frac{\delta}{\delta x(t)} f\left(\int dt' a(t') x(t')\right)=f'\left(\int dt' a(t') x(t')\right) a(t).
\ee
For example
\be\frac{\delta}{\delta \psi(t)}\int D\ulx~ e^{g(\ulx)}~e^{i\int dt' \psi(t') x(t')}=i\int D\ulx~ e^{g(\ulx)}~e^{i\int dt' \psi(t') x(t')}x(t),
\ee
and
\be\frac{\delta}{\delta \psi(t)}\ln\int D\ulx~ e^{g(\ulx)}~e^{i\int dt' \psi(t') x(t')}=\frac{i\int D\ulx~ e^{g(\ulx)}~e^{i\int dt' \psi(t') x(t')}x(t)}{\int D\ulx~ e^{g(\ulx)}~e^{i\int dt' \psi(t') x(t')}}.
\ee

With this definition, the steps in Sec.~\ref{sec:corresp} can be repeated with modest modifications (amounting to replacing sums over $t$ with integrals, and the symbol $\partial$ with the symbol $\delta$). We arrive at
\be
\avg{x(t)x(t')}=-\left.\frac{\delta^2}{\delta \psi(t)\delta \psi(t')}Z[\ulpsi]\right|_{\ulpsi=\ulnull}
\ee
as well as
\be
G(t,t')=\frac{\delta}{\delta h(t')}\avg{x(t)} = - i \avg{x(t)\widehat x(t')}.
\ee
As in discrete time, we find that averages of objects involving only the conjugate field $\{\widehat x(t)\}$ vanish.
\section{Toy example of a generating-functional calculation: Interacting processes without disorder}\label{sec:toy_model}
We'll now use generating functionals to derive a dynamic mean-field description for a problem involving interacting stochastic processes. In this initial example, there is no disorder. This is much simpler, but nevertheless instructive. The full calculation for the disordered Lotka--Volterra problem then follows in Secs.~\ref{sec:gf_lv_1}-\ref{sec:gf_lv_3}.

\medskip

\begin{figure}[h!]
\begin{center}
\includegraphics[width=1\textwidth]{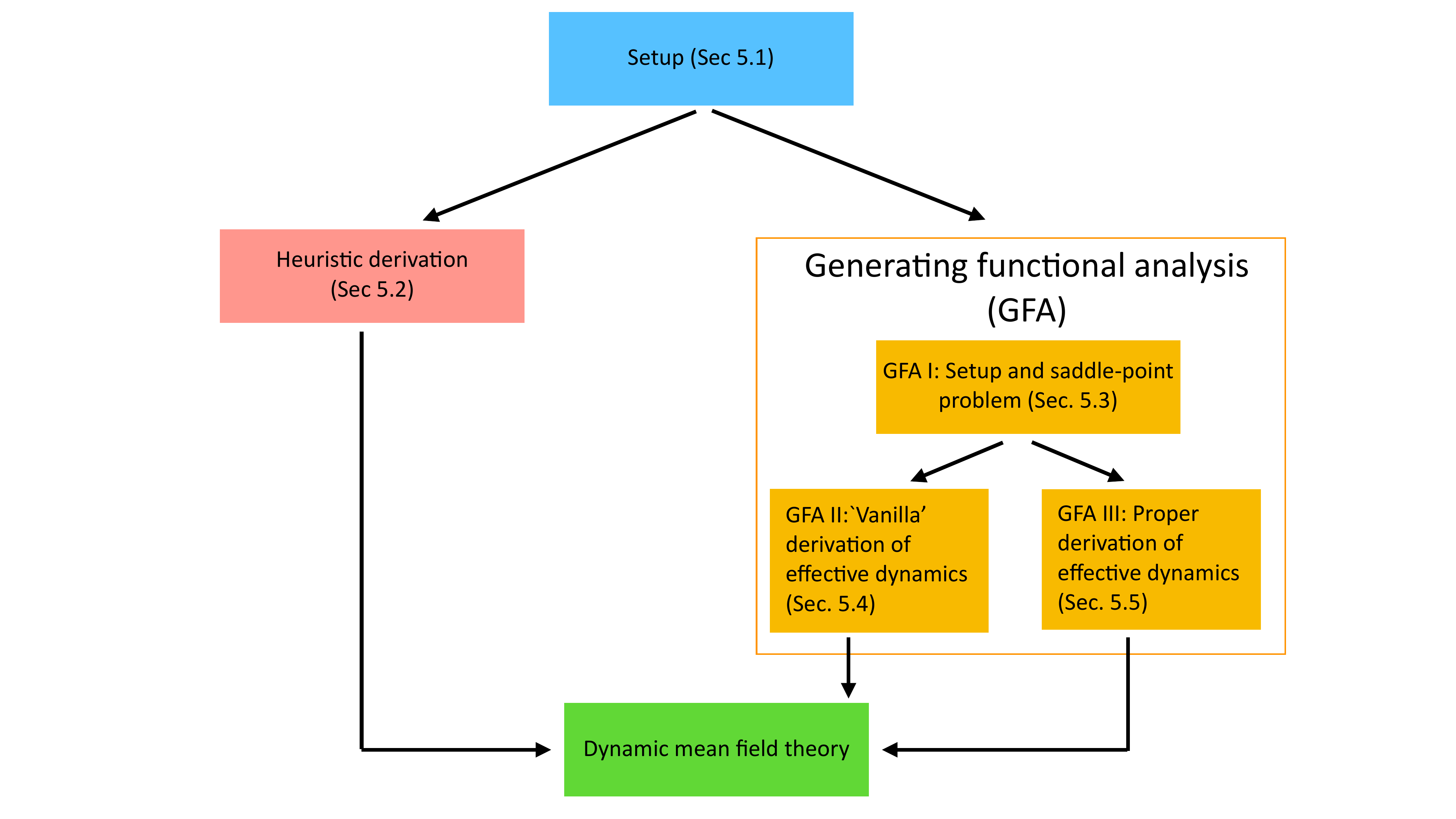}
\end{center}
\caption{Illustration of the structure of this section. We first introduce the model (\ref{sec:setup_m}). The rest of the section then describes three routes all leading to the same dynamic mean field process. The first approach is heuristic and does not use generating functionals (Sec.~\ref{sec:heuristic}). Alternatively one carries out a generating-functional analysis, leading to a saddle-point problem. In Sec.~\ref{sec:m_vanilla} we first describe a simpler, but defective method of extracting the effective dynamics from this saddle point. A more rigorous derivation can then be found in Sec.~\ref{sec:sp_M_full}. \label{fig:toy}}
\end{figure}

The structure of this section is as follows (see also the illustration in Fig.~\ref{fig:toy}). We set up the problem in Sec.~\ref{sec:setup_m}. A heuristic derivation of the dynamic mean field theory is then presented in Sec.~\ref{sec:heuristic}. Secs.~\ref{sec:gfa_1_m}-\ref{sec:sp_M_full} contain the generating-functional analysis of the model. More precisely, the generating functional is set up in Sec.~\ref{sec:gfa_1_m}. We present a first derivation of the dynamic mean-field process from the generating functional in Sec.~\ref{sec:m_vanilla}. One step of the derivation is somewhat sloppy, but nonetheless the overall direction is right, and it is instructive to look at this. A proper derivation is then presented in Sec.~\ref{sec:sp_M_full}. We note that all three approaches (heuristic derivation, sloppy analysis of the generating functional saddle point, and the proper saddle-point analysis) lead to the same dynamic mean field process.

\subsection{Setup}\label{sec:setup_m}
We choose the dynamics
\be\label{eq:intproc}
\dot x_i = a x_i + g\left(\frac{1}{N}\sum_j x_j\right)+\xi_i(t),
\ee
where $i=1,\dots,N$, and where $a$ is a constant, and $g(\cdot)$ can be any (real-valued) function. We have included Gaussian white noise variables, $\{\xi_i(t)\}$, with zero mean and variance $\sigma^2$, and no correlations between components, i.e. \be
\avg{\xi_i(t)\xi_j(t')}=\sigma^2\delta_{ij}\delta(t-t').
\ee
There is no requirement for the $\{x_i(t)\}$ to be positive.

\subsection{Heuristic derivation of the evolution equation for first moment in the thermodynamic limit}\label{sec:heuristic}
We first present a heuristic derivation of a closed equation for the relevant macroscopic dynamic order parameter. This is similar to the ad-hoc derivation of the mean-field approach in Sec.~\ref{sec:mft}. 

\medskip

We write $M(t)=\frac{1}{N}\sum_j x_j(t)$. From Eq.~(\ref{eq:intproc}) we then find
\be\label{eq:m0}
\dot M = a M (t)+g[M(t)]+\frac{1}{N}\sum_i \xi_i(t)
\ee
The quantity $M(t)$ is a random variable, because of two elements of randomness: One is the noise in Eq.~(\ref{eq:intproc}), i.e., the $\{\xi_i(t)\}$. The other comes (potentially) from the initial conditions for $\bx(t=0)$ which can also be random.

In the thermodynamic limit, $N\to\infty$, the noise in Eq.~(\ref{eq:m0}) averages out, we have $\frac{1}{N}\sum_i \xi_i(t)\to 0$ via the Central Limit Theorem. We then have the following deterministic evolution for $M(t)$,
\be\label{eq:m1}
 \dot M = a M (t)+g[M(t)],
 \ee
 with a potentially random initial condition for the $\{x_i(t=0)\}$. If the $\{x_i(t=0)\}$ are independent and identically distributed, $p[\bx(0)]=\prod_{i=1}^N p_0[x_i(0)]$, we find 
 \BE\label{eq:m2}
 M(0)&=&\lim_{N\to\infty}\frac{1}{N}\sum_i x_i(0) =\int dx~p_0(x) x,
 \EE
 via the law of large numbers, assuming the first moment of the distribution $p_0$ exists. Eqs.~(\ref{eq:m1}) and (\ref{eq:m2}) fully describe the time-evolution of the macroscopic order parameter $M(t)$.
  
\subsection{Generating-functional analysis I: Setup and saddle-point problem}\label{sec:gfa_1_m}
We now re-derive this via generating functionals. Of course, this is an overkill for this simple example, but it is useful to see the generating-functional method in action.
\subsubsection{Setting up the generating functional}
The dynamical generating functional is defined as
\be\label{eq:gf001}
Z[\ulbpsi]=\int D\ulbxi D\ulbx~ p(\ulbxi)\, p_0[\bx(0)]\,\delta(\mbox{equations of motion})e^{i\sum_i\int dt~ x_i(t)\psi_i(t) },
\ee
with $p(\ulbxi)$ the distribution of the noise variables, and $p_0[\bx(0)]$ the distribution of the initial conditions for the $\{x_i\}$ at time $t=0$. We have introduced the notation
\be
D\ulbx=\prod_{i=1}^N\prod_{t=0, 1, 2 ...} dx_i(t),
\ee
and similarly for $D\ulbxi$. When we say `equations of motion', we mean the dynamics in Eq.~(\ref{eq:intproc}), so
\be
\delta(\mbox{equations of motion})=\prod_{it}\delta\left[\dot x_i - a x_i + g\left(\frac{1}{N}\sum_j x_j\right)+\xi_i(t)\right].
\ee

\medskip

We now proceed with various manipulations in the generating functional. Expressing the delta-functions as Fourier transforms, we have
\BE
Z[\ulbpsi]  
&=& \int D\ulbxi \D\ulbx  \D\widehat\ulbx \, p(\ulbxi)\,p_0[\bx(0)]\,\exp\left(i\sum_i\int dt~ x_i(t)\psi_i(t)\right) \nonumber \\
&&\times \exp\Bigg\{i\sum_i\int dt \Bigg[\widehat x_i(t)\Big(\dot x_i(t) -ax_i(t) - g\left( \frac{1}{N}\sum_j x_j(t)\right)-\xi_i(t)\Big)\Bigg]\Bigg\},
\label{eq:gf002}
\EE
where
\be
\D\ulbx=\prod_{i=1}^N\prod_{t=0, 1, 2 ...} \frac{dx_i(t)}{\sqrt{2\pi}}.
\ee
\subsubsection{Introduction of macroscopic order parameter}
Next, for any fixed time $t$, we can insert unity in the following form
\be\label{eq:one}
1=\int_{-\infty}^\infty \frac{dM(t)
d\widehat M(t)}{2\pi/N} ~ e^{i N\widehat M(t) \left[M(t)-N^{-1}\sum_i x_i(t)\right]}
\ee
This may look a little weird at first. The purpose of this is to introduce the macroscopic order parameter $M(t)=\frac{1}{N}\sum_{i} x_i(t)$ into the calculation. It is important to realise that carrying out the integration over $\widehat M(t)$ reduces 
Eq.~(\ref{eq:one}) to $1=\int dM(t)\, \delta\left(M(t)-N^{-1}\sum_i x_i(t)\right)$ [see also Eq.~(\ref{eq:deltaexp})].

Inserting Eq.~(\ref{eq:one}) into the generating functional for all times $t$, we get
\BE
Z[\boldpsi]
&=&\int D\ulbxi \,\D\ulbx  \D\widehat\ulbx\, \D[\ulM, \widehat\ulM] \, p(\ulbxi)\,p_0[\bx(0)]\,\exp\left(i\int dt\,  \widehat M(t) \left[NM(t)-\sum_i x_i(t)\right]\right) \nonumber \\ 
&& \exp\Bigg\{i\sum_i\int dt \Bigg[\widehat x_i(t)\Big(\dot x_i(t) -ax_i(t) - g\left( \frac{1}{N}\sum_j x_j(t)\right)-\xi_i(t)\Big)\Bigg]\Bigg\} \nonumber \\
&&\times\exp\left(i\sum_i\int dt~ x_i(t)\psi_i(t)\right),
\label{eq:gf003}
\EE
with the notation
\be
\D[\ulM, \widehat\ulM]=\prod_{t=0, 1, 2, \dots} \frac{dM(t)d\widehat M(t)}{2\pi/N}.
\ee
\medskip

Given that the $\widehat \ulM$-integration leads to delta functions, $\delta\left[NM(t)-\sum_i x_i(t)\right]$, for all times $t$, we can replace $N^{-1}\sum_j x_j(t)$ with $M(t)$ in the second exponential. This gives
\BE
Z[\boldpsi]
&=&\int D\ulbxi\, \D\ulbx\,  \D\widehat\ulbx\, \D[\ulM, \widehat\ulM]\, p(\ulbxi)\,p_0[\bx(0)]\, \exp\left(iN\int dt\,  \widehat M(t) \left[M(t)-N^{-1}\sum_i x_i(t)\right]\right) \nonumber \\ 
&&\times \exp\Bigg\{i\sum_i\int dt \Bigg[\widehat x_i(t)\Big(\dot x_i(t) -a x_i(t) - g[M(t)]-\xi_i(t)\Big)\Bigg]\Bigg\}\nonumber \\
&& \times \exp\left(i\sum_i\int dt~ x_i(t)\psi_i(t)\right). \nonumber \\
\label{eq:gf004}
\EE
From now on, we assume that the random variables $x_i(0)$ ($i=1,\dots,N$) are independent and identically distributed ({\em iid}). The distribution $p_0[\bx(0)]$ then factorises, and in slight abuse of notation (double use of $p_0$), we write it in the form $p_0[\bx(0)]=\prod_i p_0[x_i(0)]$.

\subsubsection{Formulation as a saddle-point problem}
We now re-organise the terms in the generating functional in Eq.~(\ref{eq:gf004}). More precisely, we separate the contributions containing only macroscopic objects [i.e., $M(t)$ and $ \widehat M(t)$] from the terms involving individual microscopic variables, $x_i(t)$ and $\widehat x_i(t)$. 

\medskip

We write the generating functional in the following form
\BE\label{eq:gf_sp}
Z[\bulpsi]
&=&\int \D[\ulM, \widehat\ulM] \exp\left(N\left(\Psi[\ulM,\widehat \ulM]+\Omega[\ulM,\widehat \ulM,\ulbpsi]\right)\right),
\EE
with
\be\label{eq:def_Psi}
\Psi[\ulM,\widehat \ulM]=i\int dt\,  \widehat M(t)M(t),
\ee
and
\BE
\Omega[\ulM,\widehat \ulM,\bulpsi]&=&\frac{1}{N}\sum_i \ln \Bigg\{\int \D\ulxi_i \D\ulx_i \D\widehat \ulx_i ~p(\ulxi_i)\,p_0[x_i(0)] \nonumber \\
&& \times \exp\left[i\int dt \left[\widehat x_i(t)\left(\dot x_i(t) -a x_i(t) - g[M(t)]-\xi_i(t)\right)\right]\right]\nonumber \\
&&\times\exp\left[-i\int dt \widehat M(t) x_i(t)\right]\exp\left[i\int dt ~x_i(t)\psi_i(t)\right]\Bigg\},\label{eq:def_Omega}
\EE
where we have written $\D \ulx_i=\prod_{t=0, 1, 2, \dots} \frac{dx_i(t)}{\sqrt{2\pi}}$.

The variables $\xi_i(t)$ are dummy variables, to be integrated over, so we can drop the index $i$. We can then write
\BE
\Omega[\ulM,\widehat \ulM,\bulpsi]&=&\frac{1}{N}\sum_i \ln \Bigg\{\int \D\ulxi_i \D\ulx_i \D\widehat \ulx_i ~p(\ulxi)\,p_0[x_i(0)] \nonumber \\
&& \times \exp\left[i\int dt \left[\widehat x_i(t)\left(\dot x_i(t) -a x_i(t) - g[M(t)]-\xi(t)\right)\right]\right]\nonumber \\
&&\times\exp\left[-i\int dt \widehat M(t) x_i(t)\right]\exp\left[i\int dt ~x_i(t)\psi_i(t)\right]\Bigg\}.\label{eq:def_Omega_2}
\EE

\subsection{Generating functional analysis II: `Vanilla' derivation of dynamic mean field process}\label{sec:m_vanilla}
We will now proceed with the evaluation of the saddle point problem in Eq.~(\ref{eq:gf_sp}). For pedagogical reasons I will first present a simple, but slightly incorrect version of this. I call this the `vanilla' version of the calculation (if you prefer you can call it `light' or `lite'). The ideas used in this simplified calculation are broadly right, and so is the final result, but some steps along the way are not entirely justified. Nonetheless, I think it is useful to first present this simplified procedure (which, as we will see, is quite involved already). This is so that the a beginner in this area can develop a general understanding, and see where all this is going.  I'll then say what wasn't so precise (Sec.~\ref{sec:problems}), and in Sec.~\ref{sec:sp_M_full}, I'll present a refined version of the calculation.

\subsubsection{Saddle-point equations for the order parameters}

OK, here now the `light' version of the saddle-point calculation. The first simplification we make is to restrict the discussion to source fields $\psi_i(t)=\psi(t)$ for all $i$. There is nothing fishy about this, it just restricts the values of $\ulpsi$ for which we evaluate the generating functional. 

\medskip

Making this simplification makes all terms in the sum in Eq.~(\ref{eq:def_Omega_2}) equal, and we can drop the index $i$ in $x_i$ and $\widehat x_i$. The expression for $\Omega$ reduces to
\BE
\Omega[\ulM,\widehat \ulM,\bulpsi]&=&\ln  \int D\ulxi\, \D\ulx\, \D\widehat \ulx ~p(\ulxi)\,p_0[x(0)] \nonumber \\
&& \times \exp\left[i\int dt \left[\widehat x(t)\left(\dot x(t) -a x(t) - g[M(t)]-\xi(t)\right)\right]\right]\nonumber \\
&&\times\exp\left[-i\int dt \widehat M(t) x(t)\right]\exp\left[i\int dt ~x(t)\psi(t)\right].\label{eq:def_Omega_3}
\EE

\medskip

We want to compute the integral in Eq.~(\ref{eq:gf_sp}). We do this using Laplace's method (Sec.~\ref{sec:saddlepoint}). 

\medskip

We extremise the quantity $\Psi+\Omega$ in terms of $\ulM$ and $\widehat\ulM$, i.e., we set
\BE
\frac{\delta}{\delta M(t)} \left(\Psi[\ulM,\widehat \ulM]+\Omega[\ulM,\widehat \ulM,\bulpsi]\right)&=&0,\nonumber \\
\frac{\delta}{\delta \widehat M(t)} \left(\Psi[\ulM,\widehat \ulM]+\Omega[\ulM,\widehat \ulM,\bulpsi]\right)&=&0,
\EE
for all $t$.

\medskip

From these two conditions and the definitions of $\Psi$ and $\Omega$ in Eqs.~(\ref{eq:def_Psi}) and (\ref{eq:def_Omega}) we get
\BE\label{eq:scmmhat}
i\widehat M(t)&=&ig'[M(t)]\avg{\widehat x(t)}_\Omega, \nonumber \\
M(t)&=&\avg{x(t)}_\Omega,
\EE
where $g'(y)=dg(y)/dy$, and where the `average' $\avg{\cdots}_\Omega$ of any functional $f(\ulx,\widehat\ulx)$ is defined as
\BE
\avg{f(\ulx, \widehat \ulx)}_\Omega &=& \frac{1}{Z_1(\ulpsi)}\int D\ulxi \D\ulx \D\widehat \ulx ~p_0[x(0]\,p(\ulxi)~f(\ulx, \widehat\ulx) \nonumber \\
&&\times
\exp\left[i\int dt \left[\widehat x(t)\left(\dot x(t) -a x(t) - g[M(t)]-\xi(t)\right)\right]\right]\nonumber \\
&&\times\exp\left[-i\int dt \widehat M(t) x(t)\right]\exp\left[i\int dt ~x(t)\psi(t)\right],\label{eq:avomega}
\EE
with
\BE
Z_1(\ulpsi)&=&\int D\ulxi \D\ulx \D\widehat \ulx ~p_0[x(0]\,p(\ulxi)~
\exp\left[i\int dt \left[\widehat x(t)\left(\dot x(t) -a x(t) - g[M(t)]-\xi(t)\right)\right]\right]\nonumber \\
&&\times\exp\left[-i\int dt \widehat M(t) x(t)\right]\exp\left[i\int dt ~x(t)\psi(t)\right].\label{eq:zeff}
\EE
\subsubsection{Elimination of hatted order parameter leads to single effective particle process}\label{sec:vanilla_hatm_eq_0}
Next we use the result of Sec.~\ref{sec:avg_hat_eq_0}, namely that averages of hatted variables are zero. This means that $\avg{\widehat x(t)}_\Omega=0$. Using the first relation in Eq.~(\ref{eq:scmmhat}) we then find that $\widehat M(t)=0$. 

\medskip

This leads to
\BE
\left.Z_{1}(\ulpsi)\right|_{\widehat\ulM=0}&=&\int D\ulxi \D\ulx \D\widehat \ulx ~p_0[x(0)]~p(\ulxi) \nonumber \\
&&\times
\exp\left[i\int dt \left[\widehat x(t)\left(\dot x(t) -a x(t) - g[M(t)]-\xi(t)\right)\right]\right]\nonumber \\
&&\times\exp\left[i\int dt ~x(t)\psi(t)\right].\label{eq:zeff_h_M_eq_0}
\EE

This is recognised as the generating functional of the dynamics
\be\label{eq:dotx_simple}
\dot x(t) =a x(t) + g[M(t)]+\xi(t).
\ee
We will refer to this as an {\em effective single-particle process}. This is because the process describes the `effective dynamics' a typical particle `sees' in this system. It is a single-particle process, as there is no site index $i$ any longer.

\medskip

From  the second relation in Eq.~(\ref{eq:scmmhat}) we have the self-consistency relation $M(t)=\avg{x(t)}_\Omega$ . Using $Z_1[\ulpsi=\ulnull]=1$ (for $\widehat M(t)\equiv 0$) we have
\BE
\avg{f(\ulx, \widehat \ulx)}_\Omega \bigg|_{\widehat\ulM=\ulnull, ~\ulpsi=\ulnull}&=&\int D\ulxi \D\ulx \D\widehat \ulx ~p_0[x(0]\,p(\ulxi)~f(\ulx, \widehat\ulx) \nonumber \\
&&\times
\exp\left[i\int dt \left[\widehat x(t)\left(\dot x(t) -a x(t) - g[M(t)]-\xi(t)\right)\right]\right].\label{eq:avomega_hatm=0}
\EE
This describes averages over realisations of the process in Eq.~(\ref{eq:dotx_simple}).

\medskip

Averaging both sides Eq.~(\ref{eq:dotx_simple}) over realisations (of $\xi$ and of the initial value $x(0)$), i.e., carrying out the operation $\avg{\cdots}_\Omega \big|_{\widehat\ulM=\ulnull, ~\ulpsi=\ulnull}$, we find
\be\label{eq:dotM_gf}
\dot M = a M + g(M),
\ee
i.e., we recover the result of our ad-hoc calculation in Eq.~(\ref{eq:m1}).

\subsubsection{The problems with this simplified calculation}\label{sec:problems}
Now, why do I call this the `vanilla' version, or saddle point `light'? There are several issues which we have swept under the carpet.

\bigskip

\underline{Elimination of $\widehat M(t)$ not properly justified:}\\
The problem is that the justification for the step from Eq.~(\ref{eq:zeff}) to (\ref{eq:zeff_h_M_eq_0}) above is not really right. Don't get me wrong, it is true that $\widehat M(t)=0$ at the saddle point (as we will see in Sec.~\ref{sec:sp_M_full}), and the result in Eq.~(\ref{eq:dotM_gf}) is also right. But the way we justified this is not complete. 

We simply quoted Sec.~\ref{sec:avg_hat_eq_0} (`averages of hatted quantities are zero') to assert that $\avg{\widehat x(t)}_\Omega=0$, and then used Eq.~(\ref{eq:scmmhat}) to conclude that $\widehat M(t)=0$. 

The problem is that an average of a conjugate variable (hatted quantity) only vanishes if the average is over realisations of a {\em bona fide} stochastic process. But this is not clear in Eq.~(\ref{eq:avomega}). The object $\avg{\cdots}_\Omega$ in Eq.~(\ref{eq:avomega}) would be an average over an actual stochastic dynamics if the term $\exp\left[-i\int dt \widehat M(t) x(t)\right]$ wasn't there, i.e., if all $\widehat M(t)$ were zero (additionally we would have to set $\psi(t)\equiv 0$). If this was the case then indeed  $\avg{\cdots}_\Omega$ would be an average over the process in Eq.~(\ref{eq:dotx_simple}), and therefore $\avg{\widehat x(t)}_\Omega$ would be zero. But we used $\avg{\widehat x(t)}_\Omega=0$ to argue that $\widehat M(t)=0$, and we need $\widehat M(t)$ to vanish to show that $\avg{\widehat x(t)}_\Omega=0$. So our argument is circular.  

\medskip

This doesn't mean that everything is wrong though. When we make the step from Eq.~(\ref{eq:zeff}) to (\ref{eq:zeff_h_M_eq_0}) we could have simply made an ansatz and postulated that $\widehat M(t)\equiv 0 $ for all $t$. We could then have justified this {\em a posteriori}, at the end of the calculation.

\bigskip

\underline{Generating functional $Z_1[\psi]$ for single effective particle not valid for non-zero $\psi(t)$}\\
This is related to the previous issue. In order to identify the operation $\avg{\cdots}_\Omega$ as an average over a stochastic process we need the source field to vanish ($\ulpsi(t)\equiv 0$). But we then used the expression in Eq.~(\ref{eq:zeff_h_M_eq_0}) to argue that this is the generating functional of the right single effective particle dynamics. In order to make this final step, $\ulpsi$ has to be allowed to take general values, resulting in another gap in our `vanilla' derivation.
\medskip

This is not a serious issue either. The measure in Eq.~(\ref{eq:avomega_hatm=0}) does not contain the source field, and it describes the right effective process.

\bigskip

\underline{Physical meaning of $M(t)$ not fully clear}:\\
There is a further subtlety. We introduced the object $M(t)=\frac{1}{N}\sum_{i} x_i(t)$ in the course of our calculation, but this was done by inserting unity written in some complicated way  [Eq.~(\ref{eq:one})]. At that point $M(t)$ is mostly a mathematical short-hand that we used in order not have to write $\frac{1}{N}\sum_{i} x_i(t)$ all the time. It is also important to note that $M(t)$, in that form, is a stochastic quantity, and only becomes non-random at the saddle point in the limit $N\to\infty$. It is natural to assume that the saddle-point value of $M(t)$ is the same as $\lim_{N\to\infty} \frac{1}{N}\sum_i \avg{x_i(t)}$ in the original problem, where $\avg{\cdots}$ (without the subscript $\Omega$) is an average over realisations of the process in Eq.~(\ref{eq:intproc}). This is indeed the case, but so far we have not formally proved this.

\medskip

We address these points in the next section.

\subsection{Generating functional analysis III: Proper derivation of dynamic mean-field process}\label{sec:sp_M_full}
\subsubsection{Generating functional for the problem with additional perturbation fields}
We return to the original problem [Eq.~(\ref{eq:intproc})], and add perturbation fields $h_i(t)$,
\be\label{eq:intproc_h}
\dot x_i(t)= a x_i(t) + g\left(\frac{1}{N}\sum_j x_j(t)\right)+\xi_i(t)+h_i(t).
\ee
These fields are not part of the model, and we will set them to zero eventually. But we'll need them along the way as a mathematical tool (to generate moments of the conjugate field). 

\medskip

The generating functional for the system with perturbation fields is
\BE\label{eq:gf_sp_f}
Z[\bulpsi]
&=&\int \D[\ulM, \widehat\ulM] \exp\left(N\left(\Psi[\ulM,\widehat \ulM]+\Omega[\ulM,\widehat \ulM,\ulbpsi]\right)\right),
\EE
with
\be\label{eq:def_Psi}
\Psi[\ulM,\widehat \ulM]=i\int dt\,  \widehat M(t)M(t),
\ee
and
\BE
\Omega[\ulM,\widehat \ulM,\bulpsi]&=&\frac{1}{N}\sum_i \ln \Bigg\{\int D\ulxi_i \D\ulx_i \D\widehat \ulx_i ~p(\ulxi)\,p_0[x_i(0)] \nonumber \\
&& \times \exp\left[i\int dt \left[\widehat x_i(t)\left(\dot x_i(t) -a x_i(t) - g[M(t)]-\xi_i(t)-h_i(t)\right)\right]\right]\nonumber \\
&&\times\exp\left[-i\int dt \widehat M(t) x_i(t)\right]\exp\left[i\int dt ~x_i(t)\psi_i(t)\right]\Bigg\}.\label{eq:def_Omega_f}
\EE

At the saddle point we again obtain
\BE\label{eq:scmmhat_f}
i\widehat M(t)&=&ig'[M(t)]\avg{\widehat x(t)}_\Omega, \nonumber \\
M(t)&=&\avg{x(t)}_\Omega,
\EE
but the average $\avg{\cdots}_\Omega$ is now defined as
\BE
\avg{f(\ulx, \widehat \ulx)}_\Omega &=& \lim_{N\to\infty}\frac{1}{N}\sum_{i} \frac{1}{Z_1(\ulpsi_i)}\int D\ulxi \D\ulx \D\widehat \ulx ~p_0[x(0]\,p(\ulxi)~f(\ulx, \widehat\ulx) \nonumber \\
&&\times
\exp\left[i\int dt \left[\widehat x(t)\left(\dot x(t) -a x(t) - g[M(t)]-\xi(t)-h_i(t)\right)\right]\right]\nonumber \\
&&\times\exp\left[-i\int dt \widehat M(t) x(t)\right]\exp\left[i\int dt ~x(t)\psi_i(t)\right],\label{eq:avomega_f}
\EE
with
\BE
Z_1(\ulpsi_i)&=&\int D\ulxi \D\ulx \D\widehat \ulx ~p_0[x(0]\,p(\ulxi)~
\exp\left[i\int dt \left[\widehat x(t)\left(\dot x(t) -a x(t) - g[M(t)]-\xi(t)-h_i(t)\right)\right]\right]\nonumber \\
&&\times\exp\left[-i\int dt \widehat M(t) x(t)\right]\exp\left[i\int dt ~x(t)\psi_i(t)\right].\label{eq:zeff_f}
\EE

\subsubsection{Physical meaning of the order parameter $M(t)$ at the saddle point}
We pointed out earlier that the physical interpretation of the saddle-point value of the order parameter $M(t)=\frac{1}{N}\sum_{i} x_i(t)$ is not entirely clear. To understand this better we introduce
\be\label{eq:def_m}
m(t)=\lim_{N\to\infty} \frac{1}{N}\sum_{i=1}^N \avg{x_i(t)},
\ee
where the average on the right is an average over realisations of the stochastic process for the $x_i$ in Eq.~(\ref{eq:intproc_h}). This means averaging over realisations of the dynamic noise variables $\xi_i(t)$ and over the initial conditions for the $x_i(0)$, if they are random. The object $m(t)$ has a clear interpretation in the original problem. 

\medskip

We will now show that $m(t)$ and $M(t)$ are the same at the saddle point and in the thermodynamic limit.

\medskip

From Eq.~(\ref{eq:gf001}) we have
\be\label{eq:gf001_f}
Z[\ulbpsi]=\int D\ulbxi D\ulbx~ p(\ulbxi)\, p_0[\bx(0)]\,\delta(\mbox{equations of motion})e^{i\sum_i\int dt~ x_i(t)\psi_i(t) }.
\ee
This means that $m(t)$ can be expressed in the form
\be\label{eq:emm_f}
m(t)=-i\lim_{N\to\infty} \frac{1}{N}\sum_{i=1}^N \left.\frac{\delta Z[\ulbpsi]}{\delta \psi_i(t)}\right|_{\ulbpsi=\ulbnull}.
\ee
One the other hand $Z[\bulpsi]$ is given by the expression in Eq.~(\ref{eq:gf_sp_f}), so that

\BE
m(t)&=&-i\lim_{N\to\infty} \frac{1}{N}\sum_{i=1}^N \left.\frac{\delta Z[\ulbpsi]}{\delta \psi_i(t)}\right|_{\ulbpsi=\ulbnull}\nonumber \\
&=&-i\lim_{N\to\infty}\left.\frac{1}{N}\sum_i \int \D[\ulM, \widehat\ulM] \left\{\exp\left[N\left(\Psi[\ulM,\widehat \ulM]+\Omega[\ulM,\widehat \ulM,\ulbpsi]\right)\right] \,\, N \frac{\delta \Omega[\ulM,\widehat \ulM,\ulbpsi]}{\delta \psi_i(t)}\right\}\right|_{\ulbpsi=\ulbnull}\nonumber \\
&=&-i\lim_{N\to\infty}\sum_i \left.\frac{\delta \Omega[\ulM,\widehat \ulM,\ulbpsi]}{\delta \psi_i(t)}\right|_{\ulbpsi=\ulbnull, \mbox{SP}} \nonumber \\
&=&\avg{x(t)}_\Omega\Big|_{\ulbpsi=\ulbnull, \mbox{SP}}.
\EE
The subscript `SP' indicates that the relevant quantity is to be evaluated at the saddle point. In the third step we have used the rules of saddle-point integration, in particular Eq.~(\ref{eq:sp_avg_g}). In the last step we have used the expression for $\Omega$ in Eq.~(\ref{eq:def_Omega_f}), and the definition of $\avg{\cdots}_\Omega$ in Eqs.~(\ref{eq:avomega_f}) and (\ref{eq:zeff_f}).
 
 \medskip
 
 From the second relation in Eq.~(\ref{eq:scmmhat_f}) we know that $M(t)=\avg{x(t)}_\Omega$ at the saddle point. We can therefore conclude that $m(t)$ and $M(t)$ are equal to one another at the saddle point and in the limit $\ulbpsi\to \ulbnull$.

\subsubsection{Proper derivation of the fact that $\widehat M(t)=0$ at the saddle point}\label{sec:M_eq_0_proper}
The source field $\ulbpsi$ has fulfilled its purpose, and throughout this section we will set the $\psi_i(t)$ to zero for all $i$ and $t$.

\medskip

We start by defining the object
\be
p(t)=i\lim_{N\to\infty}\frac{1}{N}\sum_i  \frac{\delta}{\delta h_i(t)}\avg{1}.
\ee
The average $\avg{\cdots}$ on the right is again an average over realisations of the stochastic process in Eq.~(\ref{eq:intproc_h}). This process in turn involves the external perturbation fields $h_i(t)$, i.e., the average $\avg{\cdots}$ is different for different choices of $\ulbh$. As such one can define derivatives of the type  $\frac{\delta}{\delta h_i(t')}\avg{x_i(t)}$ for example. But of course, the average of the quantity `one' over a physical process is always one, $\avg{1}=1$. This is the case here, no matter what the fields $\ulh_i$ are. Therefore, we trivially have
\be
p(t)=0,
\ee
for all $t$. 

It may sound odd to introduce an object when we already know that it is zero. The point is though that we will now show that $p(t)=\avg{\widehat x(t)}_\Omega$ at the saddle point. This then allows us to conclude that $\avg{\widehat x(t)}_\Omega=0$ (always assuming $\ulbpsi=\ulbnull$ ). Showing that $\avg{\widehat x(t)}_\Omega=0$ was one of the problems we had identified in Sec.~\ref{sec:problems}.

\medskip

We know that $Z[\bulpsi=\ulbnull]=1$, no matter what form the perturbation fields $h_i(t)$ take. Therefore $p(t)=i\lim_{N\to\infty}\frac{1}{N}\sum_i  \frac{\delta}{\delta h_i(t)}Z[\bulpsi=\ulbnull]$ (simply because the objects on both sides of this equation are zero). 

\medskip

Using the expression for $Z[\bulpsi]$ in Eq.~(\ref{eq:gf_sp_f}), we have
\BE
p(t)&=&i\lim_{N\to\infty}\left.\frac{1}{N}\sum_i \int \D[\ulM, \widehat\ulM] \left\{\exp\left(N\left(\Psi[\ulM,\widehat \ulM]+\Omega[\ulM,\widehat \ulM,\ulbpsi]\right)\right) \times N \frac{\delta \Omega[\ulM,\widehat \ulM,\ulbpsi]}{\delta h_i(t)}\right\}\right|_{\ulbpsi=\ulbnull}\nonumber \\
&=&i\lim_{N\to\infty}\sum_i \left.\frac{\delta \Omega[\ulM,\widehat \ulM,\ulbpsi]}{\delta h_i(t)}\right|_{\ulbpsi=\ulbnull,\mbox{\footnotesize SP}} \nonumber \\
&=&\left.\avg{\widehat x(t)}_\Omega\right|_{\ulbpsi=\ulbnull,\mbox{\footnotesize SP}}.
\EE
Given that we already know that $p(t)=0$, we have thus shown that $\avg{\widehat x(t)}_\Omega=0$ at the saddle point and in absence of the source field $\ulbpsi$.  Using the first relation in Eq.~(\ref{eq:scmmhat_f}) this then also implies that $\widehat M(t)=0$ for all $t$.
\subsubsection{Single effective particle measure and effective process}
We now set  the source and perturbation fields to zero ($h_i(t)=0, \psi_i(t)=0$ for all $i,t$), as they are no longer needed. We also operate at the saddle point. We can then use $\widehat M(t)=0$ for all $t$ in Eqs.~(\ref{eq:avomega_f}) and (\ref{eq:zeff_f}). Instead of $\avg{\cdots}_\Omega$ we write $\avg{\cdots}_\star$.  

\medskip

Noting that all terms in the sum over $i$ in Eq.~(\ref{eq:avomega_f}) become the same, and that $Z_1[\ulpsi_i=\ulnull]=1$, we then have
\BE
\avg{f(\ulx, \widehat \ulx)}_\star&=& \int D\ulxi \D\ulx \D\widehat \ulx ~p_0[x(0]\,p(\ulxi)~f(\ulx, \widehat\ulx) \nonumber \\
&&\times
\exp\left[i\int dt \left[\widehat x(t)\left(\dot x(t) -a x(t) - g[M(t)]-\xi(t)\right)\right]\right], \label{eq:avstar_f}
\EE
together with the following self-consistency relation [resulting from the second relation in Eqs.~(\ref{eq:scmmhat_f})],
\be\label{eq:M_avg_proper}
M(t)=\avg{x(t)}_\star.
\ee
The integral over the conjugate variables in Eq.~(\ref{eq:avstar_f}) can, in principle, be undone to produce delta functions of the type $\delta[\dot x(t) -a x(t) - g[M(t)]-\xi(t)]$. Therefore, $\avg{\cdots}_\star$ describes an average over realisations of the process
\be\label{eq:eff_process_proper}
\dot x = a x + g(M) +\xi,
\ee
which is what we wanted to show.
\subsection{Summary of the main steps of the generating functional analysis}\label{sec:toy_summary}
\begin{figure}[h!]
\begin{center}
\includegraphics[width=1\textwidth]{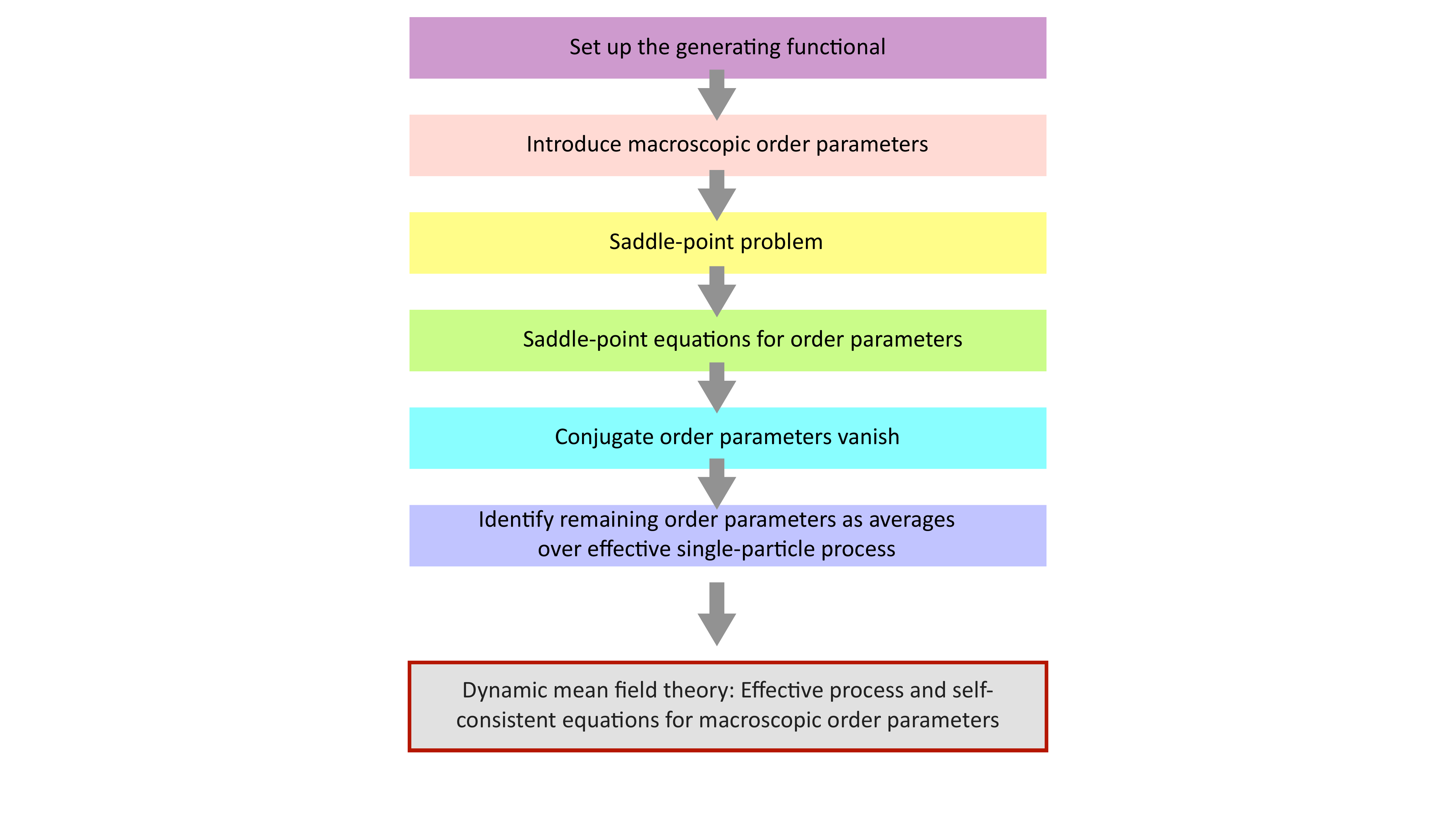}
\end{center}
\caption{Illustration of the main steps of the generating functional analysis for the toy model. \label{fig:GF_flow_toy}}
\end{figure}

We briefly summarise the main steps of the generating functional analysis of the toy model, as also illustrated in Fig.~\ref{fig:GF_flow_toy}.

\medskip

The main steps are:
\begin{enumerate} 
\item[1.] We set up the generating functional for the problem [Eq.~(\ref{eq:gf002})].
\item[2.] We then introduced the macroscopic order parameters $\ulM$ and $\widehat\ulM$ [Eqs~(\ref{eq:one}, \ref{eq:gf003})].
\item[3.] We showed that the generating functional can be written as an integral over $\ulM$ and $\widehat\ulM$, and that this integral is in saddle-point form in the limit $N\to\infty$ [Eq.~(\ref{eq:gf_sp_f})].
\item[4.] We carried out the saddle-point integration, and derived the saddle-point equations for the order parameters [Eqs.~(\ref{eq:scmmhat_f})].
\item[5.] We showed that the conjugate order parameter vanishes, i.e., $\widehat M(t)=0$ for all $t$ [Sec.~\ref{sec:M_eq_0_proper}]. Alternatively, this can be postulated and be justified retrospectively (Secs.~\ref{sec:vanilla_hatm_eq_0} and \ref{sec:problems}). 
\item[6.] We showed that the remaining order parameter, $M(t)$, can be written as the average over an effective single-particle process [Eq.~(\ref{eq:M_avg_proper}) together with Eq.~(\ref{eq:avstar_f})].
\item[7.] The resulting dynamic mean field-theory thus consists of this effective process [Eq.~(\ref{eq:eff_process_proper})] and an equation for the order parameter $M(t)$ [Eq.~(\ref{eq:M_avg_proper})], involving an average over realisations of the effective process. This combination is to be solved self-consistently. 
\end{enumerate}
In the toy problem the simple ordinary differential equation (\ref{eq:dotM_gf}) can be obtained for $M(t)$ from the effective process. This is a consequence of the fact that the only potentially nonlinear term (nonlinear in the $x_i$) on the right-hand side of the initial problem [Eq.~(\ref{eq:intproc})] can be expressed as a function of $M$ (this is the term $g[M(t)]$). We would not be able to find such simple closed equation for $M$ if the initial problem had been, say, of the type $\dot x_i = a x_i^2 + g\left(\frac{1}{N}\sum_j x_j\right)+\xi_i(t)$ [the effective single-particle process would then take the form $\dot x = a x^2 + g(M) +\xi$.]

\section{Generating functional analysis for random Lotka--Volterra equations I: Setup, disorder average and saddle-point equations} \label{sec:gf_lv_1}
\subsection{Overview of this section}
We now turn to the Lotka--Volterra problem. In Sec.~\ref{sec:LV_setup} we summarise the setup of the model, and briefly describe how realisations of the random interaction coefficients can be generated numerically.

We then present the initial steps of the generating functional calculation, namely:
\begin{enumerate} 
\item[1.] Setting up the generating functional for the problem [Eq.~(\ref{eq:gflv0})].
\item[2.] Compute the disorder average of the generating functional (Sec.~\ref{sec:disorder_average}).
\item[3.] Introduction of the macroscopic order parameters [Sec.~\ref{sec:macroscopic_op}, in particular Eq.~(\ref{eq:op})].
\item[4.] Write the disorder-averaged generating functional as a saddle-point integral over the order parameters and their conjugates [Eq.~(\ref{eq:sp})].  
\item[5.] Carry out the saddle-point integration, and derive the saddle-point equations for the order parameters [Sec.~\ref{sec:sp_analysis}, in particular Eqs.~(\ref{eq:sp_aux1}) and (\ref{eq:Var_wrt_hats0})].
\end{enumerate}
Of course, these steps are similar to steps 1-4 in the calculation of the toy model, summarised in Sec.~\ref{sec:toy_summary}. The main addition is the new step 2, the disorder average. This was not necessary earlier, as there was no disorder in the toy problem. Also, there are now significantly more order parameters than in the toy problem.

\medskip

Once step 5 above has been reached (saddle-point equations for the order parameters in the Lotka--Volterra problem), there is again a simpler 'vanilla' method of deriving the dynamical mean-field theory, or a more robust, but longer method. These are described in Secs.~\ref{sec:gf_lv_2} and \ref{sec:gf_lv_3} respectively, and are more involved than their counterparts for the toy model. Both methods lead to the same effective single-species process (the dynamical mean-field theory for the Lotka--Volterra problem).

\subsection{Setup}\label{sec:LV_setup}
\subsubsection{Model definitions}
The calculation follows \cite{Galla_2018}, and is based on the principles of \cite{dedominicis,mezard1987,coolen_dyn,Coolen_MG}. It was originally developed in the context of random replicator models in \cite{opper1992phase}, and then used for example also in \cite{galla2006,galla_farmer,sidhomgalla, froy, baron_et_al_prl, park, poley2, aguirre, azaele}. 
\bigskip

We start from the Lotka--Volterra equations
\be\label{eq:lv0}
\dot x_i=x_i\left(1-x_i+\sum_{j}\alpha_{ij}x_j\right),
\ee
with Gaussian random interaction coefficients $\alpha_{ij}$. We set $\alpha_{ii}=0$, and as in \cite{Galla_2018} we assume that for each pair $i<j$, 
\BE
\overline{a_{ij}}=\overline{\alpha_{ji}}&=&\frac{\mu}{N}, \nonumber \\
\mbox{Var}(\alpha_{ij})=\mbox{Var}(\alpha_{ji})&=&\frac{\sigma^2}{N}, \nonumber \\
\mbox{Cov}(\alpha_{ij},\alpha_{ji})&=&\Gamma. \label{eq:alpha_stat}
\EE
We have written $\overline{\cdots}$ for the average over the random couplings. $\mbox{Var}(\alpha)=\overline{\alpha^2}-(\overline\alpha)^2$ stands for the variance, and $\mbox{Cov}(\alpha,\beta)=\left[\overline{\alpha\beta}-\overline{\alpha}\overline{\beta}\right]/\sqrt{\mbox{Var}(\alpha)\mbox{Var}(\beta)}$ is the covariance. The quantities $\mu, \sigma^2$ and $\Gamma$ are the parameters of the model. (We note that the correlation parameter is called $\gamma$ in \cite{Galla_2018}. However, we here prefer to use $\Gamma$ in-line with later work, including \cite{baron_et_al_prl}.) The parameter $\mu$ can take any real value, and $\Gamma$ is restricted to the interval $[-1,1]$. We assume that elements $\alpha_{ij}$ and $\alpha_{k\ell}$ are uncorrelated, unless the $i,j,k,\ell$ are such that the two elements are identical ($i=k, j=\ell$), or diagonally opposed to each other in the matrix ($i=\ell, j=k$).

\subsubsection{Numerical generation of random matrix with correlations between diagonally opposed elements}
A common question I receive from students is how to generate instances of the random matrix $\underline{\underline{\alpha}}$ with statistics as in Eqs.~(\ref{eq:alpha_stat}). Generally, a set of correlated Gaussian random numbers with a given covariance matrix can be generated using a Cholesky decomposition of the correlation matrix. If you have not heard this, then this is probably a term worth looking up,

\medskip

We here only describe the construction for our specific model. A matrix $\underline{\underline{\alpha}}$ with the above statistics can be generated as follows (for a given finite $N$). For each pair $i<j$ proceed as follows:
\begin{enumerate}
\item[1.] Generate two independent random Gaussian variables $u$ and $v$, each with mean zero and variance one. The variables $u$ and $v$ for one pair $(i,j)$ are independent from those of any other pair.
\item[2.] Then set $\alpha_{ij}=\frac{\mu}{N}+\frac{\sigma}{\sqrt{N}}u$ and $\alpha_{ji}=\frac{\mu}{N}+\frac{\sigma}{\sqrt{N}}\left(\Gamma u + \sqrt{1-\Gamma^2}\right) v$. 
\end{enumerate}
One can then directly check that the resulting $\{\alpha_{ij}\}$ have the properties in Eqs.~(\ref{eq:alpha_stat}). For example,
\BE
\overline{(\alpha_{ji}-\mu/N)^2}&=&\frac{\sigma^2}{N}\left[\Gamma^2 \underbrace{\overline{u^2}}_{=1}+(1-\Gamma^2)\underbrace{\overline{v^2}}_{=1}+\Gamma\sqrt{1-\Gamma^2}\underbrace{\overline{uv}}_{=0} \right]\nonumber \\
&=& \frac{\sigma^2}{N}.
\EE

\subsection{Generating functional}
There is a clean way of implementing the generating functional calculation and a slightly `dirty' approach. (I stress that this distinction is different from the distinction between the `vanilla' and proper treatments of the toy problem.) The advantage of the clean implementation is, well, that it is clean. The advantage of the dirty approach is that it is perhaps easier to follow for someone who isn't familiar with dynamic mean field theory for disordered systems. I here take the `dirty' route, noting that the sloppiness of this approach is only in the first line, slightly re-phrasing the setup, and not in the actual generating-functional calculation.

\medskip

The sloppy step consists of dividing both sides of Eq.~(\ref{eq:lv0}) by $x_i$. That is, we start from
\be\label{eq:lv00}
\frac{\dot x_i}{x_i}=1-x_i+\sum_{j\neq i}\alpha_{ij}x_j.
\ee
Of course, this is only allowed if $x_i$ remains non-zero. This is technically not true, because (as we will see) some species die out in the long run. Nonetheless, we'll proceed starting from Eq.~(\ref{eq:lv00}). The clean method is described in Appendix~\ref{app:clean}. Ultimately, both routes lead to the same result, justifying the sloppiness which we exercise purely for pedagogic reasons.

\medskip

For a fixed realisation of the disorder, the dynamical generating functional of the process is then
\BE
Z[\boldpsi]
&=&\int \D\ulbx \, \D\widehat\ulbx \, p_0[\bx(0)]\, \exp\left(i\sum_i\int dt~ x_i(t)\psi_i(t)\right) \nonumber \\
&& \times \exp\Bigg(i\sum_i\int dt \Bigg[\widehat x_i(t)\Bigg(\frac{\dot x_i(t)}{x_i(t)}-[1-x_i(t)+\sum_{j\neq i}\alpha_{ij} x_j(t)+h_i(t)]\Bigg)\Bigg]\Bigg), \label{eq:gflv0}
\EE
where we have enforced the equations of motion via suitable delta-functions, written in their exponential representation, similar to the way we did this in Sec.~\ref{sec:stoch_proc}. We have also added a perturbation field $\ulbh$. As in the toy problem we note that this field is not part of the actual model, but only a mathematical device used to generate response functions. We will set $\ulbh$ to zero at the end of the calculation.

\subsection{Carrying out the disorder average}\label{sec:disorder_average}
Next, we look at the terms containing the disorder (the $\{a_{ij}\}$), and perform the Gaussian average over these random variables. For any pair $i<j$ of species we can write
\be\label{eq:alpha}
\alpha_{ij}=\frac{\mu}{N}+\frac{\sigma}{\sqrt{N}}z_{ij}, ~~~ \alpha_{ji}=\frac{\mu}{N}+\frac{\sigma}{\sqrt{N}}z_{ji}, 
\ee
where $z_{ij}$ and $z_{ji}$ are drawn from a Gaussian distribution with $\overline{z_{ij}}=0$, $\overline{z_{ij}^2}=1$, and $\overline{z_{ij}z_{ji}}=\Gamma$. 

The term in the generating functional [Eq.~(\ref{eq:gflv0})] containing the disorder is
\be\label{eq:big_X}
X\equiv\exp\left(-i\sum_{i\neq j}\int dt \, \widehat x_i(t)\alpha_{ij}x_j(t)\right).
\ee

First, we have, using Eqs.~(\ref{eq:alpha}),
\BE
\overline X &=&\overline{\exp\left(-i\sum_{i\neq j}\int dt \, \widehat x_i(t)\alpha_{ij}x_j(t)\right)}\nonumber \\ &=&\exp\left(-i\frac{\mu}{N}\sum_{i\neq j}\int dt \, \widehat x_i(t)x_j(t)\right)\times \overline{\exp\left(-i\frac{\sigma}{\sqrt{N}}\sum_{i\neq j}\int dt \, \widehat x_i(t)z_{ij}x_j(t)\right)}.\label{eq:aux_diag}
\EE
The first factor in this does not contain any disorder. We therefore look at the second factor, which we'll call $Y$.

We can write
\BE
Y&\equiv&\exp\left(-i\frac{\sigma}{\sqrt{N}}\sum_{i\neq j}\int dt \, \widehat x_i(t)z_{ij}x_j(t)\right)\nonumber \\
&=&\prod_{i<j}\exp\left(-i\frac{\sigma}{\sqrt{N}}\int dt \,\{ \widehat x_i(t)z_{ij}x_j(t)+\widehat x_j(t)z_{ji}x_i(t)\}\right).
\EE
We have split this in this way, because $z_{ij}$ and $z_{ji}$ are correlated, so we need to average over these two quantities together.

The factor
\be\label{eq:Y_ij}
Y_{ij}\equiv\exp\left(-i\frac{\sigma}{\sqrt{N}}\int dt \, \{\widehat x_i(t)z_{ij}x_j(t)+\widehat x_j(t)z_{ji}x_i(t)\}\right)
\ee
is of the form $\exp\left(z_1 A_{ij} + z_2 B_{ij}\right)$, with $z_1$, $z_2$ Gaussian random variables, with mean zero, variance one, and correlation $\overline{z_1 z_2}=\Gamma$. $A_{ij}$ and $B_{ij}$ are shorthands for
\BE\label{eq:AijBij}
A_{ij}&=&-i\frac{\sigma}{\sqrt{N}}\int dt \, \widehat x_i(t)x_j(t), \nonumber \\
B_{ij}&=&-i\frac{\sigma}{\sqrt{N}}\int dt \, \widehat x_j(t)x_i(t) ~(= A_{ji}).
\EE

Carrying out the average over the Gaussian variables $z_1$ and $z_2$ is then an exercise in bivariate Gaussian integration. This can be done using the expressions in Sec.~\ref{sec:gauss}.

We have to evaluate
\BE\label{eq:factor}
\overline{Y}_{ij}=\frac{1}{\sqrt{(2\pi)^2\det\Sigma}}\int dz_1 \int dz_2 \exp\left(-\frac{1}{2}\bz^T \Sigma^{-1} \bz\right)\exp\left(z_1 A_{ij} + z_2 B_{ij}\right),
\EE

where $\bz=(z_1, z_2)$ and where the $2\times 2$ matrix $\Sigma$ has entries $\Sigma_{ab}=\overline{z_az_b}$ for $a,b,\in\{1,2\}$, i.e., $\Sigma=\left(\begin{array}{cc} 1 & \Gamma \\ \Gamma & 1\end{array}\right)$.

To evaluate the integral we use Eq.~(\ref{eq:Gauss}). The result is  
\BE\label{eq:gaussf}
\overline{Y}_{ij}&=&\frac{1}{\sqrt{(2\pi)^2\det\Sigma}}\int dz_1 \int dz_2 \exp\left(-\frac{1}{2}\bz^T \Sigma^{-1} \bz\right)\exp\left(z_1 A_{ij} + z_2 B_{ij}\right)\nonumber \\
&=&
\exp\left(\frac{1}{2} (A_{ij}^2 +2 \Gamma A_{ij} B_{ji} + B_{ij}^2)\right).
\EE
We note that the relation in Eq.~(\ref{eq:gaussf}) is an exact identity, nothing needs to be assumed about the scaling of $A_{ij}$ and $B_{ji}$ with $N$.

\medskip
\underline{Aside (alternative way of carrying out the disorder average, assuming $N\gg1$):}\\
This method relies on $A_{ij}$ and $B_{ij}$ scaling as $N^{-1/2}$, see Eqs.~(\ref{eq:AijBij}). Eventually, we are interested in the limit of large $N$. So we expand each factor $Y_{ij}$ in Eq.~(\ref{eq:Y_ij}) in powers of $1/\sqrt{N}$, or equivalently, powers of $A_{ij}$ and $B_{ij}$. One gets
\BE\label{eq:zexpand}
\exp(z_1 A_{ij} + z_2 B_{ij})&= & 1+\left(z_1 A_{ij} + z_2 B_{ij}\right) +\frac{1}{2} (z_1 A_{ij} + z_2 B_{ij})^2 +{\cal O}(1/N^{3/2})  \nonumber \\
&=&1+z_1 A_{ij} + z_2 B_{ij} +\frac{1}{2} (z_1^2 A_{ij}^2 + 2 z_1 z_2 A_{ij} B_{ij} + z_2^2 B_{ij}^2) +{\cal O}(1/N^{3/2}). \nonumber \\
\EE
Next we average over the disorder (keeping in mind that $z_1$ and $z_2$ each have mean zero, variance one, and that their correlation is $\overline{z_1 z_2}=\Gamma$):
\BE
\overline{\exp(z_1 A_{ij} + z_2 B_{ij}})&= &1+\frac{1}{2} (A_{ij}^2 + 2 \Gamma A_{ij} B_{ij} + B_{ij}^2) +{\cal O}(1/N^{2}). 
\EE
[After averaging the correction term in the exponential is now of order $N^{-2}$. This is because terms of third order in $z_1$ and $z_2$ in the expansion in Eq.~(\ref{eq:zexpand}) average to zero.] Next we re-exponentiate and find
\BE\label{eq:gaussff}
\overline{Y}_{ij}&=& 1+\frac{1}{2} (A_{ij}^2 +2 \Gamma A_{ij} B_{ij} + B_{ij}^2) +{\cal O}(1/N^{2}) \nonumber \\
&=& \exp\left(\frac{1}{2} (A_{ij}^2 + 2 \Gamma A_{ij} B_{ij} + B_{ij}^2)+{\cal O}(1/N^{2})\right).
\EE
To leading order in $1/N$ the expression in Eq.~(\ref{eq:gaussff}) is the same as the exact result in Eq.~(\ref{eq:gaussf}). However, the derivation of Eq.~(\ref{eq:gaussff}) as just presented is based on the assumption $N\gg 1$.
\subsection{Introduction of macroscopic order parameters}\label{sec:macroscopic_op}
Using Eqs.~(\ref{eq:aux_diag}) and (\ref{eq:gaussf}) we have arrived at
\BE\label{eq:Xbar_2}
\overline X &=&\overline{\exp\left(-i\sum_{i\neq j}\int dt \, \widehat x_i(t)\alpha_{ij}x_j(t)\right)}\nonumber \\ 
&=&\exp\left(-i\frac{\mu}{N}\sum_{i\neq j}\int dt \, \widehat x_i(t)x_j(t)\right)\times \exp\left(\frac{1}{2} \sum_{i<j}(A_{ij}^2 +2 \Gamma A_{ij} A_{ji} + A_{ji}^2)\right) \nonumber \\
&=&\exp\left(-i\frac{\mu}{N}\sum_{i\neq j}\int dt \, \widehat x_i(t)x_j(t)\right)\times\exp\left(\frac{1}{2} \sum_{i\neq j}(A_{ij}^2 +\Gamma A_{ij} A_{ji})\right),
\EE
with $A_{ij}=-i\frac{\sigma}{\sqrt{N}}\int dt \, \widehat x_i(t)x_j(t)$. We note that
\BE
A_{ij}^2&=&-\frac{\sigma^2}{N}\int dt \, dt'\, \widehat x_i(t)x_j(t) \widehat x_i(t')x_j(t'), \nonumber \\
A_{ij}A_{ji}&=&-\frac{\sigma^2}{N}\int dt \, dt'\, \widehat x_i(t)x_j(t) \widehat x_j(t')x_i(t').
\EE

\medskip

Next we introduce the following short-hands,
\BE
&M(t)=\frac{1}{N}\sum_i x_i(t),\nonumber\\
&P(t)=\frac{1}{N}\sum_i  \widehat x_i(t),\nonumber\\
&C(t,t')=\frac{1}{N}\sum_i x_i(t) x_i(t'), \nonumber \\
&K(t,t')=\frac{1}{N}\sum_i x_i(t) \widehat x_i(t'),  \nonumber \\
&L(t,t')=\frac{1}{N}\sum_i \widehat x_i(t)\widehat x_i(t'). \label{eq:op}
\EE

With these definitions we can write
\BE
&&\overline{\exp\left(-i\sum_{i\neq j}\int dt \, \widehat x_i(t)\alpha_{ij}x_j(t)\right)}\nonumber\\
&=&\exp\left(-i\mu N  \int dt \, P(t) M(t)\right.\nonumber \\
&&~~~~~~\left.-\frac{1}{2}N\sigma^2\int dt ~dt' \left[L(t,t')C(t,t')+\Gamma K(t,t')K(t',t)\right]+{\cal O}(N^0)\right).
\EE
As indicated by the term ${\cal O}(N^0)$, we have left out sub-leading contributions in $N$ (i.e, terms of order $N^0$ or lower). For example we had $-i\frac{\mu}{N}\sum_{i\neq j}\int dt \, \widehat x_i(t)x_j(t)$ in the exponential in Eq.~(\ref{eq:Xbar_2}), and we have now replaced this with $-\mu N  \int dt \, P(t) M(t)=-i\frac{\mu}{N} \sum_{i,j}\int dt \widehat x_i(t) x_j(t)$. Technically, this introduces an additional contribution  $-i\frac{\mu}{N}\sum_i \int dt ~\widehat x_i(t)x_i(t)$. However, this term is of order $N^0$, and therefore a sub-leading contribution (in powers of $1/N$). We can ignore this, because we are eventually going to take the limit $N\to\infty$. We only keep terms in the exponent that are of order $N$.

\medskip

The order parameters can formally be introduced into the generating functional as delta-functions in their integral representation, e.g. 
\BE
1&=&\int\prod_{t,t'} dC(t,t')\delta\left(C(t,t')-\frac{1}{N}\sum_i x_i(t)x_i(t')\right)\nonumber \\
&=&\int \D[\ullC, \widehat\ullC]~\exp\left(iN\int dt~ dt' \widehat C(t,t') \left( C(t,t')-N^{-1} \sum_i x_i(t) x_i(t')\right)\right), 
\EE
with
\be
\D[\ullC, \widehat\ullC]=\prod_{t,t'} \frac{dC(t,t')d\widehat C(t,t')}{2\pi/N}.
\ee
We also insert similar expressions for the other order parameters. 

\medskip

The disorder-averaged generating functional can be written as follows
\be\label{eq:sp}
\overline{Z}[\ulbpsi]=\int \D[\ulM,\widehat \ulM] \D[\ulP,\widehat \ulP] \D[\ullC,\widehat\ullC]\D[\ullK,\widehat \ullK] \D[\ullL,\widehat \ullL] \,\exp\left(N\left[\Psi+\Phi+\Omega+{\cal O}(N^{-1})\right]\right).
\ee
The term
\BE\label{eq:sp_psi}
\Psi&=&i\int dt~ [\widehat M(t) M(t) +\widehat P(t) P(t)] \nonumber \\
&&+i\int dt ~dt' \left[\widehat C(t,t')C(t,t')+\widehat K(t,t')K(t,t')+\widehat L(t,t')L(t,t')\right]
\EE
results from the introduction of the macroscopic order parameters. The contribution
\BE\label{eq:sp_phi}
\Phi&=&-\frac{1}{2}\sigma^2\int dt ~dt' \left[L(t,t')C(t,t')+\Gamma K(t,t')K(t',t)\right]-i\mu\int dt ~ M(t) P(t)
\EE
comes from the disorder average, and $\Omega$ describes the details of the microscopic time evolution
\BE
\Omega&=&N^{-1}\sum_i\ln\bigg[\int \D \ulx_i \D \widehat \ulx_i\, p_{0}^{(i)}[x_i(0)]\exp\left(i\int dt~ \psi_i(t)x_i(t)\right)\nonumber \\
&&\times \exp\left(i\int dt ~\widehat x_i(t)  \left(\frac{\dot x_i(t)}{x_i(t)}-[1-x_i(t)]-h_i(t)\right)\right)\nonumber \\
&&\times \exp\left(-i\int dt ~ dt' \left[\widehat C(t,t')x_i(t)x_i(t')+\widehat L(t,t')\widehat x_i(t)\widehat x_i(t')+\widehat K(t,t') x_i(t)\widehat x_i(t')\right]\right)\nonumber \\
&&\times \exp\left(-i\int dt~  [\widehat M(t)x_i(t)+\widehat P(t) \widehat x_i(t)]\right)\bigg].
\label{eq:omega0}
\EE

The quantity $p_{0}^{(i)}[\cdot]$  describes the distribution from which the initial value of $x_i$ (at time $t=0$) is drawn. In writing down Eq.~(\ref{eq:omega0}) we have assumed that the joint distribution of all $x_i(0)$ ($i=1,\dots,N)$ factorises, i.e., $p_0[\bx(0)]=\prod_i p_0^{(i)}[x_i(0)]$. We'll now go even further and assume that the $x_i(0)$ are independent and identically distributed (iid), that is, we write $p_0^{(i)}[x_i(0)]=p_0[x_i(0)]$ for all $i$. 
\medskip

We note that the $\{x_i(t),\widehat x_i(t)\}$ in the expression for $\Omega$ are dummy variables to be integrated over. We can therefore re-write $\Omega$ in the following form
\BE
\Omega&=&N^{-1}\sum_i\ln\bigg[\int \D\ulx \D\widehat \ulx\, p_{0}[x(0)]\exp\left(i\int dt~ \psi_i(t)x(t)\right)\nonumber \\
&&\times \exp\left(i\int dt ~\widehat x(t)  \left(\frac{\dot x(t)}{x(t)}-[1-x(t)]-h_i(t)\right)\right)\nonumber \\
&&\times \exp\left(-i\int dt ~ dt' \left[\widehat C(t,t')x(t)x(t')+\widehat L(t,t')\widehat x(t)\widehat x(t')+\widehat K(t,t') x(t)\widehat x(t')\right]\right)\nonumber \\
&&\times \exp\left(-i\int dt~  [\widehat M(t)x(t)+\widehat P(t) \widehat x(t)]\right)\bigg].
\label{eq:omega}
\EE
The object inside the logarithm on the right contains $\psi_i(\cdot)$ and $h_i(\cdot)$. Therefore, the different terms in the sum over $i$ on the right are not identical at this point.

\subsection{Saddle-point analysis}\label{sec:sp_analysis}
\subsubsection{Saddle-point equations for the macroscopic order parameters}
We next use the saddle-point method (Sec.~\ref{sec:saddlepoint}) to carry out the integrals in Eq. (\ref{eq:sp}). This is valid in the limit $N\to\infty$, and amounts to finding the extrema of the term in the exponent. 

\medskip
Setting the variation with respect to components of the the integration variables $\ulM, \ulP, \ullC,\ullK$ and $\ullL$ to zero gives, respectively,
\BE
i\widehat M(t)&=&i\mu P(t), \nonumber \\
i\widehat P(t)&=&i\mu M(t), \nonumber \\
i\widehat C(t,t')&=&\frac{1}{2}\sigma^2L(t,t'),\nonumber \\
i\widehat K(t,t')&=&\Gamma \sigma^2 K(t',t),\nonumber \\
i\widehat L(t,t')&=&\frac{1}{2}\sigma^2C(t,t').\label{eq:sp_aux1}
\EE

Next we extremise with respect to components of $\widehat \ulM, \widehat \ulP, \widehat \ullC, \widehat \ullK,\widehat \ullL$. We find 
\BE
M(t)&=&i\lim_{N\to\infty}\frac{\delta \Omega}{\delta \widehat M(t)}, \nonumber \\
P(t)&=&i\lim_{N\to\infty}\frac{\delta \Omega}{\delta \widehat P(t)}, \nonumber \\
C(t,t')&=&i\lim_{N\to\infty}\frac{\delta \Omega}{\delta \widehat C(t,t')},\nonumber \\
K(t,t')&=&i\lim_{N\to\infty}\frac{\delta \Omega}{\delta \widehat K(t,t')}, \nonumber \\
L(t,t')&=&i\lim_{N\to\infty}\frac{\delta \Omega}{\delta \widehat L(t,t')}. 
\label{eq:Var_wrt_hats0}
\EE

\medskip

The object $\Omega$ in Eq.~(\ref{eq:omega}) is of the form $\Omega=\frac{1}{N}\sum_i \ln[\mbox{complicated thing}_i]$, where the subscript $i$ indicates that the object `complicated thing' is different for each $i$ (because of the $\ulpsi_i$ and $\ulh_i$ as mentioned above). Derivatives of $\Omega$ are then of the form
\BE
\frac{\delta \Omega}{\delta\mbox{something}}=\frac{1}{N}\sum_i \frac{1}{\mbox{complicated thing}_i} \frac{\delta \mbox{complicated thing}_i}{\delta\mbox{something}}.
\EE
From now on we'll use more scientific notation and write $Z_1[\ulpsi_i,\ulh_i]$ instead of `$\mbox{complicated thing}_i$'. That is, we define
\BE
Z_1[\ulpsi_i,\ulh_i]&\equiv&\int \D \ulx \D\widehat \ulx\, p_{0}[x(0)]\exp\left(i\int dt~ \psi_i(t)x(t)\right)\nonumber \\
&&\times \exp\left(i\int dt ~\widehat x(t)  \left(\frac{\dot x(t)}{x(t)}-[1-x(t)]-h_i(t)\right)\right)\nonumber \\
&&\times \exp\left(-i\int dt ~ dt' \left[\widehat C(t,t')x(t)x(t')+\widehat L(t,t')\widehat x(t)\widehat x(t')+\widehat K(t,t') x(t)\widehat x(t')\right]\right)\nonumber \\
&&\times \exp\left(-i\int dt~  [\widehat M(t)x(t)+\widehat P(t) \widehat x(t)]\right).\label{eq:z1_def}
 \EE
 Next, we note that taking a derivative of $Z_1$ with respect to one of the hatted order parameters `brings down' objects involving $x$ or $\widehat x$ in the integral. For example,
\BE
\frac{\delta Z_1[\ulpsi_i,\ulh_i]}{\delta \widehat C(t_1,t_2)}&=&-i\int \D\ulx \D\widehat \ulx\, p_{0}[x(0)] ~x(t_1) x(t_2)~\exp\left(i\int dt~ \psi_i(t)x(t)\right)\nonumber \\
&&\times \exp\left(i\int dt ~\widehat x(t)  \left(\frac{\dot x(t)}{x(t)}-[1-x(t)]-h_i(t)\right)\right)\nonumber \\
&&\times \exp\left(-i\int dt ~ dt' \left[\widehat C(t,t')x(t)x(t')+\widehat L(t,t')\widehat x(t)\widehat x(t')+\widehat K(t,t') x(t)\widehat x(t')\right]\right)\nonumber \\
&&\times \exp\left(-i\int dt~  [\widehat M(t)x(t)+\widehat P(t) \widehat x(t)]\right),
\EE
where it is important to notice the factor $x(t_1)x(t_2)$ in the first line.

With this, we have from Eqs.~(\ref{eq:Var_wrt_hats0}),

\BE
M(t)&=&\lim_{N\to\infty}N^{-1}\sum_i\avg{x(t)}_i, \nonumber \\
P(t)&=&\lim_{N\to\infty}N^{-1}\sum_i\avg{\widehat x(t)}_i, \nonumber \\
C(t,t')&=&\lim_{N\to\infty}N^{-1}\sum_i\avg{x(t)x(t')}_i,\nonumber \\
K(t,t')&=&\lim_{N\to\infty}N^{-1}\sum_i\avg{x(t)\widehat x(t')}_i, \nonumber \\
L(t,t')&=&\lim_{N\to\infty}N^{-1}\sum_i\avg{\widehat x(t)\widehat x(t')}_i, 
\label{eq:var_wrt_hats}
\EE
where $\avg{g[\ulx,\widehat \ulx]}_i$ stands for  
\BE
\avg{g[\ulx,\widehat \ulx]}_i&=&\frac{1}{Z_1[\ulpsi_i,\ulh_i]}\times \int \D\ulx \D\widehat \ulx\, p_{0}[x(0)] ~ g[\ulx,\widehat\ulx] \exp\left(i\int dt~ \psi_i(t)x(t)\right)\nonumber \\
&&\times \exp\left(i\int dt ~\widehat x(t)  \left(\frac{\dot x(t)}{x(t)}-[1-x(t)]-h_i(t)\right)\right)\nonumber \\
&&\times \exp\left(-i\int dt ~ dt' \left[\widehat C(t,t')x(t)x(t')+\widehat L(t,t')\widehat x(t)\widehat x(t')+\widehat K(t,t') x(t)\widehat x(t')\right]\right)\nonumber \\
&&\times \exp\left(-i\int dt~  [\widehat M(t)x(t)+\widehat P(t) \widehat x(t)]\right).\label{eq:avg_i}
\EE
For convenience we repeat the expression for $Z_1[\ulpsi_i,\ulh_i]$,
\BE
Z_1[\ulpsi_i,\ulh_i]&=& \int  \D\ulx \D\widehat \ulx\, p_{0}[x(0)] ~  \exp\left(i\int dt~ \psi_i(t)x(t)\right)\nonumber \\
&&\times \exp\left(i\int dt ~\widehat x(t)  \left(\frac{\dot x(t)}{x(t)}-[1-x(t)]-h_i(t)\right)\right)\nonumber \\
&&\times \exp\left(-i\int dt ~ dt' \left[\widehat C(t,t')x(t)x(t')+\widehat L(t,t')\widehat x(t)\widehat x(t')+\widehat K(t,t') x(t)\widehat x(t')\right]\right)\nonumber \\
&&\times \exp\left(-i\int dt~  [\widehat M(t)x(t)+\widehat P(t) \widehat x(t)]\right).\label{eq:Z_1_LV}
\EE

\subsubsection{Next steps}
The ultimate goal of our generating-functional calculation for the Lotka--Volterra system is to derive a dynamic mean-field theory, that is a stochastic process for an effective single species. Our next step in this direction is to simplify the expressions in Eqs.~(\ref{eq:avg_i}) and (\ref{eq:Z_1_LV}). For example, we will again show that $P(t)=0$ and $\widehat M(t)=0$ for all $t$.

\medskip

As for the toy model we first present a simplified and perhaps not fully satisfactory derivation (Sec.~\ref{sec:gf_lv_2}). However, the general idea is right, and most importantly, the effective process that comes out of this is right. A more complete derivation is given in Sec.~\ref{sec:gf_lv_3}.

\section{Generating functional analysis for random Lotka--Volterra equations II: `Vanilla' derivation of the effective single-species process} \label{sec:gf_lv_2}
\subsection{Simplification of saddle-point equations}\label{sec:simple_LV_vanilla}
We start from Eqs.~(\ref{eq:avg_i}) and (\ref{eq:Z_1_LV}). Our first move is to focus on the situation in which $\psi_i(t)=\psi(t)$ for all $i$. We also set $h_i(t)=h(t)$. There is nothing dubious about this, it is simply a restriction to a special case. We then find that $\avg{\cdots}_i$ in Eq.~(\ref{eq:avg_i}) carries no $i$-dependence, and we write $\avg{\cdots}_\star$ instead.

\medskip

The dodgy step comes now.  Both $\ulP$ and $\ullL$ are averages over objects containing only hatted quantities [as per Eqs. (\ref{eq:var_wrt_hats})]. We therefore set these to zero, appealing to Sec.~\ref{sec:avg_hat_eq_0}. For the same reasons as in Sec.~\ref{sec:vanilla_hatm_eq_0} this is not entirely legit at this point. But given the complexity of the whole procedure I think it is justified for pedagogical reasons to take the `illegal' but shorter route first, in particular given that the end result is correct. We will give a proper argument in Sec.~\ref{sec:gf_lv_3}.

\medskip

Setting  $\ulP$ and $\ullL$ to zero, together with Eqs.~(\ref{eq:sp_aux1}), allows us to make the following replacements in Eqs.~(\ref{eq:avg_i}) and (\ref{eq:Z_1_LV}):
\BE
i\widehat M(t)&\to&i\mu P(t)=0, \nonumber \\
i\widehat P(t)&\to&i\mu M(t), \nonumber \\
i\widehat C(t,t')&\to&\frac{1}{2}\sigma^2L(t,t')=0,\nonumber \\
i\widehat K(t,t')&\to&\Gamma \sigma^2 K(t',t),\nonumber \\
i\widehat L(t,t')&\to&\frac{1}{2}\sigma^2C(t,t').\label{eq:sp_aux1_v}
\EE
Additionally we introduce $\ullG \equiv -i \ullK$, i.e., we make the replacement $\ullK \to i\ullG$ (this is simply a change of notation). 

\medskip

We then have from Eq.~(\ref{eq:avg_i}) (recalling that $\avg{\cdots}_i$ no longer depends on $i$ and has been replaced by $\avg{\cdots}_\star$),
\BE
\avg{g[\ulx,\widehat \ulx]}_\star&=&\frac{1}{Z_1[\ulpsi,\ulh]}\times \int \D\ulx \D\widehat\ulx\,p_{0}[x(0)] ~ g[\ulx,\widehat\ulx] \exp\left(i\int dt~ \psi(t)x(t)\right)\nonumber \\
&&\times \exp\left(i\int dt ~\widehat x(t)  \left(\frac{\dot x(t)}{x(t)}-[1-x(t)]-h(t)\right)\right)\nonumber \\
&&\times \exp\left(-\frac{1}{2}\sigma^2\int dt ~ dt' \left[C(t,t')\widehat x(t)\widehat x(t')+2i\Gamma G(t',t) x(t)\widehat x(t')\right]\right)\nonumber \\
&&\times \exp\left(-i\mu \int dt~  M(t) \widehat x(t)\right),\label{eq:avg_star}
\EE
and
\BE
Z_1[\ulpsi,\ulh]&=& \int \D\ulx \D\widehat\ulx\, p_{0}[x(0)] ~  \exp\left(i\int dt~ \psi(t)x(t)\right)\nonumber \\
&&\times \exp\left(i\int dt ~\widehat x(t)  \left(\frac{\dot x(t)}{x(t)}-[1-x(t)]-h(t)\right)\right)\nonumber \\
&&\times \exp\left(-\frac{1}{2}\sigma^2\int dt ~ dt' \left[C(t,t')\widehat x(t)\widehat x(t')+2i\Gamma G(t',t) x(t)\widehat x(t')\right]\right)\nonumber \\
&&\times \exp\left(-i\mu \int dt~  M(t) \widehat x(t)\right).\label{eq:Z1_LV_vanilla}
\EE

\subsection{Interpretation in terms of an effective single-species process}\label{sec:eff_proc_interpret}
We now make the following statement: \\

For given $\ullC, \ullG$ and $\ulM$, the expression in Eq.~(\ref{eq:Z1_LV_vanilla}) is the generating functional of the effective dynamics
\BE
\dot x(t)=x(t)\left[1-x(t)+\Gamma \sigma^2\int dt' G(t,t') x(t')+\mu M(t)+\eta(t)+h(t)\right], \label{eq:effective}
\EE
where $\eta(t)$ is Gaussian noise of zero average ($\mathbb{E}[\eta(t)]=0$ for all $t$) and with correlations in time given by
\be
\mathbb{E}[\eta(t)\eta(t')]=\sigma^2C(t,t').
 \ee
 For the purpose of this statement (and its derivation below) we use the symbol $\mathbb{E}$ for averages over realisations of $\uleta$.
 
 \medskip
 
 \underline{Proof of the statement:}\\
 We set up the generating functional for the expression in Eq.~(\ref{eq:effective}), after dividing through by $x(t)$ (I remind the reader that Appendix~\ref{app:clean} contains a brief description of a generating-functional calculation of the Lotka--Volterra systems without this division). After enforcing the dynamics via suitable delta functions in their exponential representation, we have
 \BE
 Z[\ulpsi]&=& \int \D\ulx \D\widehat\ulx\,p_{0}[x(0)] \, \exp\left(i\int dt \,\psi(t)x(t)\right) \nonumber \\
 &&\hspace{-3em}\times\exp\left(i\int dt\, \widehat x(t)\left[\frac{\dot x(t)}{x(t)}-1+x(t)-\Gamma \sigma^2\int dt' G(t,t') x(t')-\mu M(t)-\eta(t)-h(t)\right]\right). \nonumber \\ \label{eq:gf_eff_vanilla}
\EE
This is the generating functional for the effective process, {\em given} a realisation of $\uleta$. Therefore, we next need to average over realisations of $\uleta$. The relevant term in Eq.~(\ref{eq:gf_eff_vanilla}) is $\exp\left(-i\int dt \widehat x(t)\eta(t)\right)$. Using $\mathbb{E}[\eta(t)\eta(t')]=\sigma^2C(t,t')$ and the rules of Gaussian integration (Sec.~\ref{sec:gauss}), we have
\BE
\mathbb{E}\left[\exp\left(i\int dt \widehat x(t)\eta(t)\right)\right]&=&\exp\left(-\frac{1}{2}\sigma^2 \int dt\, dt' C(t,t') \widehat x(t) \widehat x(t')\right).
\EE
Thus, we find [with $Z[\ulpsi]$ as in Eq.~(\ref{eq:gf_eff_vanilla})],
\BE
\mathbb{E}\left[ Z[\ulpsi] \right] & = & \int \D\ulx \D\widehat\ulx\,p_{0}[x(0)] \, \exp\left(i\int dt \,\psi(t)x(t)\right) \nonumber \\
 &&\hspace{-3em}\times\exp\left(i\int dt \widehat x(t)\left[\frac{\dot x(t)}{x(t)}-1+x(t)-\Gamma \sigma^2\int dt' G(t,t') x(t')-\mu M(t)-h(t)\right]\right) \nonumber \\
 &&\hspace{-3em}\times\exp\left(-\frac{1}{2}\sigma^2 \int dt\, dt' C(t,t') \widehat x(t) \widehat x(t')\right).
\EE

This, in turn, is exactly the generating functional in Eq.~(\ref{eq:Z1_LV_vanilla}), which is what we wanted to show.

\qed

\medskip

The statement we have just proved means that the average $\mathbb{E}$ (over realisations of $\uleta$, i.e., realisations of the process in Eq.~(\ref{eq:effective})] is the same as the average $\avg{\dots}_\star$ in Eq.~(\ref{eq:avg_star}) [with $\ulpsi$ set to zero in Eq.~(\ref{eq:avg_star})].

Using Eq.~(\ref{eq:var_wrt_hats}) (and recalling that $\avg{\dots}_i$ has turned into $\avg{\cdots}_\star$, and that we have made the replacement $\ullK \to i\ullG$), we then have
\BE
C(t,t')&=&\avg{x(t)x(t')}_\star, \nonumber \\
G(t,t')&=&-i\avg{x(t)\widehat x(t')}_\star.
\EE
With the definition of $\avg{\cdots}_\star$, and recalling that $Z_1[\ulpsi=\ulnull,\ulh]=1$ in Eq. (\ref{eq:avg_star}) for any $\ulh$, we find
\be
G(t,t')=\frac{\delta}{\delta h(t')} \avg{x(t)}_\star.
\ee
This means that the self-consistent problem for the macroscopic order parameters $\ulM$, $\ullC$ and $\ullG$ is given by
\BE
\dot x(t)&=&x(t)\left[1-x(t)+\Gamma \sigma^2\int dt' G(t,t') x(t')+\mu M(t)+\eta(t)+h(t)\right],\nonumber \\
M(t)&=&\avg{x(t)}_\star, \nonumber \\
C(t,t')&=&\avg{x(t)x(t')}_\star, \nonumber \\
G(t,t')&=&\frac{\delta}{\delta h(t')} \avg{x(t)}_\star,
\EE
where $\avg{\cdots}_\star$ is an average over realisation of the process in the first line. We have used the first relation in Eq.~(\ref{eq:var_wrt_hats}) to write down the self-consistency relation for $\ulM$.

\medskip

We will discuss the interpretation of this effective single-species process and the associated self-consistency relations further in Sec.~\ref{sec:gf_lv_4}.
 
\section{Generating functional analysis for random Lotka--Volterra equations III: Proper derivation of single effective-species process} \label{sec:gf_lv_3}

[For the purposes of this chapter, we return to the situation in which the $\ulpsi_i$ and $\ulh_i$ can (at least potentially) be different for different values of $i$.]

\medskip

The dodgy step of our vanilla derivation consisted in simply setting `averages' of hatted quantities to zero, at a point in the calculation where we could not be sure that these were averages over a {\em bona fide} stochastic process. More precisely, we set $\ulP=\ulnull$ and $\ullL=\ullnull$ at the beginning of Sec.~\ref{sec:simple_LV_vanilla}. Secondly, from the vanilla derivation alone, we cannot be sure what exactly the order parameters $\ulM, \ullC$ and $\ullG$ mean in terms of the original Lotka--Volterra process. Of course it is natural to assume that $C(t,t')=\lim_{N\to\infty}\frac{1}{N}\sum_i \olavg{x_i(t)}$ for example, where the overbar describes average over realisation of the disorder in the original problem, and $\avg{\dots}$ an average over initial conditions (if they are random). We'd also suspect that $M(t)=\lim_{N\to\infty}\frac{1}{N}\sum_i \olavg{x_i(t)}$, and $G(t,t')=\lim_{N\to\infty}\frac{1}{N}\sum_i \frac{\delta}{\delta h_i(t')} \olavg{x_i(t)}$. But we have not formally proved any of this. The main purpose of Sec.~\ref{sec:id_op_LV} is to show precisely this. Along the way we also provide a proper justification for setting $\ulP=\ulnull$ and $\ullL=\ullnull$. This then results in an effective single-species measure and associated single-species process, as described in Sec.~\ref{sec:effective_process}

\subsection{Identification of order parameters}\label{sec:id_op_LV}

Our derivation can be divided into three steps:
\begin{itemize}
\item In step 1, we define order parameters for the {\em original} Lotka--Volterra problem (as opposed to the effective process resulting from the saddle-point calculation). These original order parameters will be called $\ulm, \ulp, \ullc, \ullk$ and $\ullell$. 
\item In step 2 we then write these original order parameters as derivatives of the disorder-averaged generating functional.
\item In step 3 we then show that $\ulm, \ulp, \ullc, \ullk$ and $\ullell$ can be expressed in exactly the same way as the order parameters $\ulM, \ulP, \ullC, \ullK$ and $\ullL$ which we introduced when we carried out the saddle-point calculation. Thus, both sets of order parameters are identical at the saddle point.
\end{itemize}
\medskip

\subsubsection{Step 1: Definition of order parameters from the original Lotka--Volterra problem}
The objects $\ulM, \ulP, \ullC, \ullK$ and $\ullL$ were introduced in Eqs.~(\ref{eq:op}) and we have shown that they can be written in the form of Eqs.~(\ref{eq:var_wrt_hats}) in the thermodynamic limit.

\medskip

Let us now define the following
\BE
m(t)&=&\lim_{N\to\infty}\frac{1}{N}\sum_i \olavg{x_i(t)},\nonumber\\
c(t,t')&=&\lim_{N\to\infty}\frac{1}{N}\sum_i \olavg{x_i(t) x_i(t')}, \nonumber \\
k(t,t')&=&i\lim_{N\to\infty}\frac{1}{N}\sum_i \frac{\delta}{\delta h_i(t')} \olavg{x_i(t)},\nonumber \\
p(t)&=&i\lim_{N\to\infty}\frac{1}{N}\sum_i  \frac{\delta}{\delta h_i(t)}\olavg{1},\nonumber\\
\ell(t,t')&=&\lim_{N\to\infty}\frac{1}{N}\sum_i \frac{\delta}{\delta h_i(t)\delta h_i(t')}\olavg{1},
\label{eq:op0}
\EE
where the angle brackets stand for an average over realisations of the original Lotka--Volterra dynamics in Eq.~(\ref{eq:lv})  for given disorder. In our case the only randomness of the dynamics (for fixed disorder) is in the potentially random initial conditions. However, one could, in principle, also include dynamic noise. The overbar describes an average over realisations of the disorder, i.e., the interaction matrix elements $\alpha_{ij}$ in Eq.~(\ref{eq:lv}).

\medskip

We note the obvious fact $\olavg{1}=1$ for all choices of $h_i(\cdot)$. This means that $p(t)$ and $\ell(t,t')$ are both zero by construction. We will use this to show that $P(t)=0$ and $L(t,t')=0$.

\medskip

We re-iterate that the objects in Eqs.~(\ref{eq:op0}) are constructed from the original dynamics. This is clear immediately for $m(t)$ and $c(t,t')$. These two objects can, in principle, be measured from numerical simulations, modulo the fact that simulations will necessarily be for finite $N$ and that any measurements will have to rely on a finite number of realisation runs. For a given (large) number of species $N$ one runs a large number of realisations of the random Lotka--Volterra system (each realisation with a new draw of the interaction coefficients, and with a new initial condition).  The object $\frac{1}{N}\sum_i \olavg{x_i(t) x_i(t')}$ for example can then be obtained from these simulations, as an average of the object $\frac{1}{N}\sum_i x_i(t) x_i(t')$ over realisations of the disorder and the initial condition. Measuring $k(t,t')$ is a little more complicated (one would have to run the dynamics for different $\ulh_i$), but nonetheless, $\ullk$ can (at least in principle) be obtained from simulating the original Lotka--Volterra system.

\medskip

\subsubsection{Step 2: Order parameters for the original problem as derivatives of the disorder-averaged generating functional}
We know that moments of the $\{x_i(t)\}$ can be expressed as derivatives of the generating functional with respect to the $\{\psi_i(t)\}$, evaluated at $\ulbpsi=0$. We start with a fixed realisation of the disorder in mind, so the word `moment' refers to an average over initial conditions only at this point. For example
\be
\avg{x_i(t) x_i(t')}=-\left.\frac{\delta^2 Z[\ulbpsi]}{\delta \psi_i(t)\delta \psi_i(t')}\right|_{\ulbpsi=0},
\ee
where $Z[\ulbpsi]$ is the generating functional of the system for the relevant fixed realisation of the interaction matrix [Eq.~(\ref{eq:gflv0})].

\medskip

Taking derivatives of the {\em disorder-averaged} generating functional instead implements an additional average over the interaction coefficients. We have
\BE\label{eq:equiv_aux}
m(t)&=&-i\lim_{N\to\infty} \frac{1}{N}\sum_i\frac{\delta}{\delta\psi_i(t)}\left.\overline{Z[\ulbpsi]}\right|_{\ulbpsi=0}, \nonumber \\
c(t,t')&=&-\lim_{N\to\infty} \frac{1}{N}\sum_i\frac{\delta^2}{\delta\psi_i(t)\delta\psi_i(t')}\left.\overline{Z[\ulbpsi]}\right|_{\ulbpsi=0}, \nonumber \\
k(t,t')&=&\lim_{N\to\infty} \frac{1}{N}\sum_i  \frac{\delta^2}{\delta\psi_i(t)\delta h_i(t')}\left.\overline{Z[\ulbpsi]}\right|_{\ulbpsi=0}, \nonumber \\
p(t)&=&i\lim_{N\to\infty} \frac{1}{N}\sum_i  \frac{\delta}{\delta h_i(t')}\left.\overline{Z[\ulbpsi]}\right|_{\ulbpsi=0},\nonumber\\
\ell(t,t')&=&\lim_{N\to\infty} \frac{1}{N}\sum_i  \frac{\delta^2}{\delta h_i(t)\delta h_i(t')}\left.\overline{Z[\ulbpsi]}\right|_{\ulbpsi=0}.
\EE

\subsubsection{Step 3: Equivalence to order parameters at the saddle point}
We now use the saddle-point expression for $\overline Z$ to show that the expressions in Eqs.~(\ref{eq:equiv_aux}) are the same as those in Eqs.~(\ref{eq:var_wrt_hats}) at the saddle point. We recall the definition of the average $\avg{\cdots}_i$ in Eq.~(\ref{eq:avg_i}), as well as the definition of the object $\Omega$ in Eq.~(\ref{eq:omega}). We repeat the latter here
\BE
\Omega&=&N^{-1}\sum_i\ln\bigg[\int \D\ulx \D\widehat \ulx\, p_{0}[x(0)]\exp\left(i\int dt~ \psi_i(t)x(t)\right)\nonumber \\
&&\times \exp\left(i\int dt ~\widehat x(t)  \left(\frac{\dot x(t)}{x(t)}-[1-x(t)]-h_i(t)\right)\right)\nonumber \\
&&\times \exp\left(-i\int dt ~ dt' \left[\widehat C(t,t')x(t)x(t')+\widehat L(t,t')\widehat x(t)\widehat x(t')+\widehat K(t,t') x(t)\widehat x(t')\right]\right)\nonumber \\
&&\times \exp\left(-i\int dt~  [\widehat M(t)x(t)+\widehat P(t) \widehat x(t)]\right)\bigg].
\label{eq:omega_2}
\EE

From Eq.~(\ref{eq:sp}) we have
\be
\overline{Z}[\ulbpsi]=\int D[\ulM,\widehat \ulM] \D[\ulP,\widehat \ulP] \D[\ullC,\widehat\ullC]\D[\ullK,\widehat \ullK] \D[\ullL,\widehat \ullL] \,\exp\left(N\left[\Psi+\Phi+\Omega+{\cal O}(N^{-1})\right]\right),
\ee
noting that only $\Omega$ contains the $\{\psi_i(t)\}$ and $\{h_i(t)\}$. 

\medskip

Let us focus on the expression for $c(t,t')$ in Eqs.~(\ref{eq:equiv_aux}) as an example.  We have
\BE
c(t,t')&=&-\lim_{N\to\infty} \frac{1}{N}\sum_i\frac{\delta^2}{\delta\psi_i(t)\delta\psi_i(t')}\left.\overline{Z[\ulpsi]}\right|_{\ulbpsi=0}, \nonumber \\
&=&-\lim_{N\to\infty} \frac{1}{N}\sum_i \frac{\delta}{\delta\psi_i(t)}\int D[\ulM,\widehat \ulM] \D[\ulP,\widehat \ulP] \D[\ullC,\widehat\ullC]\D[\ullK,\widehat \ullK] \D[\ullL,\widehat \ullL] \, \nonumber \\
&&\times\bigg\{\left.N \frac{\delta\Omega}{\delta \psi_i(t')} \exp\left(N\left[\Psi+\Phi+\Omega+{\cal O}(N^{-1})\right]\right)\bigg\}\right|_{\ulbpsi=0}\nonumber \\
&=&-\lim_{N\to\infty} \frac{1}{N}\sum_i \int D[\ulM,\widehat \ulM] \D[\ulP,\widehat \ulP] \D[\ullC,\widehat\ullC]\D[\ullK,\widehat \ullK] \D[\ullL,\widehat \ullL] \, \bigg\{\nonumber \\
&&\left.\left[ N\frac{\delta^2\Omega}{\delta\psi_i(t)\delta \psi_i(t')} +N^2 \left(\frac{\delta\Omega}{\delta \psi_i(t)}\right)\left(\frac{\delta\Omega}{\delta \psi_i(t')}\right)\right]\right.\nonumber \\
&& \left.\times\exp\left(N\left[\Psi+\Phi+\Omega+{\cal O}(N^{-1})\right]\right)\bigg\}\right|_{\ulbpsi=0}. 
\EE
Following the rules of saddle-point integration (and using the normalisation of the generating functional, $\left.\overline{Z[\ulpsi]}\right|_{\ulbpsi=0}=1$), the integral is given by the the object $\left.N\frac{\delta^2\Omega}{\delta\psi_i(t)\delta \psi_i(t')} +N^2 \left(\frac{\delta\Omega}{\delta \psi_i(t)}\right)\left(\frac{\delta\Omega}{\delta \psi_i(t')}\right)\right|_{\ulbpsi=0}$, evaluated at the saddle point. We therefore find
\BE
c(t,t')&=&-\lim_{N\to\infty} \frac{1}{N}\sum_i\left.\left[ N\frac{\delta^2\Omega}{\delta\psi(t)\delta\psi_i(t')} +N^2 \left(\frac{\delta\Omega}{\delta \psi_i(t)}\right)\left(\frac{\delta\Omega}{\delta \psi_i(t')}\right)\right]\right|_{\mbox{SP}, \ulbpsi=0},~~~~
\EE
where the subscript `SP' indicates that this is to be evaluated at the saddle point.

We next use 
\be
\Omega=\frac{1}{N}\sum_j \ln Z_1[\ulpsi_j, \ulh_j],
\ee
with $Z_1[\ulpsi_j, \ulh_j]$ given in Eq.~(\ref{eq:z1_def}). Therefore (and leaving out the argument $\ulh_i$ inside $Z_1$), we have identities such as $N\frac{\delta\Omega}{\delta \psi_i(t)}=\frac{\delta\ln Z_1[\ulpsi_i]}{\delta \psi_i(t)}$ for any fixed $i$. We can then write
\BE
c(t,t')&=&-\lim_{N\to\infty} \frac{1}{N}\sum_i\left.\left[ \frac{\delta^2\ln\,Z_1[\ulpsi_i]}{\delta\psi_i(t)\delta \psi_i(t')} +\left(\frac{\delta\ln\,Z_1[\ulpsi_i]}{\delta \psi_i(t)}\right)\left(\frac{\delta\ln\,Z_1[\ulpsi_i]}{\delta \psi_i(t')}\right)\right]\right|_{\mbox{SP}, \ulbpsi=0}.\nonumber \\
\EE
The first term in the square brackets is
\BE
\frac{\delta^2\ln\,Z_1[\ulpsi_i]}{\delta\psi_i(t)\delta \psi_i(t')}  &=&\frac{\delta}{\delta\psi_i(t)}\frac{1}{Z_1[\ulpsi_i]}\frac{\delta Z_1[\ulpsi_i]}{\delta \psi_i(t')} \nonumber \\
&=&-\underbrace{\frac{1}{Z_1[\ulpsi_i]^2}\frac{\delta Z_1[\ulpsi_i]}{\delta \psi_i(t)}\frac{\delta Z_1[\ulpsi_i]}{\delta \psi_i(t')}}_{=\left(\frac{\delta\ln\,Z_1[\ulpsi_i]}{\delta \psi_i(t)}\right)\left(\frac{\delta\ln\,Z_1[\ulpsi_i]}{\delta \psi_i(t')}\right)} +\frac{1}{Z_1[\ulpsi_i]}\frac{\delta^2 Z_1[\ulpsi_i]}{\delta \psi_i(t) \delta \psi_i(t')}.
 \EE
 Therefore, 
 \BE
c(t,t')&=&-\lim_{N\to\infty} \frac{1}{N}\sum_i\left.\frac{1}{Z_1[\ulpsi_i,\ulh_i]}\frac{\delta^2 Z_1[\ulpsi_i,\ulh_i]}{\delta \psi_i(t) \delta \psi_i(t')}\right|_{\mbox{SP}, \ulbpsi=0}.
\EE
Using the definition of $Z_1$ in Eq.~(\ref{eq:z1_def}) as well as that of the average $\avg{\cdots}_i$ in Eq.~(\ref{eq:avg_i}), we can then take the remaining derivatives, and find
\BE\label{eq:final}
c(t,t')=\lim_{N\to\infty}\frac{1}{N}\sum_i\left.\avg{x(t)x(t')}_i\right|_{\mbox{SP},\ulpsi_i=0}
\EE
Comparing with Eq.~(\ref{eq:var_wrt_hats}), we thus conclude that the quantity $c(t,t')$, defined from the original Lotka--Volterra problem, is the same as the order parameter $C(t,t')$ we introduced in the course of the generating functional calculation, provided the expressions in Eq.~(\ref{eq:var_wrt_hats}) are evaluated at the saddle-point and at $\mathbf{\ulpsi}=0$. Analogous equivalences for the remaining order parameters can be demonstrated along similar lines.

\medskip

We can therefore summarise (all valid at the saddle point and for $\mathbf{\ulpsi}=0$),
\BE\label{eq:fina2l}
\lim_{N\to\infty} M(t)&=&m(t)=\lim_{N\to\infty}\frac{1}{N}\sum_i\left.\avg{x(t)}_i\right|_{\mbox{SP},\ulpsi_i=0}, \nonumber \\
\lim_{N\to\infty} P(t)&=&p(t)=\lim_{N\to\infty}\frac{1}{N}\sum_i\left.\avg{\widehat x(t)}_i\right|_{\mbox{SP},\ulpsi_i=0}=0, \nonumber \\
\lim_{N\to\infty} C(t,t')&=&c(t,t')=\lim_{N\to\infty}\frac{1}{N}\sum_i\left.\avg{x(t)x(t')}_i\right|_{\mbox{SP},\ulpsi_i=0}, \nonumber \\
\lim_{N\to\infty} K(t,t')&=&k(t,t')=\lim_{N\to\infty}\frac{1}{N}\sum_i\left.\avg{x(t)\widehat x(t')}_i\right|_{\mbox{SP},\ulpsi_i=0}, \nonumber \\
\lim_{N\to\infty} L(t,t')&=&\ell(t,t')=\lim_{N\to\infty}\frac{1}{N}\sum_i\left.\avg{\widehat x(t)\widehat x(t')}_i\right|_{\mbox{SP},\ulpsi_i=0}=0.
\EE
We have therefore demonstrated two relevant facts: (i) In the thermodynamic limit and at the saddle point, the order parameters $\ulM, \ullC$ and $\ullK$, initially introduced as bookkeeping variables in the generating-functional calculation have an interpretation in terms of averages in the original Lotka--Volterra problem. (ii) The order parameters $\ulP$ and $\ullL$ vanish at the saddle point.

\subsection{Effective process for a representative species}\label{sec:effective_process}
The fields $\ulpsi_i$ have now served their purpose and we can set them to zero. We also restrict the perturbation fields $h_i(t)$ to be uniform across species, that is, replace $h_i(t)\to h(t)$. Recalling that the only $i$-dependence of the average $\avg{\dots}_i$ is through the fields $h_i(t)$ and $\ulpsi_i$ [Eq.~(\ref{eq:avg_i})] we then find that the operation $\avg{\dots}_i$ becomes independent of $i$. We simply write $\avg{\dots}_*$ for this average, and include in the definition of $\avg{\cdots}_\star$ that this is to be evaluated at the saddle point. 

We then have 
\be\label{eq:av_g}
\avg{g[\ulx,\widehat\ulx]}_*= \frac{\int \D\ulx \D\widehat\ulx \, g[\ulx,\widehat\ulx]\, {\cal M}[\ulx,\widehat\ulx]}{\int  \D\ulx \D\widehat\ulx \, \,  {\cal M}[\ulx,\widehat\ulx]}
\ee
with
\BE
{\cal M}[\ulx,\widehat\ulx]&=&  p_{0}[x(0)]\, \times \exp\left(i\int dt ~\widehat x(t)  \left(\frac{\dot x(t)}{x(t)}-[1-x(t)]-h(t)\right)\right)\nonumber \\
&&\times \exp\left(-i\int dt ~ dt' \left[\widehat C(t,t')x(t)x(t')+\widehat L(t,t')\widehat x(t)\widehat x(t')+\widehat K(t,t') x(t)\widehat x(t')\right]\right)\nonumber \\
&&\times \left.\exp\left(-i\int dt~  [\widehat M(t)x(t)+\widehat P(t) \widehat x(t)]\right)\right|_{\mbox{SP}}.
\EE
At the saddle point we have from Eq.~(\ref{eq:sp_aux1}) and using $\ulP=\ulnull, \ullL=\ullnull$,
\BE
i\widehat M(t)&=&i\mu P(t)=0, \nonumber \\
i\widehat P(t)&=&i\mu M(t), \nonumber \\
i\widehat C(t,t')&=&\frac{1}{2}\sigma^2L(t,t')=0,\nonumber \\
i\widehat K(t,t')&=&\Gamma \sigma^2 K(t',t),\nonumber \\
i\widehat L(t,t')&=&\frac{1}{2}\sigma^2C(t,t').\label{eq:sp_aux1_3}
\EE
Using this, and again implementing the change of notation $\ullK\to i\ullG$, we find
\BE
{\cal M}[\ulx,\widehat\ulx]&=&p_{0}[x(0)] \, \exp\left(i\int dt ~\widehat x(t)  \left(\frac{\dot x(t)}{x(t)}-[1-x(t)]-\mu M(t)-h(t)\right)\right)\nonumber \\
&&\times \exp\left(-\sigma^2\int dt ~ dt' \left[ \frac{1}{2}C(t,t')\widehat x(t)\widehat x(t')+i\Gamma G(t',t) x(t)\widehat x(t')\right]\right).\label{eq:m_effective}
\EE
Following the lines of Sec.~\ref{sec:eff_proc_interpret}, the single effective species measure ${\cal M}[\ulx,\widehat\ulx]$ again describes the effective process
\BE
\dot x(t)=x(t)\left[1-x(t)+\Gamma \sigma^2\int dt' G(t,t') x(t')+\mu M(t)+\eta(t)+h(t)\right]. \label{eq:effective_copy_2}
\EE
From Eq.~(\ref{eq:var_wrt_hats}) (and recalling that $\lim_{N\to\infty}N^{-1}\sum_i\avg{\cdots}_i$ has turned into $\avg{\cdots}_\star$), we also have the self-consistency relations
\BE
M(t)&=&\avg{x(t)}_*, \nonumber \\
\avg{\eta(t)\eta(t')}_\star&=&\sigma^2 C(t,t')~\mbox{with}~ C(t,t')=\avg{x(t)x(t')}_*,\nonumber \\
G(t,t')&=&-i\avg{x(t)\widehat x(t')}_*,
\label{eq:var_wrt_hats_2}
\EE
and where the last relation can be written in the form $G(t,t')=\frac{\delta}{\delta h(t')}\avg{x(t)}_\star$.

\subsection{Comments on the legitimacy of the vanilla approach}
The main deficiency of the vanilla approach was to set the order parameters $\ulP$ and $\ullL$ to zero without proper justification. Nonetheless, the vanilla method leads to the right effective process. Once this process is established, the self-consistency relations for $\ulP$ and $\ullL$ result in the conclusion that $\ulP$ and $\ullL$ must vanish (because they are averages of conjugates field over a {\em bona fide} stochastic dynamics, namely the effective process). The vanilla approach effectively amounts to simply asserting $\ulP=\ulnull$ and $\ullL=\ullnull$, and then justifying them retrospectively from the effective process. This is not incorrect, albeit a little weaker than the bottom-up derivation from first principles in Sec.~\ref{sec:gf_lv_3}.

\medskip

The second issue is that we have kept the source fields $\psi_i(t)$ general throughout Sec.~\ref{sec:gf_lv_2}, even if setting $\ulP$ and $\ullL$ to zero is at best justified in the limit $\ulbpsi\to\ulbnull$. This is not a serious problem though. In fact we are not using the source fields in Sec.~\ref{sec:gf_lv_2}, so no harm would have been done if we had simply set them to zero from the beginning. From the vanilla approach we would then not obtain a generating functional, but instead the single effective particle measure ${\cal M}[\ulx,\widehat\ulx]$ in Eq (\ref{eq:m_effective}). But of course the generating functional and the single effective particle measure are really two sides of the same coin, and describe the same effective process (the generating functional is the Fourier transform of the probability measure in the space of paths).

\medskip

As an overall conclusion, it is therefore fair to say that the vanilla approach is OK, but only if it is used responsibly and if the user is aware of the subtleties. 

\section{Generating functional analysis for random Lotka--Volterra equations IV: Dynamic mean-field process and its interpretation}\label{sec:gf_lv_4}

We have now arrived at the dynamic mean-field description of the Lotka--Volterra system. This description consists of the effective single-species process
\BE
\dot x(t)=x(t)\left[1-x(t)+\Gamma \sigma^2\int dt' G(t,t') x(t')+\mu M(t)+\eta(t)+h(t)\right], \label{eq:effective_copy}
\EE
along with the self-consistency relations
\BE
 \sigma^2C(t,t')&=&\sigma^2\avg{x(t)x(t')}_*=\avg{\eta(t)\eta(t')}_*,\nonumber \\
 G(t,t')&=&\frac{\delta }{\delta h(t')} \avg{x(t)}_*,\nonumber \\
M(t)&=&\avg{x(t)}_*.\label{eq:sc_copy}
\EE
The notation $\avg{\cdots}_*$ denotes an average over realizations of the effective dynamics (\ref{eq:effective_copy}). 

\medskip

Technically, the first line in Eqs.~(\ref{eq:sc_copy}) entails two different statements: One is the actual self-consistency relation $\sigma^2\avg{x(t)x(t')}_*=\avg{\eta(t)\eta(t')}_*$. The second is the fact that the order parameter $C(t,t')$ ($=c(t,t')$ in the original problem) is obtained from an average of the object $x(t)x(t')$ over realisations of the effective process, i.e., $C(t,t')=\avg{x(t)x(t')}_*$.
\medskip

We'll now briefly talk about the mathematical structure of this self-consistent problem, and then about its interpretation and relation to the original Lotka--Volterra system.

\subsection{Mathematical structure}
First of all one needs to realise that Eqs.~(\ref{eq:effective_copy},\ref{eq:sc_copy}) are primarily a set of equations determining the dynamical order parameters $M(t), C(t,t')$ and $G(t,t')$. They are the analog of Eq.~(\ref{eq:dyn_m}) in the kinetic Ising model that we discussed as a warm-up, and of Eqs.~(\ref{eq:M_avg_proper}) and (\ref{eq:eff_process_proper}) in the toy problem in Sec.~\ref{sec:toy_model}.

Eqs.~(\ref{eq:effective_copy},\ref{eq:sc_copy}) do not provide explicit expressions for the order parameters. Instead this is an implicit system, that needs to be solved self-consistently. This means that the stochastic (integro-) differential equation for the effective species abundance $x(t)$ involves the order parameters -- the quantities $M(t)$ and $G(t,'t)$ appear on the right-hand side of Eq.~(\ref{eq:effective_copy}). The properties of the noise $\eta(t)$, also on the right-hand side of the effective process, are determined by $\ullC$, where $C(t,t')=\avg{x(t)x(t')}_*$.

Conversely, Eqs.~(\ref{eq:sc_copy}) indicate that $C(t,t'), G(t,t')$ and $M(t)$ are to be obtained as averages over realisations of the effective process. So we have here a Catch-22 situation, we need the order parameters to construct realisations of the effective process, and at the same time, we need these realisations to know what the order parameters are. This is what we mean by a problem that has to be solved {\em self-consistently}.

\medskip

How could one go about solving this set of equations? We'll look for fixed points of the effective dynamics in Sec.~\ref{sec:fixed_point} -- this can be done analytically. Other than that, one would have to do this numerically. One possible method is to proceed along the following steps:
\begin{enumerate}
\item[1.] Make an initial guess for the functions $C(t,t'), G(t,t')$ and $M(t)$.
\item[2.] Use $C(t,t')$ to generate realisations of the noise $\eta(t)$.
\item[3.] Use these realisations, and the guesses for $G(t,t')$ and $M(t)$ to generate realisations of the process $x(t)$.
\item[4.] From these realisations of $x(t)$, obtain new functions $C(t,t'), G(t,t')$ and $M(t)$ via the self-consistently relations.
\item[5.] Repeat steps 2-4 until convergence (i.e., until the order parameters do not change any more).
\end{enumerate}
I would not say that these steps are easy, and I am not really going to go into {\em how} one would actually do this (determining the response function $G(t,t')$ in step 4 for example is non-trivial). The point is here just to explain the nature of the self-consistent problem. A procedure along those lines is described in \cite{froy}. 

Alternatively, there is a celebrated method by Eissfeller and Opper \cite{oppereissfeller}. This is for discrete-time processes, and builds up the matrices $C(t,t')$ and $G(t,t')$ gradually, time step by time step. Realisations of $x(t)$ and $\eta(t)$ are also constructed time step by time step in the Eissfeller--Opper scheme.

\subsection{Interpretation}

This is a good point to briefly think about the interpretation of the effective process (the mean field dynamics). Suppose we have found the order parameters self-consistently. With those $\ullC, \ullG$ and $\ulM$ the effective process in Eq.~(\ref{eq:effective_copy}) then describes what a typical or representative species `experiences'. The right-hand side is the change of abundance in time of such a `typical' species.  The word `typical' does not mean an `average' over species. The right-hand side of the effective dynamics contains randomness (the noise $\eta(t)$). This noise comes from the original problem -- the quenched disorder. For fixed disorder, different species in the original problem will experience different right-hand sides of Eq.~(\ref{eq:lv}). The noise $\eta(t)$ in the effective process reflects this.

To be a little more precise, the statement is the following. Eqs.~(\ref{eq:effective_copy},\ref{eq:sc_copy}), if solved self-consistently, generate an ensemble of realisations $x(t)$ of the effective process. The statistics of this ensemble is the same as that of trajectories $x_i(t)$ one obtains in the original Lotka--Volterra problem (for large $N$). We can phrase this slightly more formally as follows:

\medskip

{\em Equivalence of averages in original problem and at the level of single effective species.}\\
For any function $f[\ulx_i,\widehat\ulx_i]$ we have (at the saddle point and for $\ulbpsi=\ulbnull$),
\BE
\lim_{N\to\infty}\frac{1}{N}\sum_i \olavg{f[\ulx_i,\widehat\ulx_i]}=\lim_{N\to\infty}\frac{1}{N}\sum_i\left.\avg{f[\ulx,\widehat \ulx]}_i\right|_{\mbox{SP},\ulpsi_i=0}.\label{eq:general_equivalence}
\EE
This is demonstrated in Appendix~\ref{app:equivalence}. 

\medskip

We can say the same thing again in different words. Suppose I fix $\mu, \Gamma$ and $\sigma^2$, and I choose a sufficiently large $N$. I then draw a realisation of the disorder, and integrate the Lotka--Volterra equations (\ref{eq:lv}). This coupled system produces trajectories $\ulx_1, \ulx_2, \dots, \ulx_N$, one for each species. One can therefore think of a distribution of single-species trajectories. Modulo finite-size effects ($N$ will obviously have to be finite for the numerical integration of the Lotka--Volterra equations) this distribution is the same as the distribution of trajectories for the effective process (at the self-consistent solution).
\begin{figure}[h!]
\begin{center}
\includegraphics[width=1\textwidth]{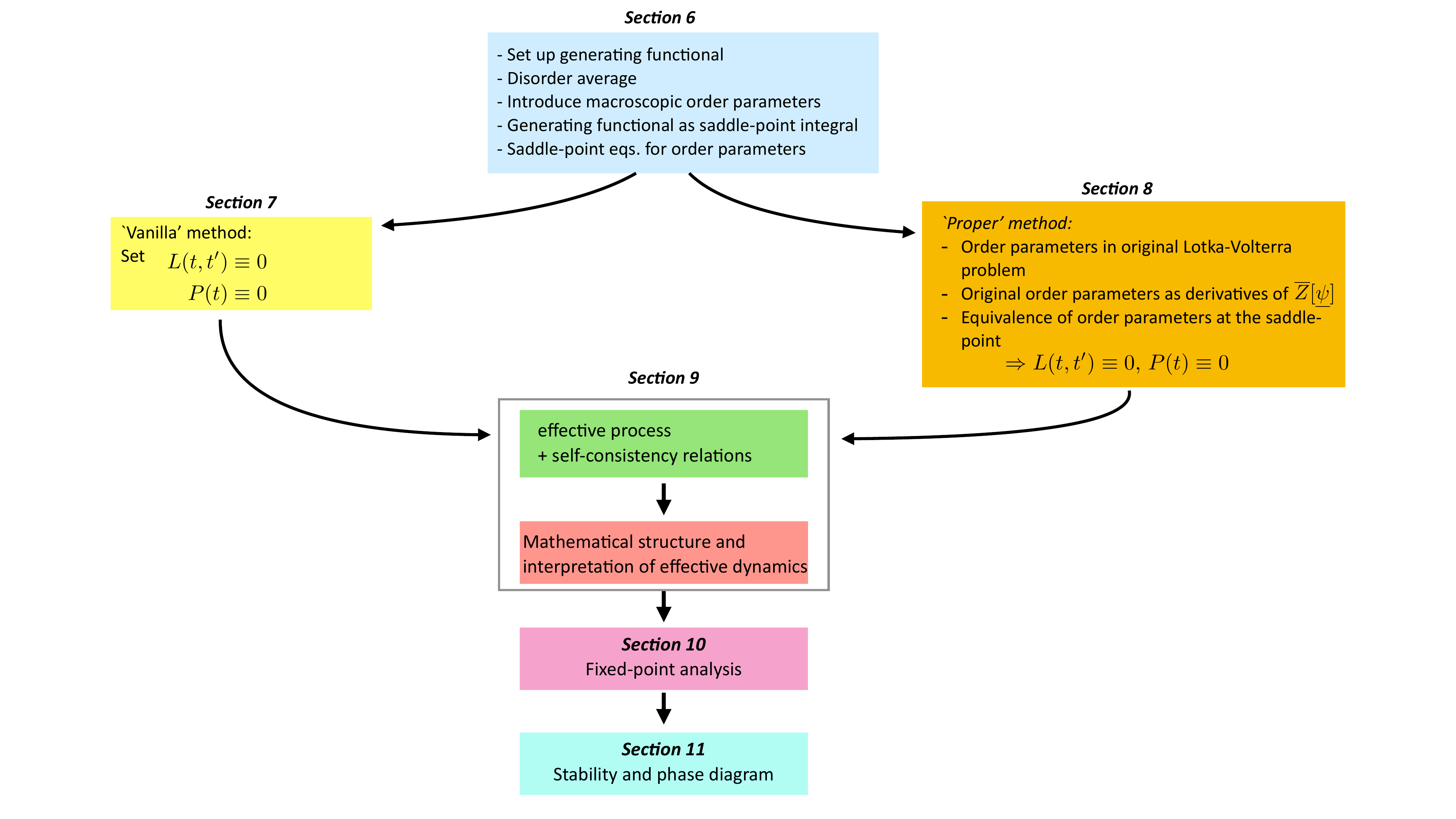}
\end{center}
\caption{Illustration of the relation of the different sections on the Lotka--Volterra problem. \label{fig:LV_flow}}
\end{figure}

\subsection{Summary so far}
In Sec.~\ref{sec:gf_lv_1} we set up the generating functional for the Lotka--Volterra problem, and then wrote the disorder-averaged generating functional in the form of a saddle-point integral. The steps involved carrying out the disorder average, and the introduction order parameters as book-keeping variables. We then derived the saddle-point relations for the order parameters. 

\medskip

There are then two ways to proceed. Along the easier route (Sec.~\ref{sec:gf_lv_2}) we simply assert that the order parameters $\ulP$ and $\ullL$ vanish, and we then arrive directly at the effective dynamics. A more complete and longer route (Sec.~\ref{sec:gf_lv_3}) involves the definition of order parameters at the level of the original problem, identifying these with the book-keeping variables at the saddle point, and explicitly showing that averages of hatted variables do indeed vanish (i.e., $\ulP=\ulnull$ and $\ullL=\ullnull$).

\medskip

Either way one arrives at the same dynamical mean field theory (Sec.~\ref{sec:gf_lv_4}).

\medskip

Our next steps are now as follows: In Sec.~\ref{sec:fixed_point} we make a fixed-point ansatz for the effective process and derive analytical relations for the resulting order parameters at the fixed point. In Sec.~\ref{sec:stability_pg} we then study the stability of the fixed point, and obtain a phase diagram for the Lotka--Volterra problem, indicating for what combinations pf $\mu, \sigma^2$ and $\Gamma$ the system is stable, and what types of instability occur where in parameter space.

\medskip

The relation of these different sections to one another is also illustrated in Fig.~\ref{fig:LV_flow}.

\section{Fixed point analysis}\label{sec:fixed_point}
The purpose of the perturbation fields $h_i(t)$ was to generate response functions. The fields are not part of the actual model. Therefore, with the exception of Sec.~\ref{sec:chi}, we always set $h_i(t)\equiv 0$ throughout this section.

\subsection{Fixed-point ansatz and its consequences}
\subsubsection{Fixed-point ansatz for effective abundances}
We now assume that the system reaches a stationary state and that this stationary state does not depend on the initial condition (i.e., we assume absence of long-term memory, see also \cite{galla2006}). The response function $G$ is then a function of time differences only, i.e. $G(t,t')=G(\tau)$, where $\tau=t-t'$. Causality dictates that $G(\tau<0)=0$. 

We assume further that the dynamics reaches a fixed point. In the original system this means that all $x_i$ tend to fixed-point values $x_i^*$ at long times. At the level of the effective dynamics, all trajectories of the single-species process tend to constant values $x(t)\to x^*$, but it is important to realise that different realisations of the effective process will have different fixed-point values. The quantity $x^*$ is therefore a static (time-independent) but random quantity. For given model parameters $\mu, \Gamma$ and $\sigma^2$ the distribution of $x^*$ corresponds to the distribution of the $x_i^*$ in the original problem. 

As a consequence the order parameter $C(t,t')=\avg{x(t)x(t')}_*=\avg{(x^*)^2}$ is constant (independent of $t$ and $t'$). We write 
\be
C(t,t')\equiv q.
\ee
\subsubsection{The noise variable in the effective process becomes static}
We next note that the relation
\be
 \avg{\eta(t)\eta(t')}_*=\sigma^2C(t,t')
\ee
in Eq.~(\ref{eq:sc_copy}) means that $\avg{\eta(t)\eta(t')}_*=\sigma^2 q$ for all $t,t'$. The only way this can be the case is if the random variable $\eta(t)$ itself is independent of time. To see this, look at
\BE
\avg{[\eta(t)-\eta(t')]^2}_*&=&\avg{(\eta(t))^2}_* + \avg{(\eta(t'))^2}_*-2\avg{\eta(t)\eta(t')}_* \nonumber \\
&=& C(t,t)+C(t',t')-2C(t,t').
\EE
If $C(t,t)$, $C(t',t')$ and $C(t,t')$ are all equal to $\sigma^2 q$ for times $t,t'$ in the stationary regime, then $\avg{[\eta(t)-\eta(t')]^2}_*=0$. If the average of a square is zero, the object that is squared must be zero itself with probability one. In other words, we find that $\eta(t)=\eta(t')$ with probability one. Therefore, $\eta(t)$ is in effect a static random variable, we write $\eta(t)=\eta^*$. The variance of $\eta^*$ is $\sigma^2 q$. The mean is zero (we already knew that $\avg{\eta(t)}_*=0$ for all $t$).

\subsubsection{Distribution of fixed-point abundances}
 We now introduce the fixed-point ansatz into the effective process in Eq.~(\ref{eq:effective_copy}). In doing this we note that $\int dt' G(t,t')x(t')$ becomes $\int_{-\infty}^t dt' G(t-t') x^*$, where we have sent the starting point of the dynamics to time $-\infty$ (this has the advantage that any finite time $t$ is now in the fixed-point regime). We can then replace 
 \be
 \int dt' G(t,t') x(t') \to x^* \int_0^\infty d\tau \,G(\tau) = x^* \chi,
 \ee
where we have defined
 \be
 \chi \equiv \int_0^\infty d\tau \,G(\tau).
 \ee
We then find from Eq.~(\ref{eq:effective_copy}), setting $h(t)\equiv 0$ (as explained at the beginning of Sec.~\ref{sec:fixed_point}),
\be \label{eq:fp}
x^*\left[1-x^*+\Gamma\sigma^2\chi x^*+\mu M^*+\eta^*\right]=0.
\ee
 Keeping in mind that $\avg{(\eta^*)^2}_*=\sigma^2 q$ we write $\eta^*=\sqrt{q}\sigma z$ with $z$ a static Gaussian random variable of mean zero and unit variance. 
\\

Eq. (\ref{eq:fp}) always (for all $z$) has the solution $x^*=0$. The second solution, 
\be\label{eq:xstar}
x^*=\frac{1+\mu M^*+\sqrt{q}\sigma z}{1-\Gamma\sigma^2\chi}
\ee
can only be physical for those values of $z$ for which this expression is non-negative ($x^*$ is a fixed-point abundance, and must be non-negative). In the following we use
\be\label{eq:max}
x(z)=\mbox{max}\left(0, \frac{1+\mu M^*+\sqrt{q}\sigma z}{1-\Gamma\sigma^2\chi}\right),
\ee
that is to say, we use the expression in Eq.~(\ref{eq:xstar}) whenever this expression is positive, and $x^*=0$ for values of $z$ for which the right-hand side of Eq.~(\ref{eq:xstar}) is negative (or equal to zero).
The zero solution can be seen to be unstable when the right-hand side of Eq.~(\ref{eq:xstar}) is positive, see Sec.~\ref{sec:lsa} below.
\subsection{Self-consistency relations for order parameters}
\subsubsection{Summary of status so far}
We have now achieved the following: 
\begin{enumerate}
\item[1.] We have made a fixed-point ansatz for the effective process.
\item[2.] We have established the consequences for the macroscopic order parameters. Only the quantities $q, M^*$ and $\chi$ (which are derived from the order parameters) are relevant at the fixed point.
\item[3.] We have expressed the fixed points of the effective process in terms of the order parameters and a static noise variable $z$ (Gaussian of mean zero and variance one) [Eq.~(\ref{eq:max})].
\end{enumerate}
This almost completes our fixed-point simplifications of the self-consistent problem, but not quite. We still have to express the quantities $M^*, q$ and $\chi$ as averages over the fixed-point abundances $x(z)$, or objects derived from these abundances. This is what we will do next.
\subsubsection{Self-consistent expressions for order parameters $M^*$ and $q$}
The order parameter $M(t)$ is equal to $\avg{x(t)}_*$ at finite times, so it is clear that $M^*$ must be the average over the fixed-point abundances. We have
\be
M^*=\avg{x(z)}_*,
\ee
where we note that, at the fixed-point, the average $\avg{\cdots}_*$ is really an average over the Gaussian random variable $z$. The functional form of $x(z)$ is given in Eq.~(\ref{eq:max}). Similarly, the quantity $q$ is the value the correlation function $C(t,t')=\avg{x(t)x(t')}_*$ takes in the fixed-point regime (where  $C(t,t')$ becomes independent of $t$ and $t'$ as discussed above). We therefore have
\be
q=\avg{x(z)^2}_*.
\ee
\subsubsection{Self-consistent relation for the integrated response $\chi$} \label{sec:chi}
[For the purposes of this subsection (and only this subsection) we allow non-zero perturbation fields again. We impose $h_i(t)\equiv h(t)$ for all $i$.]

\medskip

\begin{figure}[h!]
\begin{center}
\hspace{-0.2\textwidth}\includegraphics[width=1.2\textwidth]{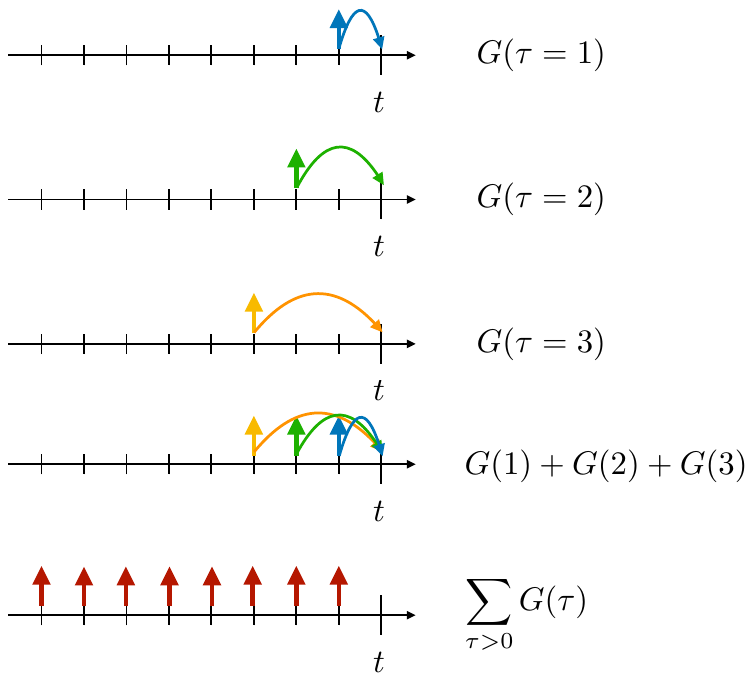}
\vspace{-10em}
\end{center}
\caption{Illustration of the interpretation of the integrated response $\chi=\sum_{\tau>0} G(\tau)$ in a discrete-time system. $G(\tau=1)$ describes the average effect on $x(t)$ of a small perturbation one time step earlier (upper panel), $G(2)$ is the average effect two time steps after a perturbation has been applied, and $G(3)$ three steps after the perturbation. Thus, $G(1)+G(2)+G(3)$ describes the linear response at a time $t$ when a perturbation is applied simultaneously at times $t-1, t-2$ and $t-3$ (fourth panel). The integrated response $\chi$ is then the response of the system if a constant perturbation is applied at {\em all} earlier times. Now you may ask:  OK, if I perturb the system at some time $t_1$ and then I perturb the same system again at some later time $t_2$, then surely the later perturbation at $t_2$ acts on an already perturbed signal. Don't I have to take this into account somehow? Why are all derivatives in Eq.~(\ref{eq:chi_discrete}) to be taken at $\ulh=\ulnull$? The answer is that we are looking at linear response, so we are not considering perturbations of perturbations. Instead we are summing up the small effects of perturbations at different times independently.\label{fig:chi}.}
\end{figure}

We have to work a little harder to express $\chi$ in terms of an average over something to do with $x(z)$. First we remember that the response function $G(t,t')$ depends only on time differences in the stationary state, and in absence of memory effects. We can write $G(\tau)=G(t,t-\tau)$, that is $G(\tau)$ describes the average linear response of abundances at a time $t$ in the stationary state to a (small) perturbation which is applied $\tau$ units of time earlier,
\be\label{eq:g_eff}
G(\tau)=\left.\frac{\delta}{\delta h(t-\tau)} \avg{x(t)}_*\right|_{\ulh=\ulnull}~\mbox{(with $t$ in the stationary regime)}.
\ee
The quantity $\chi$ is the integrated response function in the stationary state, $\chi=\int_0^\infty d\tau\, G(\tau)$. It is helpful to briefly imagine what this would look like in discrete time (with time step equal to one). We'd have $\chi=\sum_{\tau=1}^\infty G(\tau)$, and $G(\tau)$ describes the response of $x(t)$ to a perturbation at $t-\tau$ (with $t$ sufficiently long so that all this happens in the stationary state). The point is that for the discrete-time system
\be\label{eq:chi_discrete}
\chi=\sum_{\tau>0}^\infty G(\tau)=\sum_{\tau>0} \left.\frac{\delta}{\delta h(t-\tau)}\avg{x(t)}_*\right|_{\ulh=\ulnull} ~\mbox{(with $t$ in the stationary regime)},
\ee
so $\chi$ sums up the effects on $x(t)$ of a perturbation at time $t-1$, a perturbation at time $t-2$, at $t-3$ and so on. So one would have to apply a small perturbation at {\em all} times. The integrated response, in the fixed-point regime, tells us how the fixed point value for $x$ is affected by a small constant perturbation, acting at {\em all} times. This is illustrated in Fig.~\ref{fig:chi}.
\medskip

OK, so what we need to do it to apply a constant perturbation $h(t')\equiv h$. Instead of Eq.~(\ref{eq:fp}) we the find

\be \label{eq:fp_h}
x^*\left[1-x^*+\Gamma\sigma^2\chi x^*+\mu M^*+\eta^*\textcolor{blue}{+h}\right]=0,
\ee
and accordingly,
\be\label{eq:max_h}
x(z)=\mbox{max}\left(0, \frac{1+\mu M^*+\sqrt{q}\sigma z\textcolor{blue}{+h}}{1-\Gamma\sigma^2\chi}\right),
\ee
instead of Eq.~(\ref{eq:max}).

Following the above argument, we then have
\be
\chi=\left.\frac{d}{dh}\avg{x(z)}_*\right|_{h=0},
\ee
with $x(z)$ given by Eq.~(\ref{eq:max_h}). The quantity $1-\Gamma\sigma^2\chi$ will always come out positive, so we proceed here assuming that this will be the case. The second argument in the maximum in Eq.~(\ref{eq:max_h}) is then positive if and only if $z>-\frac{1+\mu M^*\textcolor{blue}{+h}}{\sigma\sqrt{q}}$. Therefore
\be
\avg{x(z)}_*=\int_{-\frac{1+\mu M^*\textcolor{blue}{+h}}{\sigma\sqrt{q}}}^\infty dz\, e^{-z^2/2} \frac{1+\mu M^*+\sqrt{q}\sigma z\textcolor{blue}{+h}}{1-\Gamma\sigma^2\chi}.\label{eq:chi_aux}
\ee
From this we find
\BE
\left.\frac{d}{dh}\avg{x(z)}_*\right|_{h=0}&=&\frac{1}{\sigma\sqrt{q}}\times \left(\mbox{integrand in Eq.~(\ref{eq:chi_aux}) evaluated at $h=0$ and~} z=-\frac{1+\mu M^*}{\sigma\sqrt{q}}\right) \nonumber \\
&&+\frac{1}{1-\Gamma\sigma^2\chi} \int_{-\frac{1+\mu M^*}{\sigma\sqrt{q}}}^\infty dz\, e^{-z^2/2}.
\EE
The integrand in Eq.~(\ref{eq:chi_aux}) vanishes when $h=0$ and $z=-\frac{1+\mu M^*}{\sigma\sqrt{q}}$, therefore we are left with
\be
\chi=\frac{1}{1-\Gamma\sigma^2\chi} \int_{-\frac{1+\mu M^*}{\sigma\sqrt{q}}}^\infty dz\, e^{-z^2/2}.
\ee
\medskip

\subsubsection{Summary of self-consistency relations at the fixed point}
[We now set $h(t)\equiv 0$ again.]
\medskip

We had already established that $M^*=\avg{x(z)}_*$ and $q=\avg{(x(z))^2}_*$ at the fixed point. Only non-zero $x(z)$ contribute to these averages. It is now helpful to introduce
\be
\Delta \equiv \frac{1+\mu M^*}{\sigma\sqrt{q}}.
\ee
We then have
\BE
M^*&=&\int_{-\Delta}^\infty Dz~x(z),\nonumber \\
q&=&\int_{-\Delta}^\infty Dz~ x(z)^2, \label{eq:fpsc_qM}
\EE
and 
\be
\chi=\frac{1}{1-\Gamma\sigma^2\chi} \int_{-\Delta}^\infty Dz
\ee
with the definition $Dz=e^{-z^2/2}/\sqrt{2\pi}$. In the integration range $z>-\Delta$ we also have [from Eq.~(\ref{eq:max})]
\be
x(z)=\sqrt{q}\sigma \frac{\Delta+z}{1-\Gamma\sigma^2\chi}.
\ee
Explicitly writing things out we therefore have
\BE
\chi&=&\frac{1}{1-\Gamma\sigma^2\chi}\int_{-\Delta}^\infty Dz, \nonumber \\
M^*&=&\sqrt{q}\sigma \frac{1}{1-\Gamma\sigma^2\chi}\int_{-\Delta}^\infty Dz~ (\Delta+z),\nonumber\\
1&=&\frac{\sigma^2}{(1-\Gamma\sigma^2\chi)^2}\int_{-\Delta}^\infty Dz~ (\Delta+z)^2.\label{eq:sc_int}
\EE

\underline{Remark:}\\
It isn't particularly important for the further analysis, but nonetheless useful to note that, in short form, the self-consistency relations can be written as
\BE
\chi&=&\frac{1}{\sqrt{q}\sigma}\avg{\frac{\partial x(z)}{\partial z}}_*,\nonumber \\
M^*&=&\avg{x(z)}_*,\nonumber\\
q&=&\avg{(x(z))^2}_*.
\EE
with $x(z)$ given in Eq.~(\ref{eq:max}).
 
 \medskip

\underline{Further remark (assumptions we have made for the fixed-point ansatz)}\\
 Eqs.~(\ref{eq:sc_int}) describe the system in the phase of unique stable fixed points. What this means is that (i) the original Lotka--Volterra system approaches a stable fixed point, and (ii) for any fixed realisation of the disorder this fixed point does not depend on initial conditions (otherwise the system would not be memory-less, and there could be different solutions for the order parameters for different initial conditions). As we will describe below these assumptions do not hold for all choices of the model parameters $\mu,\sigma^2$ and $\Gamma$. 
 
 But even within the phase of unique stable fixed points some qualifications are required. In this phase the above assumptions are valid in the thermodynamic limit, i.e., for infinitely many species $N\to \infty$. In practice, any numerical integration of the Lotka--Volterra equations (\ref{eq:lv}) will necessarily have to be carried out for a finite $N$. This means that in simulations for parameters $\mu,\sigma^2,\Gamma$ in the phase of unique fixed points one may not always find a unique fixed point for all samples of the disorder. Some samples might not converge to a fixed point, or perhaps one finds multiple fixed points. But the fraction of these samples of the disorder will go to zero for larger and larger system sizes $N$.

\subsection{Further analysis}\label{sec:further_analysis}
We now change integration variable $z$ into $-\mz$ (the symbol $\mz$ is also pronounced `zet'). Doing this in each integral in Eqs.~(\ref{eq:sc_int}) we have
\BE\label{eq:sc_3}
\chi&=&\frac{1}{1-\Gamma\sigma^2\chi}\int_{-\infty}^\Delta D\mz, \nonumber \\
M^*&=&\sqrt{q}\sigma \frac{1}{1-\Gamma\sigma^2\chi}\int_{-\infty}^\Delta D\mz ~(\Delta-\mz),\nonumber\\
1&=&\frac{\sigma^2}{(1-\Gamma\sigma^2\chi)^2}\int_{-\infty}^\Delta D\mz~ (\Delta-\mz)^2. 
\EE
These relations are for example given in Eqs. (9-11) of \cite{Galla_2018}. Together with 
\be\label{eq:Delta_2}
\Delta=\frac{1+\mu M^*}{\sqrt{q}\sigma}
\ee
this is a set of four equations for the four unknown quantities $q,\chi,M^*$, and $\Delta$.

This can be reduced to a system of three equations for the three quantities $q, M^*,\chi$ by using Eq.~(\ref{eq:Delta_2}) in Eqs.~(\ref{eq:sc_3}), but it will turn out convenient for us not to eliminate $\Delta$ at this stage. This system is to be solved as a function of the model parameters $\mu, \sigma^2$, and $\Gamma$.

\medskip
 
 Along the way, we have made the assumption $1-\Gamma\sigma^2\chi>0$. This can be checked retrospectively from the numerical solution. It is also required self-consistently in the second relation in Eq.~(\ref{eq:sc_3}), as $M^*$ is an average of non-negative abundances and therefore must be non-negative itself, $M^*\geq0$. (The integral in the second relation is clearly positive, as the integrand is positive in the integration range.) We also note that the first relation in Eq. (\ref{eq:sc_3}) then implies $\chi>0$. This makes sense intuitively, as the perturbation field in Eq.~(\ref{eq:effective_copy}) has a positive effect on the growth rate of $x$.
 
  \medskip
 
For future reference we also introduce
\be
\phi=\int_{-\Delta}^\infty Dz=\int_{-\infty}^{\Delta} D\mz.
\ee
This means that $\phi$ is the probability that the standard Gaussian variable $z$ takes values above $-\Delta$. This is the range in which $x(z)$ in Eq.~(\ref{eq:max}) is positive. Therefore $\phi$ is the fraction of surviving species in the stable phase of the Lotka--Volterra system.

\medskip
For future use, we further define
\be
w_\ell(\Delta)\equiv  \int_{-\infty}^\Delta D\mz~ (\Delta-\mz)^\ell,
\ee
for $\ell=0,1,2$. Using the definition of the error function,
\be
\erf(x)=\frac{2}{\sqrt{\pi}}\int_0^x dt\, e^{-t^2}
\ee
and recalling $D\mz = \frac{1}{\sqrt{2\pi}}e^{-\mz^2/2}d\mz$ we have the following explicit expressions for the $w_\ell(\Delta)$,
\BE\label{eq:w}
w_0(\Delta) & = & \frac{1}{2}\left[1+\erf\left(\frac{\Delta}{\sqrt{2}}\right)\right], \nonumber \\
w_1(\Delta) & = & \frac{1}{2}\left[e^{-\Delta^2/2}\sqrt{\frac{2}{\pi}}+\Delta\left(1+\erf\left(\frac{\Delta}{\sqrt{2}}\right)\right)\right], \nonumber \\
w_2(\Delta) & = & \frac{1}{2} \left(1+\Delta^2\right) \left[1+\erf\left(\frac{\Delta}{\sqrt{2}}\right)\right] +\frac{1}{\sqrt{2\pi}}e^{-\Delta^2/2}\Delta.
\EE
We also have
\BE
w_2(\Delta) & = & w_0(\Delta)+\Delta w_1(\Delta).
\EE

 \subsection{Solution procedure}\label{sec:proc}
We now proceed to solve the system of Eqs.~(\ref{eq:sc_3}) in combination with Eq.~(\ref{eq:Delta_2}). 
\medskip

In principle one can do this as follows: Substitute Eq.~(\ref{eq:Delta_2}) into Eqs.~(\ref{eq:sc_3}), and then numerically solve the resulting system of three equations for $q,\chi$ and $M^*$ for given $\sigma^2, \mu$ and $\Gamma$. This is possible, for example using a Newton--Raphson scheme. To do this one can use the explicit forms of the integrals in Eqs.~(\ref{eq:sc_3}) given in Eqs.~(\ref{eq:w}). Nonetheless, the resulting Newton--Raphson problem shows some degree of dependence on initial guesses for $q, M^*$ and $\chi$, and often turns out to be rather unstable. So I do not recommend this `brute-force' route.

\medskip

Luckily, there is a much more efficient way of doing this -- one can find a parametric solution in closed form. The idea is that we fix $\mu$ and $\Gamma$, but not $\sigma^2$. Then we use $\Delta$ as a parameter. This means that we vary $\Delta$, and then obtain $\sigma^2, q, \chi$ and $M$ as functions of $\Delta$. In practice this produces a table of the following type:

\begin{center}
\begin{tabular}{ccccc}
$\Delta^{(1)}$ & $(\sigma^2)^{(1)}$ & $(M^*)^{(1)}$ & $q^{(1)}$ & $\chi^{(1)}$ \\
$\Delta^{(2)}$ & $(\sigma^2)^{(2)}$ & $(M^*)^{(2)}$ & $q^{(2)}$ & $\chi^{(2)}$ \\
$\Delta^{(3)}$ & $(\sigma^2)^{(3)}$& $(M^*)^{(3)}$ & $q^{(3)}$ & $\chi^{(3)}$ \\
$\vdots$ & $\vdots$  & $\vdots$  & \vdots &  $\vdots$
\end{tabular}
\end{center}
Here, $\Delta^{(k)}$ is the $k$-th value of our parameter $\Delta$ that we use in our scheme, and $(\sigma^2)^{(k)}$, $(M^*)^{(k)}$, $q^{(k)}$ and $\chi^{(k)}$ are the corresponding values for $\sigma^2, M^*, q$ and $\chi$. Once this table has been obtained, one can the ignore the first column (values of $\Delta$) and plot, for example, $q$ as a function of $\sigma^2$.
\medskip

To achieve the parametric solution we first square both sides of the first relation in Eq.~(\ref{eq:sc_3}), and then divide by the expressions in the third equation. We find
\be\label{eq:help1}
\sigma^2\chi^2=\frac{w_0(\Delta)^2}{w_2(\Delta)}.
\ee
From the first relation in Eq.~(\ref{eq:sc_3}) on the other hand we have
\be
\chi-\Gamma\sigma^2\chi^2=w_0(\Delta).
\ee
Putting this together we find
\be
 \chi=w_0(\Delta)+\Gamma\frac{w_0(\Delta)^2}{w_2(\Delta)}.\label{eq:chi_ex}
\ee
For given $\mu$ and $\Gamma$ we have therefore found an explicit expression for $\chi$ as a function of $\Delta$.
We now use this in Eq.~(\ref{eq:help1}) and arrive at
\BE
\sigma^2&=&\frac{w_0(\Delta)^2}{w_2(\Delta)\chi^2}\nonumber \\
&=&\frac{w_2(\Delta)}{[w_2(\Delta)+\Gamma w_0(\Delta)]^2}.\label{eq:sig_ex}
\EE
For given $\mu$ and $\Gamma$, this is an explicit expression for $\sigma^2$ as a function of $\Delta$.
\medskip

It remains to find $q$ and $M^*$ as functions of $\Delta$. To do this, it is useful to introduce the following shorthand
\be
b\equiv 1-\Gamma\sigma^2\chi.
\ee
Using Eqs.~(\ref{eq:chi_ex}) and (\ref{eq:sig_ex}) we then have
\BE
b&=&1-\Gamma\frac{w_2}{(w_2+\Gamma w_0)^2}\left(w_0+\Gamma \frac{w_0^2}{w_2}\right)\nonumber \\
&=& 1-\Gamma\frac{w_0}{w_2+\Gamma w_0} \nonumber \\
&=& \frac{w_2}{w_2+\Gamma w_0},
\EE
where we don't write out the argument $\Delta$ of the $w_\ell$. For fixed $\Gamma$, this is an explicit expression for $b$ as a function of $\Delta$.
\medskip

From the second relation in Eq.~(\ref{eq:sc_3}) we then have
\be
\sigma\sqrt{q}=\frac{bM^*}{w_1},
\ee
which, combined with $\Delta=\frac{1+\mu M}{\sigma\sqrt{q}}$, gives
\BE
\frac{1}{M^*}&=&\frac{\Delta}{w_1}b-\mu\nonumber \\
&=&\frac{\Delta}{w_1}\frac{w_2}{w_2+\Gamma w_0}-\mu.
\EE
For given $\Gamma$ and $\mu$, this is an explicit expression for $1/M^*$ as a function of $\Delta$ (again we have not written out the argument $\Delta$ of the $w_\ell$ on the right-hand side).
\medskip

Finally, we have
\be
  q=\left(\frac{bM^*}{\sigma w_1}\right)^2.
  \ee
Given $w_1=w_1(\Delta)$ and that we know $b, \sigma^2$ and $M^*$ as functions of $\Delta$, we therefore have $q$ as a function of $\Delta$ (for fixed $\Gamma$ and $\mu$).
  
  \medskip
  
{\bf  Summary:}\\
For given $\Gamma$ and $\mu$, Eqs.~(\ref{eq:sc_3},\ref{eq:Delta_2}) can be solved parametrically as follows:  Vary $\Delta$ (in practice for example in range from $-10$ to $10$). For each value of $\Delta$ calculate $w_0(\Delta), w_1(\Delta)$ and $w_2(\Delta)$ from Eq.~(\ref{eq:w}). Then:
\BE
        \chi&=&w_0+\Gamma\frac{w_0^2}{w_2}, \nonumber \\
	\sigma^2&=&\frac{w_2}{(w_2+\Gamma w_0)^2}, \nonumber \\
	\frac{1}{M^*}&=&\frac{\Delta}{w_1}\frac{w_2}{w_2+\Gamma w_0}-\mu, \nonumber \\
         q&=& \left( \frac{w_2}{w_2+\Gamma w_0}\frac{M^*}{\sigma w_1}\right)^2,\nonumber\\
         \phi&=&w_0. \label{eq:procedure}
         \EE
         For fixed $\Gamma$ and $\mu$, this gives $M^*, q, \phi$ and $\chi$ as a function of $\sigma^2$ in parametric form.

 \subsection{Interpretation of the order parameters}
 We now briefly turn to an interpretation of these order parameters in the fixed-point phase in terms of the original Lotka--Volterra system. The notation $x_i^*$ indicates abundances at stable fixed points.
 \begin{itemize}
 \item The quantity $\phi=w_0(\Delta)$ is the fraction of surviving species, i.e., 
 \be
 \phi=\lim_{N\to\infty}\frac{1}{N}\sum_i \overline{\Theta(x_i^*)},
 \ee
 where $\Theta(\cdot)$ is the Heaviside function, $\Theta(x)=1$ for $x>0$, and $\Theta(x)=0$ for $x\leq 0$. In the original Lotka--Volterra equations (\ref{eq:lv}) extinctions can never occur at finite times. Instead, an abundance $x_i$ can only go to zero asymptotically at infinite times if it is initially positive. Obviously we cannot integrate the Lotka--Volterra system to infinite times numerically. To measure $\phi$ from numerical integration (in a system with finite $N$ obviously), one therefore has to apply a threshold, $\vartheta$, at the finite end-time, $t_f$, of the integration, and identify species $i$ as surviving if $x_i(t_f)\geq \vartheta$. 
 \item The quantity $M^*$ is the mean abundance per species, 
 \be
 M^*=\lim_{N\to\infty}\frac{1}{N}\sum_{i=1}^N \overline{x_i^*},
  \ee
 i.e., the first moment of abundances. (The sum includes the species that have died out.).
 \item The quantity $q$ is the second moment of the distribution of abundances,
\be
q=\lim_{N\to\infty}\frac{1}{N}\sum_{i=1}^N \overline{(x_i^*)^2},
\ee
where the sum again includes species that have died out.
 \item The quantity $\chi$ is a susceptibility, and measures how strongly species abundances change if an external perturbation is applied. It is not easy to measure this directly in simulations (but not impossible either).
 \end{itemize}
\underline{Remark:}\\
We re-iterate that Eqs.~(\ref{eq:procedure}) were derived assuming that (i) the Lotka--Volterra dynamics converges to a fixed point, and (ii) that this fixed point is, for a given realisation of the disorder, independent of initial conditions (we assumed the absence of memory effects). As we will see below, this is not the case for all choices of the model parameters $\mu,\sigma^2$ and $\Gamma$. The procedure in Eqs.~(\ref{eq:procedure}) only describes the outcome of the system in the phase of unique stable fixed points.

 \subsection{Species abundance distribution}
 \begin{figure}[t]
    \centering
    \includegraphics[width=0.75\textwidth]{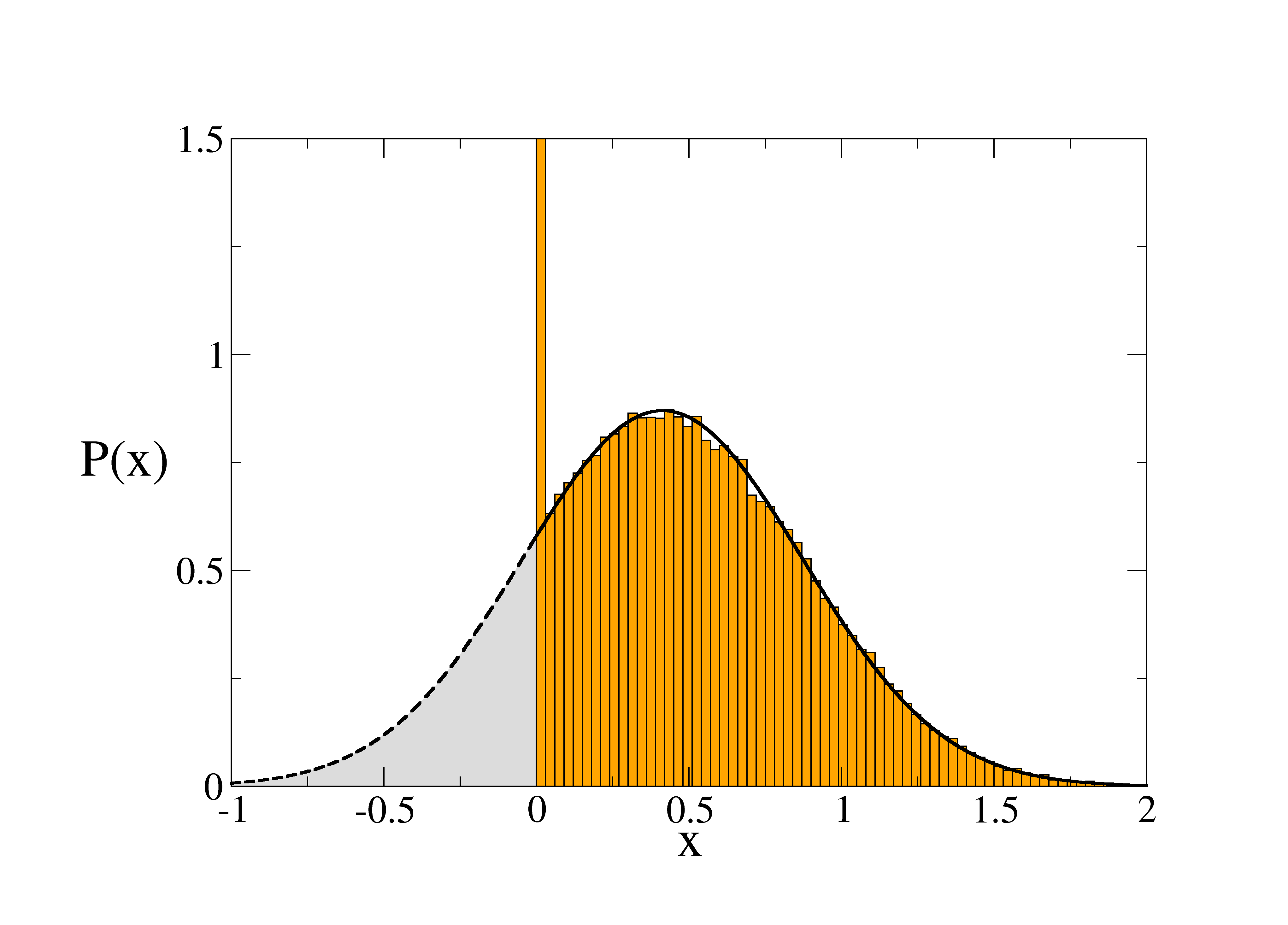}
     
    \caption{{\bf Species abundance distribution $P(x)$ [Eq.~(\ref{eq:pofx})].}   Solid black curve shows $p_{\rm surv}(x)$ in for $x>0$ [Eq.~(\ref{eq:psurv})]. The dashed line shows the extension of this curve to negative $x$. The orange histogram is from numerical integration of the Lotka--Volterra system. Here all abundances are non-negative by construction. The peak at $x=0$ are extinct species. Parameters are $\mu=-1$, $\Gamma=-0.5$ and $\sigma^2=1$. Numerical data is from $400$ realisations of the disorder, and $N=200$. The numerical integration is performed using an Euler scheme with adaptive time step.}
    \label{fig:sad}
\end{figure}
 Suppose now are at a point $\mu,\sigma^2,\Gamma$ in the phase with unique fixed points. We can then determine the order parameters $q,M^*$ and $\chi$ from Eqs.~(\ref{eq:procedure}). The species abundance distribution (distribution of the $x_i^*$'s) can then be deduced from Eq.~(\ref{eq:max}), which we repeat here,
 \be\label{eq:max_2}
x(z)=\mbox{max}\left(0, \frac{1+\mu M^*+\sqrt{q}\sigma z}{1-\Gamma\sigma^2\chi}\right),
\ee
  and where we recall that $z$ is a static Gaussian random variable with mean zero and variance one.
 
Thus the abundance $x(z)$ is zero whenever the second argument inside the maximum is zero or negative. Consequently, the distribution of abundances has a delta peak at zero, plus a part which describes the surviving species,
 \be\label{eq:pofx}
 P(x)=(1-\phi)\delta(x) + p_{\rm surv}(x)
 \ee
 The weight of the delta-peak at $x=0$ is $1-\phi$, and we have $\int_0^\infty dx ~p_{\rm surv}(x)=\phi$. 
 \medskip
 
 The functional form of $p_{\rm surv}$ is a clipped Gaussian. More specifically, 
 \be\label{eq:psurv}
 p_{\rm surv}(x)=\frac{1}{\sqrt{2\pi\Sigma^2}}\exp\left(-\frac{(x-x_S)^2}{2\Sigma^2}\right)\Theta(x),
 \ee
with the Heaviside function $\Theta(x)$, and 
\BE
x_S&=&\frac{1+\mu M^*}{1-\Gamma\chi\sigma^2}, \nonumber \\
\Sigma^2&=&\frac{q\sigma^2}{(1-\Gamma\sigma^2\chi)^2}.
\EE
These can be read off directly from Eq.~(\ref{eq:max_2}), as the mean and variance of the object $\frac{1+\mu M^*+\sqrt{q}\sigma z}{1-\Gamma\sigma^2\chi}$, keeping in mind that $z$ is a standard Gaussian random variable.

\medskip

 The quantities $x_S$ and $\Sigma^2$ are the mean and variance respectively of an unclipped Gaussian, and they are not to be confused with the moments of the clipped Gaussian $p_{\rm surv}(\cdot)$. That is to say in general we have $M=\int_0^\infty dx\, p_{\rm surv}(x) x \neq x_S$ and $q=\int_0^\infty dx\, p_{\rm surv}(x) x^2 \neq \Sigma^2$. Using the definitions of $x_S$ and $\Sigma^2$, and the relation $\phi=w_0(\Delta)$ with $\Delta=(1+\mu M^*)/(\sigma\sqrt{q})$ one can directly check that $\int_0^\infty dx ~p_{\rm surv}(x)=\phi$.

\medskip

An example is shown in Fig.~\ref{fig:sad}. This is for specific parameters $\mu=-1$, $\Gamma=-0.5$ and $\sigma^2=1$. The solid curve shows $p_{\rm surv}(x)$ for $x>0$. The figure also shows $p_{\rm surv}(x)$ for $x<0$ (dashed black line). This is the part of the Gaussian that gets clipped, i.e., the fraction of extinct species ($x^*=0$) is equal to they shaded grey area under $p_{\rm surv}$ at negative $x$. The orange histogram is from numerical integration of the Lotka--Volterra system. More precisely the data is from $400$ independent realisations of the interaction matrix, and for a system size of $N=200$ species. The numerical integration is performed using an Euler scheme with adaptive time step.

 \subsection{Test against numerical integration of the Lotka--Volterra dynamics}
\begin{figure}[t]
    \centering
    \includegraphics[width=0.45\textwidth]{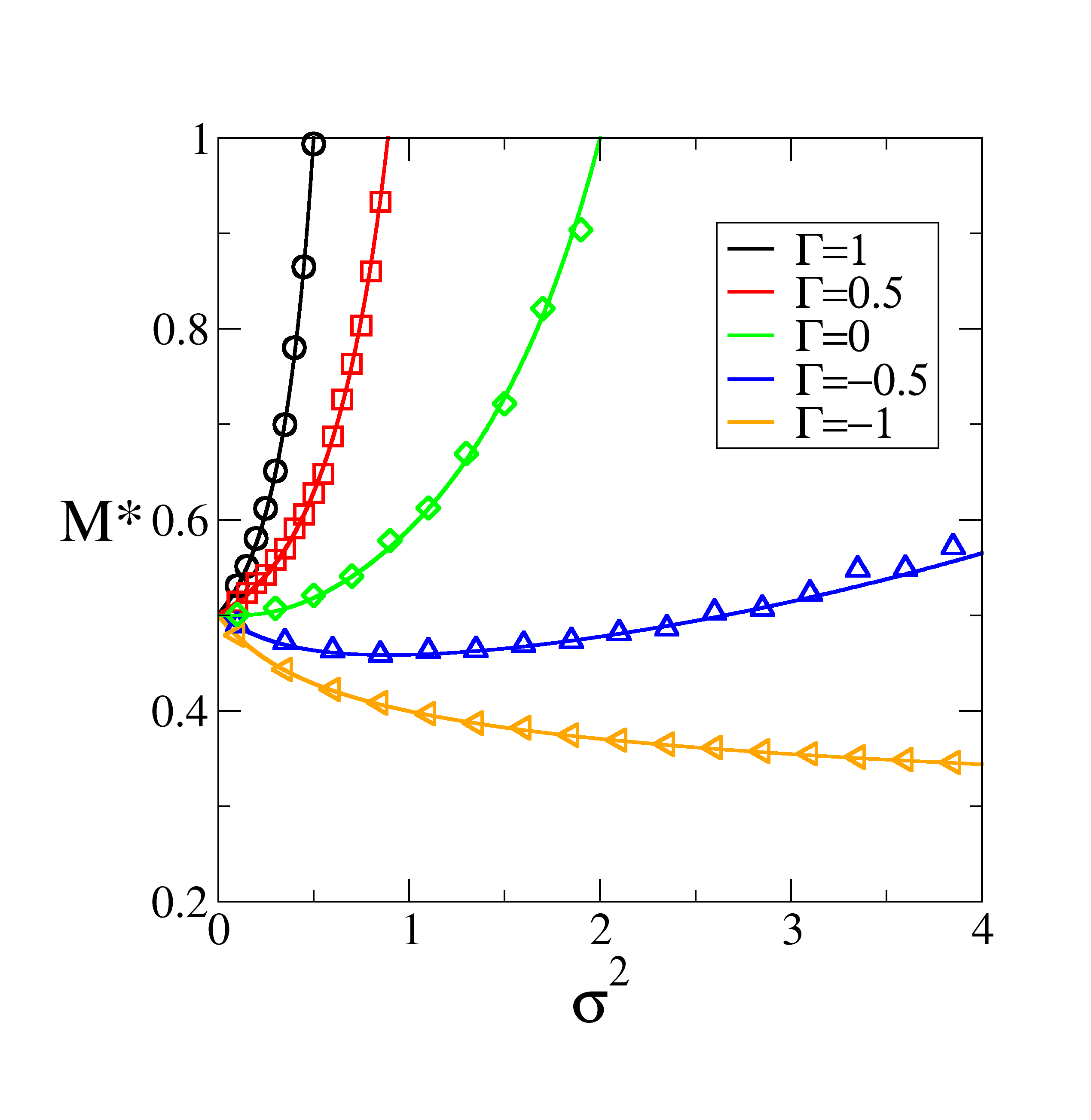}
         \includegraphics[width=0.45\textwidth]{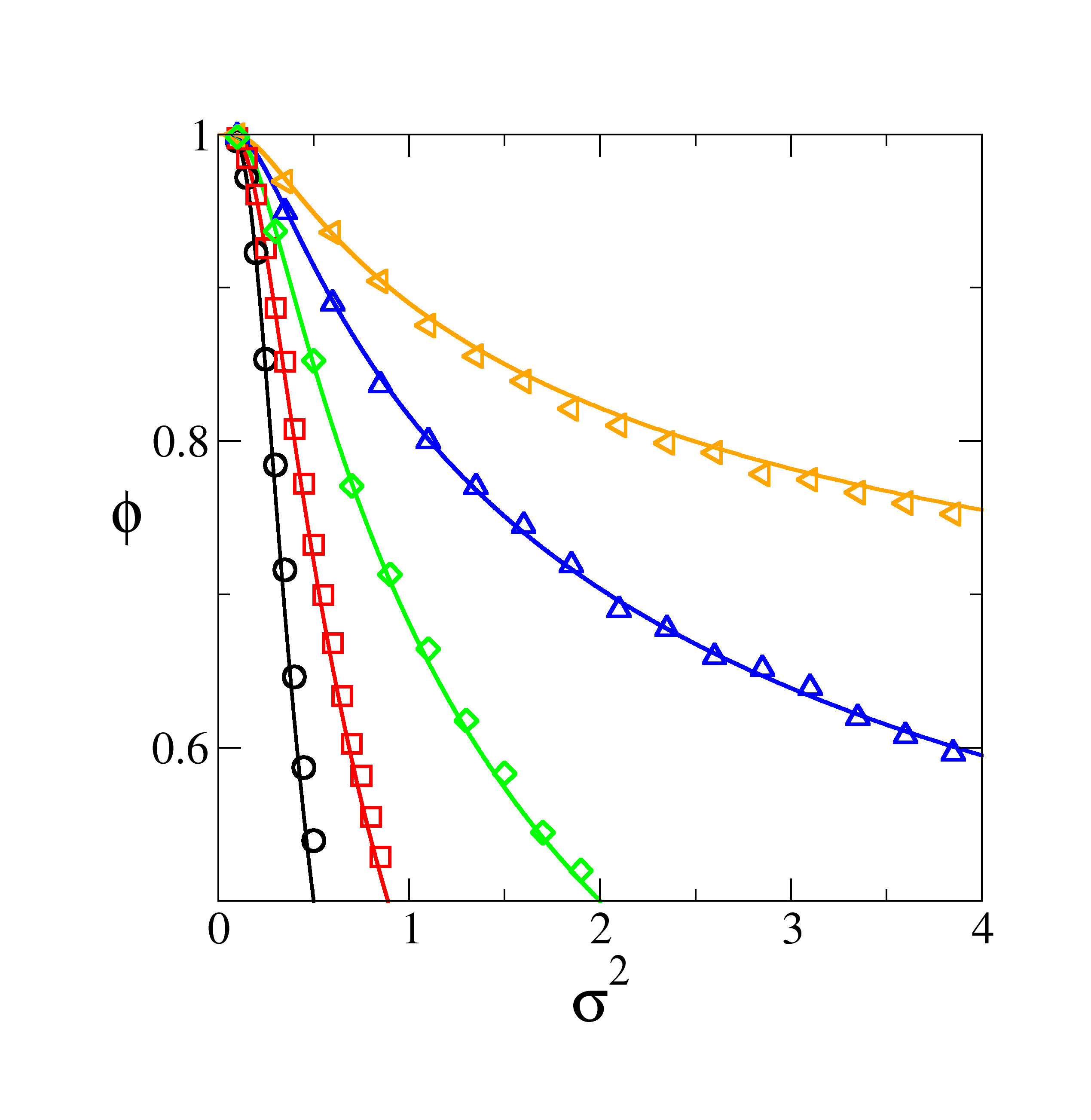}
    \caption{{\bf Static order parameters in the unique fixed point phase.}  We show the mean abundance $M^*$ and the fraction of surviving species $\phi$ as a function of the variance of interactions $\sigma^2$. The parameter $\mu$ is set to $\mu=-1$, the value of $\Gamma$ is as indicated in the legend. Solid lines are from the theory in the unique fixed-point phase [Eqs.~(\ref{eq:procedure})]. The markers are from a numerical integration of the Lotka--Volterra system ($N=200$, averaged over $100$ realisations of the disorder). In the numerical integration surviving species are identified as those with $x_i>0.01$ at the end of the integration.}
    \label{fig:sim}
\end{figure}

We test the predictions of the theory against results from a numerical integration of the Lotka--Volterra equations in Fig.~\ref{fig:sim}. We show both the mean abundance $M^*$ and the fraction of surviving species $\phi$. Results are only shown in the phase of unique stable fixed points, as the theory is otherwise not applicable.

 \section{Stability analysis and phase diagram}\label{sec:stability_pg}
We will now ask when the validity of the self-consistent equations (\ref{eq:sc_int}), or equivalently, Eqs.~(\ref{eq:sc_3}) breaks down. In principle this could happen in several different ways. For example the equations could simply not have any physically meaningful solutions at all for some combinations of $\mu,\sigma^2$ and $\Gamma$. Or the Lotka--Volterra system could show memory, i.e., there could be multiple physically meaningful solutions for $C(t,t'), G(t,t'), M(t)$, depending on initial conditions for example. 

As we will see, the relevant ways in which the assumptions of unique fixed points break down in the Lotka--Volterra system are: (i) an onset of divergent abundances, and (ii) a linear instability.  These have been identified for example in \cite{bunin2016interaction, bunin2017, Galla_2018,froy}. We now discuss the two types of transition in turn.

 \subsection{Onset of divergent solutions}
 The onset of divergent abundances is signalled by a divergent mean abundance $M^*$. That is, if we start in the stable phase and then modify parameters $\mu,\sigma^2$ and $\Gamma$, we may approach combinations of these parameters at which $M^*$ diverges. 
 
Setting $1/M^*=0$ in the third relation in Eqs.~(\ref{eq:procedure}) we find that this occurs when
 \be
 \mu=\frac{\Delta}{w_1(\Delta)}\frac{w_2(\Delta)}{w_2(\Delta)+\Gamma w_0(\Delta)}. \label{eq:div1}
 \ee
 We also have from the second relation in Eqs.~(\ref{eq:procedure}), 
\be
\sigma^2=\frac{w_2(\Delta)}{[w_2(\Delta)+\Gamma w_0(\Delta)]^2}. \label{eq:div2}
\ee
 For a given choice of $\Gamma$, Eqs.~(\ref{eq:div1},\ref{eq:div2}) provide a parametric description of a line in the $(\mu,\sigma^2)$-plane (the parameter being $\Delta$), telling us where the onset of divergence of the mean divergence $M^*$ occurs. At (and beyond) this line a non-zero fraction of species abundances diverge, and the fixed-point assumptions no longer hold.
 
\subsection{Linear stability analysis}\label{sec:lsa}
The Lotka--Volterra system also shows a linear instability. This was first identified in a similar replicator system in \cite{opper1992phase}, and we proceed along the lines of this reference. The analysis is divided into six steps.

\subsubsection{Step 1: Add small perturbation to effective single-species dynamics}\label{sec:linearisation}
Assume we choose parameters such that the system is in the phase of unique stable fixed points. We then add white noise $\varepsilon \xi_i(t)$ to the Lotka--Volterra equations, i.e., we have
\be\label{eq:lv_perturb}
\dot x_i=x_i\left(1-x_i+\sum_{j}\alpha_{ij}x_j+\varepsilon\xi_i(t)\right).
\ee
We assume $\avg{\xi_i(t)}=0$, and $\avg{\xi_i(t)\xi_j(t')}=\delta_{ij}\delta(t-t')$. 

At the level of the effective process, this corresponds to adding white noise $\xi(t)$ of zero mean and unit variance [$\avg{\xi(t}=0, \avg{\xi(t)\xi(t')}=\delta(t-t')$],
\BE
\dot x(t)=x(t)\left[1-x(t)+\Gamma \sigma^2\int dt' G(t,t') x(t')+\mu M(t)+\eta(t)+\varepsilon\xi(t)\right]. \label{eq:effective_xi}
\EE
We now study fluctuations $y(t)$ about a fixed point of Eq.~(\ref{eq:effective}). Specifically, we assume that we run the unperturbed system ($\varepsilon=0$) to a fixed point and then imagine we `switch' the perturbation $\xi(t)$ on after the fixed point has been reached. As a consequence the abundance $x$ will be perturbed about the fixed point. We write 
\be
x(t)=x^*+\varepsilon y(t).
\ee
In the fixed-point phase and when there are no perturbations, the noise $\eta(t)$ in the effective process becomes a static random variable at large times. This is no longer the case in the perturbed system, and we denote the resulting additional term in the self-consistent noise by $\varepsilon v(t)$, i.e., 
\be
\eta(t)=\eta^*+\varepsilon v(t).
\ee 
\medskip

We then expand both sides of Eq.~(\ref{eq:effective_xi}). The leading-order term ($\varepsilon=0$) is simply the fixed-point relation [Eq.~(\ref{eq:fp})],
\BE
0=x^*\left[1-x^*+\Gamma \sigma^2\chi x^*+\mu M^*+\eta^*\right]. 
\EE
Comparing terms of linear order in $\varepsilon$ we find,
\BE\label{eq:linearised}
\varepsilon \dot y(t)&=&\varepsilon y(t) \left[1-x^*+\Gamma \sigma^2\chi x^*+\mu M^*+\eta^*\right] \nonumber \\
&&+ x^* \left[-\varepsilon y(t)+\Gamma \sigma^2\int dt' \{G(t,t') \varepsilon y(t')\}+\varepsilon v(t)+\varepsilon\xi(t)\right].
\EE

\subsubsection{Step 2: Independence of random variables}\label{sec:indep}
In our analysis, we assume that
\begin{enumerate}
\item[1.] The random variables $\xi(t)$ are independent of the static random variable $x^*$ (equivalently, $\xi(t)$ is independent of $\eta^*$).
\item[2.] The random variables $v(t)$ are independent of the static random variable $x^*$.
\item[3.] The random variables $\ulv$ and $\ulxi$ are independent.
\end{enumerate}

The first statement is clear, as $x^*$ is the fixed point of the effective process {\em in absence} of the perturbation $\xi(t)$, which is only added {\em after} the fixed point has been reached. 

\medskip

The assumption that $v(t)$ and $x^*$ are independent from one another is a little more intricate to justify. It helps to remember that the noise $\uleta=\eta^*+\varepsilon\ulv$ in the effective process reflects the disorder in the original Lotka--Volterra problem, and ultimately results from the term $\sum_{j\neq i} \alpha_{ij}x_j$ in the equation for $\dot x_i$. This is also transparent from the derivation of the effective process via the cavity method in Sec.~\ref{sec:cavity}. As such, the noise variable $v(t)$ results from the fluctuations of  $\sum_{j\neq i} \alpha_{ij}x_j(t)$ in the dynamics for $x_i$ away from the fixed point value $\sum_{j\neq i} \alpha_{ij}x_j^*$. Within a linear approximation it is then sensible to assert that $v(t)$ is independent of $x^*$.

\medskip

The argument for the third item is similar. The term $\ulv$ results from the disordered interactions of one given species with all other species. The noise $\ulxi$ acts directly on that focal species itself and, within the linear approximation, has nothing to do with the other species. Therefore $\ulv$ is independent of $\ulxi$.
\medskip

We now proceed to linearise Eq.~(\ref{eq:effective_xi}) in $\varepsilon$, i.e. in $y(t), v(t)$ and $\xi(t)$. 
We first do this for effective species with zero abundance (step 3), and then subsequently for effective species with a positive abundance (step 4).

\subsubsection{Step 3: Further analysis for species with vanishing fixed-point abundance}
Linearising Eq.~(\ref{eq:effective_xi}) about a fixed point $x^*=0$ one has from Eq.~(\ref{eq:linearised}) and writing again $\eta^*=\sqrt{q}\sigma z$,
\be\label{eq:linear_0}
\frac{dy(t)}{dt}=y(t)\left[1+\mu M^*+\sqrt{q}\sigma z\right].
\ee

Within our ansatz, the object in the square brackets is negative for fixed points at zero [see Eq. (\ref{eq:max}), and recall that $1-\Gamma\sigma^2\chi>0$]. We conclude that $y(t)\to 0$ for $t \to\infty$. We note explicitly that the presence of the noise $\varepsilon \xi(t)$ makes no difference for $\dot y$ to linear order in $\varepsilon$, i.e., Eq.~(\ref{eq:linear_0}) is valid (to linear order in $\varepsilon$) for the model with noise $\varepsilon\xi(t)$ in Eq.~(\ref{eq:effective_xi}).  This implies the following: Imagine we run the unperturbed Lotka--Volterra system for a particular realisation of the disorder, and a particular species $i$ goes extinct at fixed point of this system. If we then switch on the noise and perturb the fixed point, then species $i$ will still remain extinct (provided the noise strength is small enough for the linear approximation to work). 

In essence the result in this step means, that once a species has gone extinct, the system is stable against re-invasion of that species.

\medskip

\underline{Remark:}\\
We also note at this point that Eq.~(\ref{eq:linear_0}) indicates that a zero fixed point is not stable if the object in the square brackets on the right-hand side is positive. This retrospectively justifies the use the non-zero solution $x^*$ in Eq.~(\ref{eq:max}) when $1+\mu M^*+\sqrt{q}\sigma z>0$.

\subsubsection{Step 4: Further analysis for non-zero fixed-point abundances}
For non-zero $x^*$ we have $1-x^*+\Gamma \sigma^2\chi x^*+\mu M^*+\eta^*=0$ [see Eq.~(\ref{eq:fp})], and hence Eq.~(\ref{eq:linearised}) turns into 
\BE
\frac{dy(t)}{dt}&=&x^*\left[-y(t)+\Gamma\sigma^2\int_{0}^t G(t,t')y(t')dt'+v(t)+\xi(t)\right]. \label{eq:pert}
\EE
Recalling $G(t,t')=G(t-t')$ in the fixed-point phase, we transform into Fourier space (Fourier transforms are indicated by tildes), and obtain
\BE
\frac{i\omega\widetilde y(\omega)}{x^*}&=&\left(\Gamma\sigma^2 \widetilde G(\omega)-1\right) \widetilde y(\omega)+\widetilde v(\omega)+\widetilde \xi(\omega). \label{eq:pert2}
\EE
This leads to
\be\label{eq:help7}
 \widetilde y(\omega)=\frac{\widetilde v(\omega)+\widetilde \xi(\omega)}{\frac{i\omega}{x^*}+\left(1-\Gamma\sigma^2 \widetilde G(\omega)\right)}, 
\ee
and we remember that this is valid only for extant species (species that have not gone extinct at the fixed point of the unperturbed system). 

\medskip

The noise $\ulxi$ has mean zero, and within the linear approximation in Eq.~(\ref{eq:pert}) it is then reasonable to conclude that $\avg{y(t)}_y=0$ and $\avg{v(t)}_v=0$, for any fixed $x^*$.

\subsubsection{Step 5: Self-consistency relations}

The self-consistency relation $\avg{\avg{\eta(t)\eta(t')}}_{*,v}=\sigma^2 \avg{\avg{x(t)x(t')}}_{*,y}$ translates into
\BE
&&\avg{(\eta^*)^2}_*+\varepsilon \avg{\avg{\eta^* v(t)}}_{*,v}+\varepsilon \avg{\avg{\eta^* v(t')}}_{*,v}+\varepsilon^2 \avg{v(t)v(t')}_{*,v} \nonumber \\
&=&\sigma^2\left(\avg{(x^*)^2}_*+\varepsilon \avg{\avg{x^* y(t)}}_{*,y}+\varepsilon \avg{\avg{x^* y(t')}}_{*,y}+\varepsilon^2 \avg{y(t)y(t')}_{*,y}\right)\label{eq:averages}
\EE
We have used the notation $\avg{\cdots}_*$ for averages over $\eta^*$ or $x^*$, and $\avg{\avg{\cdots}}_{*,y}$ stands for a combined average over $x^*$ (or $\eta^*$) and $y$, and similarly for $\avg{\avg{\cdots}}_{*,v}$.

We know that $\avg{(\eta^*)^2}_*=\sigma^2\avg{(x^*)^2}_*$ (self-consistency relation in the earlier fixed-point analysis). The linear terms in $\varepsilon$ in Eq.~(\ref{eq:averages}) vanish because $v(t)$ and $y(t)$ each have zero average for any fixed $x^*$. Thus we are left with the requirement
\be\label{eq:sc_vv}
\avg{v(t)v(t')}_{*,v}=\sigma^2 \avg{y(t)y(t')}_{*,y}.
\ee
Given the independence of $\ulv$ from $x^*$ we can drop the average over $x^*$ on the left. We therefore have
\be\label{eq:sc_vv2}
\avg{v(t)v(t')}_{v}=\sigma^2 \avg{y(t)y(t')}_{*,y}.
\ee
\medskip

\subsubsection{Step 6: Criterion for linear instability}
We now proceed to calculate $\avg{|\widetilde y(\omega)|^2}_{*,y}$. This is the Fourier transform of the correlation function $\avg{y(t+\tau)y(\tau)}_{*,y}$ in the stationary state ($t\to\infty$), see also Sec.~\ref{sec:ft_corr_fct}. We recall that $\uly$ is a perturbation about $x^*$, and that the statistics of $y(t)$ (equivalently, those of $\widetilde y(\omega)$) depend on the fixed-point value $x^*$. Specifically $\lim_{t\to\infty} y(t)=0$, when $x^*=0$, whereas $\widetilde y(\omega)$ is given by Eq.~(\ref{eq:help7}) when $x^*>0$. In other words

\be
|\widetilde y(\omega)|^2=\left\{\begin{array}{cl} \left|\frac{\widetilde v(\omega)+\widetilde \xi(\omega)}{\frac{i\omega}{x^*}+\left(1-\Gamma\sigma^2 \widetilde G(\omega)\right)}\right|^2 & \mbox{for surviving species}, \\
~&~\\
0 & \mbox{for extinct species}.\end{array}\right.
\ee
We now carry out an average on both sides. This involves a three-fold averaging procedure, over fixed-point abundances $x^*$, realisations of the noise $\ulv$ and over realisations of $\ulxi$. Using the independence properties from Step 2, we have
\BE\label{eq:help8}
\avg{ |\widetilde y(\omega)|^2}_{*,y}
&=&\left(\int dx^* \, p_{\rm surv}(x^*)\left|\frac{i\omega}{x^*}+\left(1-\Gamma\sigma^2 \widetilde G(\omega)\right)\right|^{-2}\right)\nonumber \\
&&\times \left(\avg{|\widetilde v(\omega)|^2}_v+ \avg{|\widetilde \xi(\omega)|^2}_\xi\right).
\EE

Next we use the self-consistency relation in Eq.~(\ref{eq:sc_vv2}), which translates into 
\be
\avg{|\widetilde v(\omega)^2|}_v=\sigma^2\avg{|\widetilde y(\omega)|^2}_{*,y}.
\ee
Thus, we find
\BE\label{eq:help8}
\avg{ |\widetilde y(\omega)|^2}_{*,y}
&=&\left(\int dx^* \, p_{\rm surv}(x^*)\left|\frac{i\omega}{x^*}+\left(1-\Gamma\sigma^2 \widetilde G(\omega)\right)\right|^{-2}\right)\nonumber \\
&&\times \left(\sigma^2 \avg{|\widetilde y(\omega)|^2}_{*,y}+ \avg{|\widetilde \xi(\omega)|^2}_\xi\right).
\EE
Following \cite{opper1992phase} we focus on $\omega=0$. We have $\widetilde G(0)=\chi$, and the term inside the integral over $x^*$ becomes independent of $x^*$. Noting that $\int_0^\infty dx^* p_{\rm surv}(x^*)=\phi$, we then arrive at
\be
\avg{|\widetilde y(0)|^2}_{*,y}=\phi \left(\Gamma\sigma^2 \chi-1\right)^{-2}\left[\sigma^2\avg{|\widetilde y(0)|^2}_{*,y} +1\right]. \label{eq:pert3}
\ee

Eq. (\ref{eq:pert3}) can be re-written as
\be
\avg{|\widetilde y(0)|^2}_{*,y}=\frac{\phi}{\left(\Gamma\sigma^2 \chi-1\right)^2-\phi\sigma^2}.
\ee
This indicates that $\avg{|\widetilde y(0)|^2}_{*,y}$ diverges when $\phi\sigma^2=(1-\Gamma\sigma^2\chi)^2$.  When $\phi\sigma^2>(1-\Gamma\sigma^2\chi)^2$, the right-hand side is negative, leading to a contradiction (and therefore a breakdown of the fixed-point ansatz). One has $\phi\sigma^2<(1-\Gamma\sigma^2\chi)^2$ in the stable phase, consistent with a well-defined (positive) quantity $\avg{|\widetilde y(0)|^2}$.  

\medskip

The onset of linear instability is thus signalled by the condition $\phi\sigma^2=(1-\Gamma\sigma^2\chi)^2$. This in turn leads to $\Delta=0$ in Eqs.~(\ref{eq:sc_3}). To see this we substitute $\phi\sigma^2=(1-\Gamma\sigma^2\chi)^2$ into $1=\frac{\sigma^2}{(1-\Gamma\sigma^2\chi)^2}\int_{-\infty}^\Delta D\mz (\Delta-\mz)^2$ [the third relation in Eq.~(\ref{eq:sc_3})], and find
\be
\phi=\int_{-\infty}^\Delta D\mz~ (\Delta -\mz)^2.
\ee
On the other hand we also have $\phi=\int_{-\infty}^\Delta D\mz$. Comparing the two expressions gives $\Delta=0$, and hence $\phi=1/2$.

\medskip

Using this in the first and third relations in Eqs.~(\ref{eq:sc_3}), we have, respectively,
\BE\label{eq:helppp}
2\chi&=&\frac{1}{1-\Gamma\sigma^2\chi}, \nonumber \\
2&=& \frac{\sigma^2}{(1-\Gamma\sigma^2\chi)^2}.
\EE
From this we find $\chi^2=1/(2\sigma^2)$. Using $\chi>0$ (which we know from Sec.~\ref{sec:further_analysis}) we have $\chi=1/(\sqrt{2}\sigma)$. Substituting this in the first relation in Eq. (\ref{eq:helppp}) in turn leads to
 
\be\label{eq:stab}
\sigma_c^2(\Gamma)=\frac{2}{(1+\Gamma)^2}.
\ee
This is the linear instability condition reported for example in \cite{bunin2016interaction,bunin2017,Galla_2018}.
\subsection{Stability diagram}
\begin{figure}[t]
    \centering
    \includegraphics[width=0.45\textwidth]{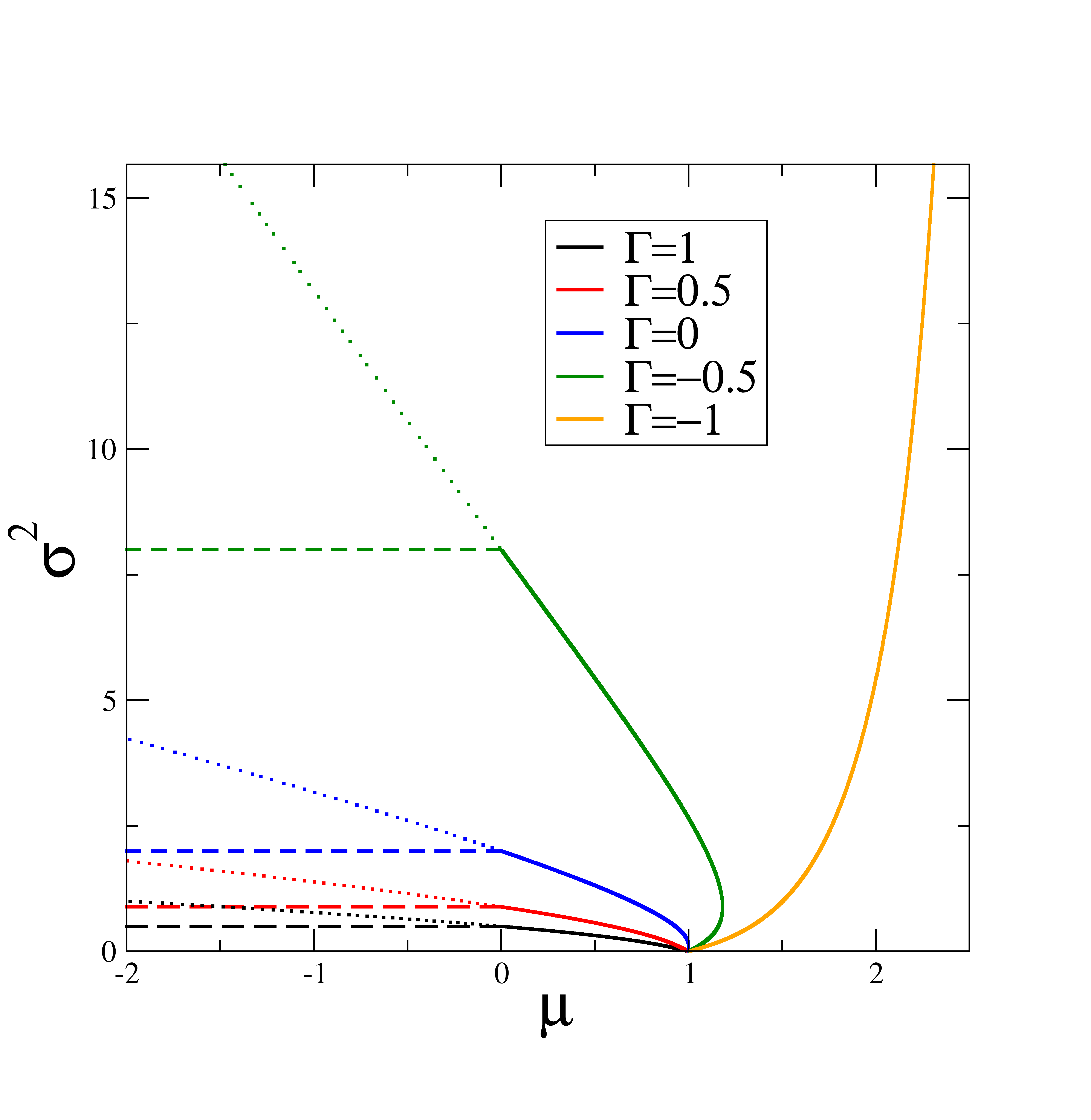}
        \includegraphics[width=0.45\textwidth]{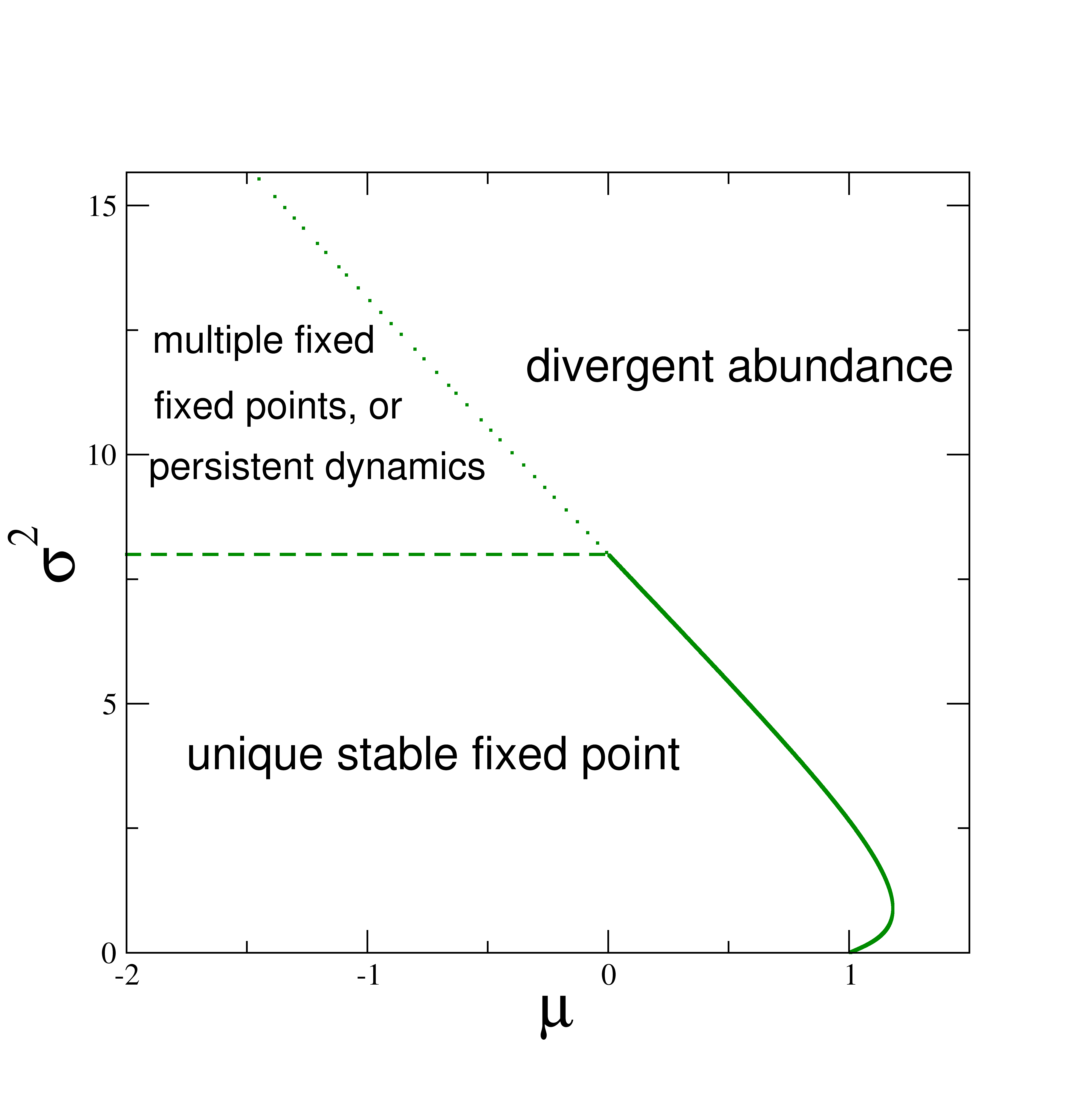}
    \caption{{\bf Stability diagram for the random Lotka-Volterra system in the $(\mu,\sigma)$ plane.} Solid lines mark the onset of diverging abundances [$M\to\infty$, Eqs.~(\ref{eq:div1},\ref{eq:div2})], dashed lines are the onset of linear instability [Eq.~(\ref{eq:stab})]. The dotted lines also show the condition of diverging abundances. However, the fixed-point equations are no longer valid in this part of the phase diagram, as the linear instability has already set in. On the left we show the phase diagram for different values of the symmetry parameter $\Gamma$. On the right we fix $\Gamma=-0.5$ and indicate the typical behaviour observed in the different parts of the phase plane (see text for further details).}
    \label{fig:pg}
\end{figure}

The stability diagram is shown in Fig.~\ref{fig:pg} in the $(\mu,\sigma^2)$ plane. On the left I show the lines of divergent abundance (solid) and of the onset of linear instability (dashed) for different values of $\Gamma$. The linear instability sets in at $\sigma^2=2/(1+\Gamma^2)$ [Eq.~(\ref{eq:stab})], the onset of diverging abundance is given by Eqs.~(\ref{eq:div1},\ref{eq:div2}). On the right-hand panel I indicate the typical behaviour of the system in the different phases. 
\medskip

There are three different phases:

\medskip

{\bf Unique stable fixed point phase.} The system converges to one single fixed point for any fixed realisation of the disorder. That is to say, once the interaction matrix has been drawn, the Lotka--Volterra dynamics always goes to the same stable fixed point, regardless of the initial conditions (provided all $x_i$ are positive at the beginning). This is the phase described by our fixed-point theory [for example Eqs.~(\ref{eq:sc_3}) and (\ref{eq:procedure})]. This phase is realised in the part of the phase diagram below the linear instability line (dashed) and to the left of the (solid) line marking the onset of diverging abundance.
\medskip

{\bf Diverging abundance.} Starting in the phase with unique stable fixed points and increasing the mean of interactions $\mu$ (keeping $\sigma^2$ fixed) the system eventually crosses into the phase of diverging abundance. Within the stable fixed point phase $M$ remains finite and there is a unique stable fixed point. However, as one approaches the solid line in the stability diagram from the left, the fixed point abundances gradually increase, and eventually the fixed point `escapes' to infinity, so that $M^*\to \infty$.

\medskip

{\bf Multiple fixed points or persistent dynamics.} In the part of the phase diagram labelled `multiple fixed points or persistent dynamics' there are several possible types of behaviour. For example, it is possible that the system never really settles down, the abundances keep fluctuating indefinitely. This is what I mean by 'persistent dynamics' (the dynamics persists forever). It is also possible that the system has many stable fixed points for a given realisation of the random interactions. Which one is reached, depends on initial conditions. This is what we mean by `multiple fixed points'. I add that this phase is sometimes also called `multiple attractor' (MA) phase, see e.g. \cite{bunin2016interaction,bunin2017}.  The `multiple fixed points or persistent dynamics'/MA phase is usually observed in a relatively small part of the phase diagram just above the linear instability line (dashed). If the variance of interactions ($\sigma^2$) is increased further, abundances typically diverge at some point. In the phase diagram we indicate the boundary between the `multiple fixed points or persistent dynamics' phase and the phase with diverging abundances by a dotted line. This line is obtained from Eqs.~(\ref{eq:div1},\ref{eq:div2}). However, it is important to realise that the linear instability line has been crossed at this point (the dotted line is above the dashed line in the phase diagram). So technically, our fixed-point equations (\ref{eq:sc_3}) do not hold any longer. In this sense, the dotted line is only an approximation. Starting from the multiple-attractor phase, the line indicates `roughly' where abundances start to diverge.

\begin{figure}[t]
    \centering
    \includegraphics[width=0.3\textwidth]{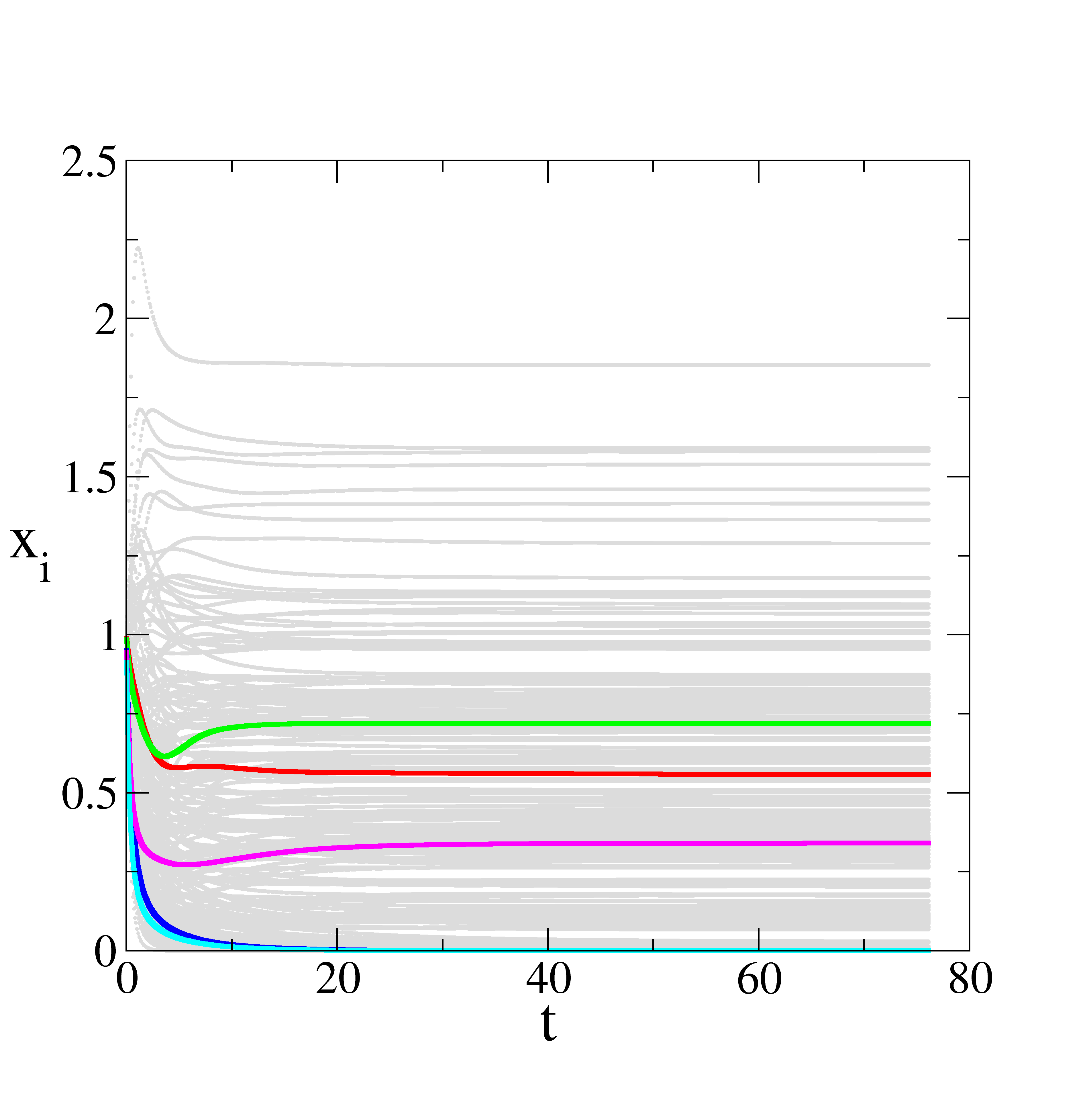}
        \includegraphics[width=0.3\textwidth]{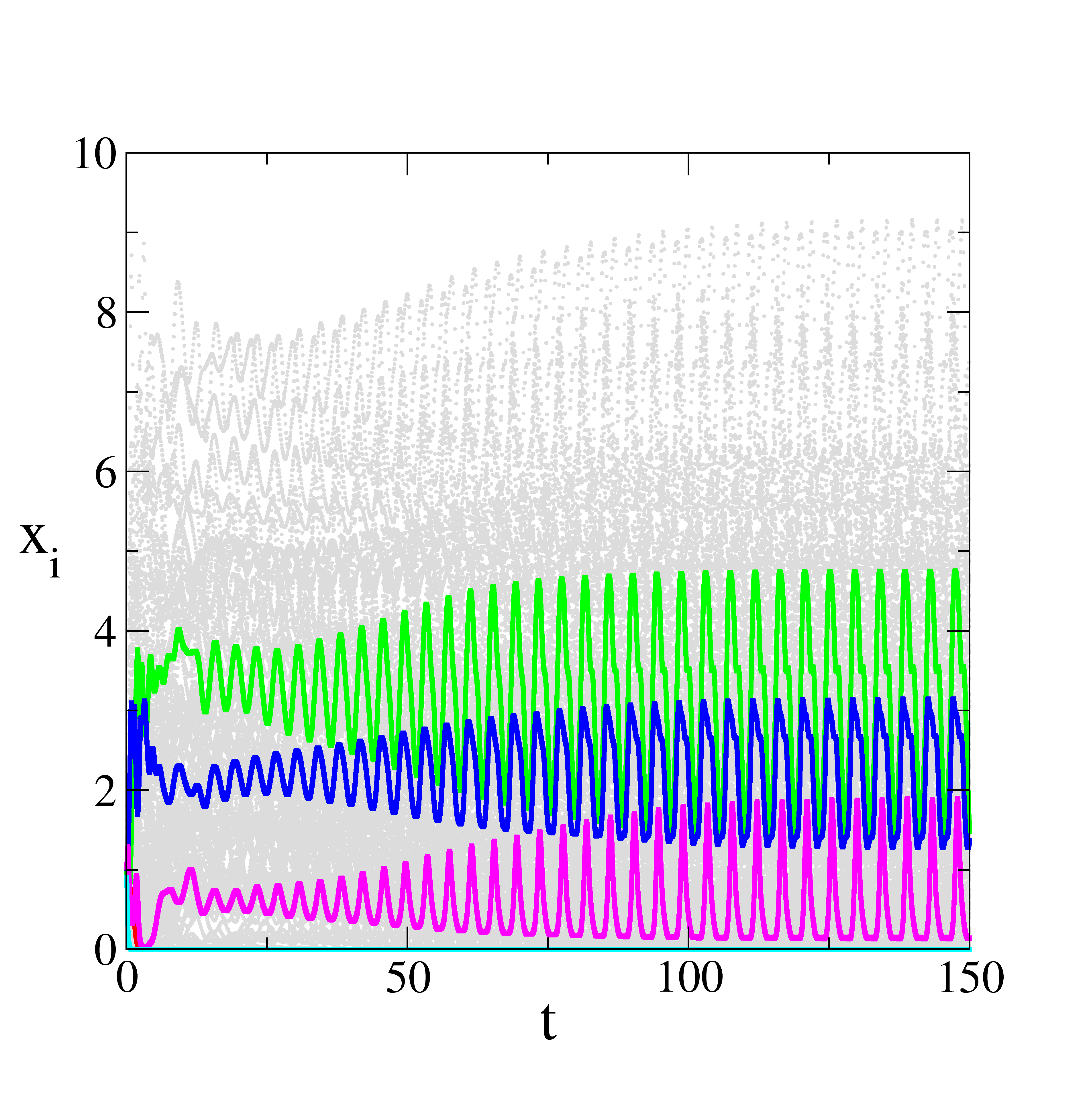}
         \includegraphics[width=0.3\textwidth]{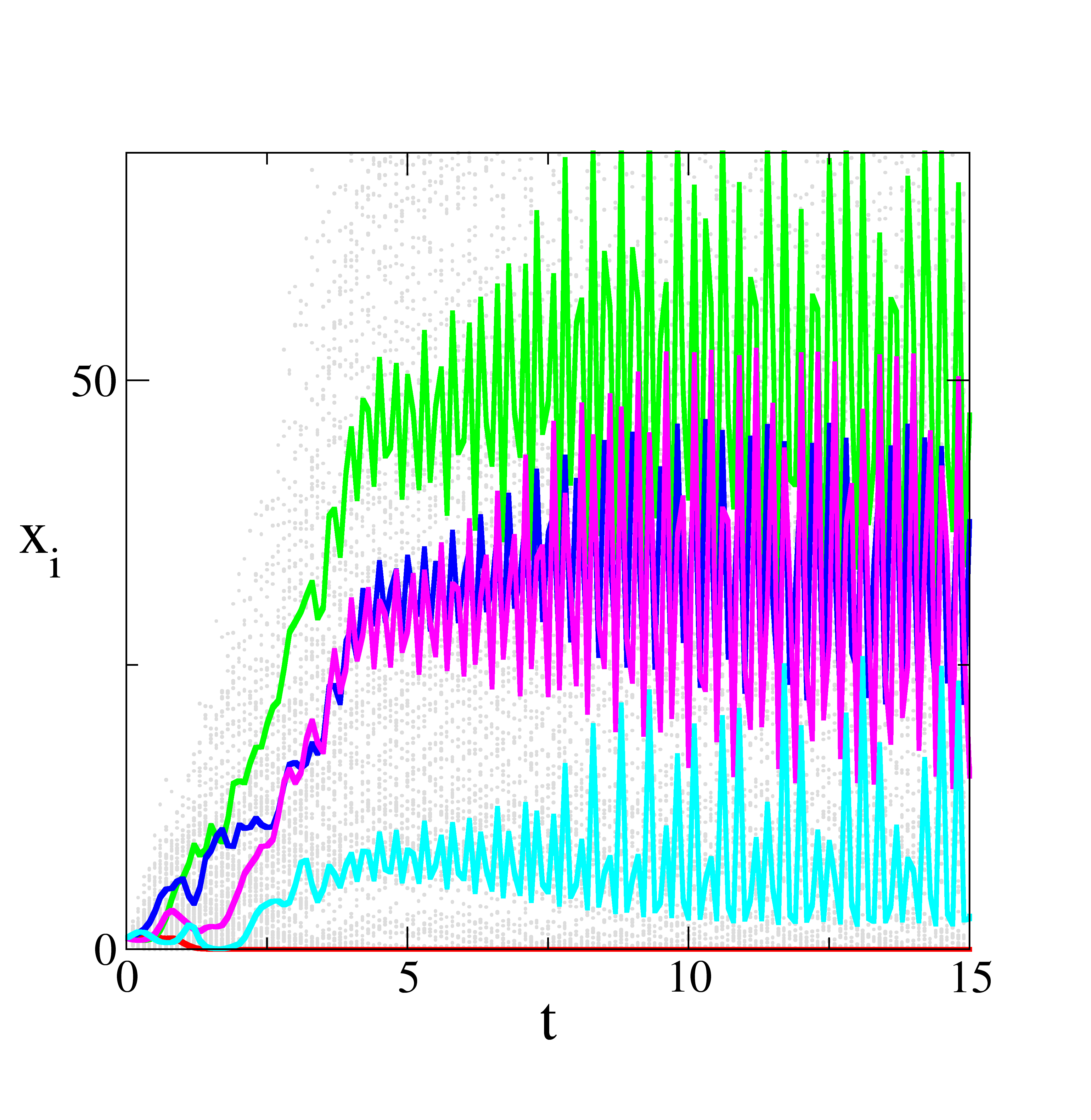}
            \caption{{\bf Sample trajectories.} The figure shows single trajectories of the system in different phases of the stability diagram. On the right: stable phase ($\mu=-1, \Gamma=-0.5, \sigma^2=1$). Middle: Oscillatory behaviour ($\mu=-1, \Gamma=-0.5, \sigma^2=10$). Right: Irregular, potentially chaotic dynamics ($\mu=-0.1, \Gamma=-0.5, \sigma^2=10$). Each plot is for a single realisation with $N=200$ species. Grey dots in the background show all species abundances. A selection of trajectories $x_i(t)$ is highlighted.}
    \label{fig:traj}
\end{figure}

\medskip

Examples of possible trajectories are shown in Fig.~\ref{fig:traj}. For further examples, see also \cite{FerranTFM}.

\underline{Further remark:}\\
The phase diagram is obtained from the analysis in the thermodynamic limit. I stress again that not {\em all} realisations for finite $N$ show the same behaviour in a given part of the phase diagram. For example, it is possible that individual samples diverge in the phase labelled `unique fixed point', or that some samples converge to single stable fixed points above the line marking the linear instability. The stability diagram describes the behaviour for $N\to\infty$, which we would also expect to be dominant (but not fully exclusive) for sufficiently large $N$.

\medskip

It is possible to systematically classify the different behaviours of realisations from a numerical integration of the Lotka--Volterra system. One can then systematically identify the dominant behaviour in the different parts of the stability diagram. Details can be found for example in \cite{sidhomgalla,park}.

\section{Cavity approach to dynamic mean-field theory}\label{sec:cavity}
In this section we describe an alternative derivation of the effective single-species process. This is based on the `cavity method'. For background material see e.g. \cite{mezard1987}. The method has been used for the study of Lotka--Volterra systems for example in \cite{bunin2016interaction,bunin2017,froy}. A an overview of applications of the cavity method to models in community ecology can be found in \cite{barbier_cavity}.

\medskip

Assume we have a solution $x_1(t), \dots, x_N(t)$ for the original Lotka--Volterra problem
\be\label{eq:lv_cav}
\dot x_i=x_i\left(1-x_i+\sum_{j}\alpha_{ij}x_j+h_i(t)\right),
\ee
for given parameters $\mu, \Gamma,\sigma^2$ and a fixed large, but finite $N$. Now we introduce an additional species $i=0$. To do this we need additional interaction coefficients $\alpha_{i,0}$ and $\alpha_{0,i}$ ($i=1,\dots, N$. These are chosen with the same statistics as the $\alpha_{ij}$ ($i,j=1,\dots,N$). We have included perturbation fields $h_i(t)$. The reasons for this will become clear below.

\medskip

The introduction of the new species will change the dynamics for $x_1,\dots,x_N$, and we have
\be\label{eq:lv_cav2}
\frac{d}{dt}{\widetilde x_i}(t)=\widetilde x_i(t)\left(1-\widetilde x_i(t)+\sum_{j=1}^N\alpha_{ij}\widetilde x_j(t)+\alpha_{i0} x_0(t)+h_i(t)\right),
\ee
We have used the notation $\widetilde x_i$ for species abundances in the system with the additional species. (I am aware that we used tildes for Fourier transforms in earlier sections. Unfortunately the selection of possible diacritics that can be added to letters is finite. The risk of confusion should be minimal though, throughout this section tildes are not used for Fourier transforms.) We do not give $x_0$ a tilde, as it is clear that $x_0$ only exists in the system with species $0$ added.

\medskip

Given that the effect of adding one single species to a system with $N\gg 1$ existing species is small, we can work in the linear-response regime. We then have
\be\label{eq:tilde_x}
\widetilde x_i(t)=x_i(t)+\int_0^t dt' \frac{\delta x_i(t)}{\delta h_i(t')}\alpha_{i0}x_0(t').
\ee
The time-evolution of $x_0$ on the other hand is given by
\be\label{eq:ddtx0_0}
\frac{d}{dt} x_0(t) =  x_0(t) \left[1-x_0(t)+\sum_{i=1}^N \alpha_{0,i} \widetilde x_i(t)+h_0(t)\right].
\ee

\medskip

Using Eq.~(\ref{eq:tilde_x}) the dynamics for $x_0$ can be written as
\BE\label{eq:ddtx0}
\frac{d}{dt} x_0(t) &=& x_0(t) \left[1-x_0(t)+\sum_{i=1}^N \alpha_{0,i}  x_i(t)+h_0(t)\right. \nonumber \\
&&+\left. \sum_{i=1}^N \alpha_{0,i}\alpha_{i,0}\int_0^t dt' \frac{\delta x_i(t)}{\delta h_i(t')}x_0(t')\right].
\EE
We first deal with the last term on the right-hand side. This term can be approximated as follows for large $N$,
\be\label{eq:cav1}
\sum_{i=1}^N \alpha_{0,i}\alpha_{i,0}  \frac{\delta x_i(t)}{\delta h_i(t')}\approx \Gamma\sigma^2 G(t,t')
\ee
This is justified because $\overline{\alpha_{0,i}\alpha_{i,0}}=\Gamma\frac{\sigma^2}{N}+ \left(\frac{\mu}{N}\right)^2$ and $G(t,t')=\frac{1}{N}\sum_{i=1}^N  \frac{\delta x_i(t)}{\delta h_i(t')}$.

\medskip

Making this replacement in Eq.~(\ref{eq:ddtx0}) we arrive at
\BE\label{eq:ddtx0_2}
\frac{d}{dt} x_0(t) &=& x_0(t) \left[1-x_0(t)+\Gamma\sigma^2 \int_0^t dt' \,G(t,t') x_0(t')+\sum_{i=1}^N \alpha_{0,i}  x_i(t)\right],
\EE
where we have set the perturbation fieldf $h_0(t)$ to zero (the field is no longer required, now that we have generated the response function $\ullG$).

\medskip

Next we look at the remaining term $\sum_i \alpha_{0,i}  x_i(t)$ in Eq.~(\ref{eq:ddtx0_2}), noting that $\alpha_{0,i}$ has mean $\mu/N$ and variance $\sigma^2/N$, and can therefore be written in the form $\alpha_{0,i}=\frac{\mu}{N}+\frac{\sigma}{\sqrt{N}}u_{0,i}$, with a Gaussian random variable $u_{0,i}$ of mean zero and variance one, and with no correlations between $u_{0,i}$ and $u_{0,j}$ for $i\neq j$. We then have
\be\label{eq:cav_aux1}
\sum_{i=1}^N \alpha_{0,i}  x_i(t)=\frac{\mu}{N}\sum_i x_i(t)+ \frac{\sigma}{\sqrt{N}}\sum_i u_{0,i}  x_i(t).
\ee
Writing $M(t)=\frac{1}{N}\sum_i x_i(t)$, we therefore find that the Gaussian random variable $\sum_i \alpha_{0,i}  x_i(t)$ has mean $\mu M(t)$. The variance and correlations in time of $ \frac{\sigma}{\sqrt{N}}\sum_i u_{0,i}  x_i(t)$ are,
\BE
\frac{\sigma^2}{N}\overline{\left(\sum_i u_{0,i}  x_i(t)\right)\left(\sum_i u_{0,i}  x_i(t')\right)} &=& \frac{\sigma^2}{N}\sum_{i,j=1}^N \overline{u_{0,i}u_{0,j}} x_i(t)x_j(t') \nonumber \\
&=& \frac{\sigma^2}{N}\sum_{i=1}^N x_i(t)x_i(t'), \nonumber \\
&=&\sigma^2 C(t,t')\label{eq:cav_aux2}
\EE
We have used the fact that $u_{0,i}$ and $u_{0,j}$ are uncorrelated for $i\neq j$, and that the variance of each $u_{0,i}$ is $\sigma^2/N$, so that $ \overline{u_{0,i}u_{0,j}}=\delta_{ij}\sigma^2/N$. In the last step, we have then written $C(t,t')=\frac{1}{N}\sum_{i=1}^N x_i(t)x_i(t')$.

Putting things together, we have found that we can write
\be
\sum_i \alpha_{0,i}  x_i(t)= \mu M(t)+\eta_0(t),
\ee
where $\eta_0(t)$ is Gaussian noise with mean zero, and with correlations in time given by
\be
\avg{\eta_0(t)\eta(t')_0}=\sigma^2 C(t,t').
\ee
Using this, we recover the effective dynamics in Eq.~(\ref{eq:effective_copy}), namely,
\BE\label{eq:ddtx0_3}
\frac{d}{dt} x_0(t) &=& x_0(t) \left[1-x_0(t)+\mu M(t)+ \Gamma\sigma^2 \int_0^t dt' \,G(t,t') x_0(t')+\eta_0(t)\right].
\EE
This derivation [Eqs.~(\ref{eq:cav_aux1}) and (\ref{eq:cav_aux2}) in particular] highlights the origin of the coloured noise in the effective process. The term $\eta_0(t)$ in the dynamcics for $x_0$ results from the disordered interaction $\sum_i \alpha_{0,i}  x_i(t)$ of species $0$ with all other species $i=1,\dots,N$. We used this in Sec.~\ref{sec:indep} for the linear stability analysis of fixed points of the Lotka--Volterra system.

\medskip

The reader may legitimately ask why we went through dozens pages of material to introduce the generating-functional method, when the same effective process can be derived in about two pages using the cavity method. This is partly a matter of taste. Personally I'd argue that the generating-functional approach, while longer, is clearer and leaves less space for ambiguity. Once the initial problem is defined, the sequence of steps one needs to go through is well defined, and in a sense `fail safe'. It is like an locomotive that you put on the tracks. Once you have done this, it is just a matter of having enough `steam' to plough ahead. The cavity path seems a little more cumbersome, for example the replacements after Eq.~(\ref{eq:cav1}) and those near Eq.~(\ref{eq:cav_aux2}), while sensible, are not justified to the same level as the steps in the generating-functional calculation leading to the effective single-species process. The generating functional method can also be used to derive further (non-Gaussian) properties of the surviving community, and its stability, see \cite{baron_et_al_prl}.
\section{Summary and discussion}\label{sec:summary}
In summary, we have presented a detailed step-by-step guide through the generating-functional analysis for a Lotka--Volterra system with random interactions. We first reviewed the idea of mean-field theory (Sec.~\ref{sec:mft}) and some of the central mathematical tools (Sec.~\ref{sec:maths}). Then we discussed the concept of generating functions and functionals (Sec.~\ref{sec:gf}). In Sec.~\ref{sec:toy_model} we then looked at a relatively simple toy model, with no disorder. Sections~\ref{sec:gf_lv_1} to \ref{sec:gf_lv_4} then contained the actual generating-functional analysis of the Lotka--Volterra problem. We first derived the saddle-point equations, and then proceeded along the `vanilla' route, before we presented a more complete derivation of the dynamic mean-field theory for the Lotka--Volterra system (the effective single-species process and associated self-consistency relations). Sections~\ref{sec:fixed_point} and \ref{sec:stability_pg} then focused on a fixed-point ansatz and a stability analysis of this fixed point respectively. From this we were able to establish the stability diagram. In Sec.~\ref{sec:cavity} finally, we presented an alternative derivation of the effective dynamics, using the so-called `cavity' method.

\medskip

I hope the notes are sufficiently self-contained so that beginners do not have to compile the relevant pieces from a  number of different sources. I also hope that the length of this paper is not too daunting. I tried to be as clear as I could about the structure of the document as a whole (how the different sections relate to one another), and about the contents of each section. I also tried to balance pedagogical efficiency with analytical rigour (the rigour of a theoretical physicist, not that of a pure mathematician). One of my main aims was to tell the reader at any point of the calculation what the assumptions and simplifications are that lead to a certain result, what earlier results are used to make the next step, and to make sure that it is possible keep track the flow of the overall argument.

\medskip

There is much that is not said in these notes. As mentioned in the introduction I don't really motivate {\em why} one would be interested in the disordered Lotka--Volterra model (I point to the relevant literature though). Further, I restrict the notes to the `standard' generalised Lotka--Volterra equations (\ref{eq:lv}). There are many extensions one could study and that have been studied in the literature, either with generating functionals or with other methods. This includes more structured interaction matrices and networks \cite{poley1,poley2,park,aguirre}, intrinsic noise \cite{altieri2021properties,ferran}, MacArthur type resource models \cite{yoshino,bunin_macarthur} and the study of the detailed properties of the interactions between surviving species \cite{bunin2016interaction,baron_et_al_prl}. One should also note the complementary physics literature using equilibrium methods (e.g. the replica approach) to Lotka--Volterra systems with symmetric interactions (see e.g. \cite{birolibunin,ros1,ros2}). Exciting recent work uses dynamic mean field theory to study the unstable phase \cite{Pirey_Bunin}.

\medskip

As a final note, I'd like to re-iterate that the generating-functional approach to disordered systems has been used and continues to be used in a range of different areas, including spin glass physics \cite{mezard1987}, random-matrix theory \cite{baron_sparse}, neural networks \cite{coolen_kuehn_sollich,coolen_dyn}, game theory and related topics \cite{Coolen_MG,galla_farmer,jerome}. While these notes focus on the Lotka--Volterra model, the main steps in particular in Sections~\ref{sec:gf_lv_1}-\ref{sec:gf_lv_4} can be transferred without much change to other disordered dynamical system. Because of this I hope that these notes are useful to a broader set of readers.

\section*{Acknowledgements}
I am grateful to David Sherrington and Ton Coolen who introduced me to generating functionals in the first place (more than twenty years ago during my PhD). I thank Andrea De Martino, Reimer K\"uhn and in particular Joseph W. Baron for collaboration and discussions which sharpened my understanding of the method. I also thank (in alphabetical order) Sebastian Castedo, Alaina Cockerell, Mark Crumpton, Joshua Holmes, Thomas Jun Jewell, Carlotta Nunzi, Lyle Poley, Enrique Rozas, Christopher Ryder, and Laura Sidhom for useful comments on earlier versions of these notes. 

\paragraph{Funding information} 
I acknowledge partial financial support from the Agencia Estatal de Investigaci\'on and Fondo Europeo de Desarrollo Regional (FEDER, UE) under project APASOS (PID2021-122256NB-C21, PID2021-122256NB-C22), and the Mar\'ia de Maeztu programme for Units of Excellence, CEX2021-001164-M funded by MICIU/AEI/10.13039/501100011033.

\begin{appendix}

\section{Derivation of the effective process for the Lotka--Volterra problem without division by species abundances}\label{app:clean}
In this appendix we describe how the generating-functional analysis would have to be modified in order to avoid having to divide both sides of the Lotka--Volterra equations (\ref{eq:lv}) by $x_i$.

\medskip

Instead of Eq.~(\ref{eq:gflv0}) we start from
\BE
Z[\boldpsi]
&=&\int \D\ulbx \D\widehat\ulbx \exp\Bigg(i\sum_i\int dt \Bigg[\widehat x_i(t)\Bigg(\dot x_i(t)-x_i(t)[1-x_i(t)+\sum_{j\neq i}\alpha_{ij} x_j(t)+h_i(t)]\Bigg)\Bigg]\Bigg)\nonumber\\
&&\times \exp\left(i\sum_i\int dt~ x_i(t)\psi_i(t)\right),\label{eq:gflvp0}
\EE
We now introduce the following short-hand
\be
f_i[\bx(t)]=1-x_i(t)+\sum_{j\neq i}\alpha_{ij} x_j(t)+h_i(t).
\ee
This can be introduced into the generating functional via suitable delta functions, and we find 
\BE
Z[\boldpsi]
&=&\int \D\ulbx \D\widehat\ulbx ~\D\ulbf \D\widehat\ulbf ~\exp\Bigg(i\sum_i\int dt \Bigg[\widehat x_i(t)\Bigg(\dot x_i(t)-x_i(t)f_i(t)\Bigg)\Bigg]\Bigg)\nonumber\\
&&\times\exp\left(i\sum_i\int dt~\widehat f_i(t)\left\{f_i(t)-[1-x_i(t)+\sum_{j\neq i}\alpha_{ij} x_j(t)+h_i(t)]\right\}\right)\nonumber \\
&&\times \exp\left(i\sum_i\int dt~ x_i(t)\psi_i(t)\right),\label{eq:gflvp1}
\EE
The disorder is now contained in the term $\exp\left(i\sum_{i\neq j}\int dt~\widehat f_i(t)\alpha_{ij} x_j(t)\right)$. This is the equivalent of the expression in Eq.~(\ref{eq:big_X}).  The calculation then proceeds following the same steps as in Sec.~\ref{sec:disorder_average}, with the replacement $\widehat x_i(t)\to \widehat f_i(t)$.

\medskip

Following these steps, one ends up with the following disorder-averaged generating functional of the same form as in Eq.~(\ref{eq:sp}),
\be\label{eq:sp_p}
\overline{Z}[\ulbpsi]=\int D[\ulM,\widehat\ulM] D[\ulP,\widehat\ulP] D[\ullC,\widehat\ullC] D[\ullK,\widehat\ullK] D[\ullL,\widehat\ullL] \exp\left(N\left[\Psi+\Phi+\Omega+{\cal O}(N^{-1})\right]\right).
\ee
The expressions for $\Psi$ and $\Phi$ are as before [see Eqs.~(\ref{eq:sp_psi}) and (\ref{eq:sp_phi})]. We have 
\BE
\Psi&=&i\int dt~ [\widehat M(t) M(t) +\widehat P(t) P(t)] \nonumber \\
&&+i\int dt ~dt' \left[\widehat C(t,t')C(t,t')+\widehat K(t,t')K(t,t')+\widehat L(t,t')L(t,t')\right],
\EE
and
\BE
\Phi&=&-\frac{1}{2}\sigma^2\int dt ~dt' \left[L(t,t')C(t,t')+\Gamma K(t,t')K(t',t)\right]\nonumber \\
&&-\mu\int dt ~ M(t) P(t).
\EE
The contribution $\Omega$, however, now takes a slightly different form, 
\BE
\Omega&=&N^{-1}\sum_i\log\bigg[\int \D\ulx_i \D \widehat \ulx_i \D\ulf_i\D\widehat \ulf_i p_{0}[x_i(0)]\exp\left(i\int dt~ \psi_i(t)x_i(t)\right)\nonumber \\
&&\times \exp\left(i\int dt ~\widehat x_i(t)  \left(\dot x_i(t) - x_i(t) f_i(t) \right)\right)\exp{\left(i\int dt ~\widehat f_i(t)\{f_i(t)-[1-x_i(t)+h_i(t)]\}\right)} \nonumber \\
&&\times \exp\left(-i\int dt ~ dt' \left[\widehat C(t,t')x_i(t)x_i(t')+\widehat L(t,t')\widehat f_i(t)\widehat f_i(t')+\widehat K(t,t') x_i(t)\widehat f_i(t')\right]\right)\nonumber \\
&&\times \exp\left(-i\int dt~  [\widehat M(t)x_i(t)+\widehat P(t) \widehat f_i(t)]\right)\bigg].
\label{eq:omega_p_0}
\EE
Variation with respect to the unhatted order parameters gives the same results as before [Eq.~(\ref{eq:sp_aux1})], 
\BE
i\widehat M(t)&=&i\mu P(t), \nonumber \\
i\widehat P(t)&=&i\mu M(t), \nonumber \\
i\widehat C(t,t')&=&\frac{1}{2}\sigma^2L(t,t'),\nonumber \\
i\widehat K(t,t')&=&\Gamma \sigma^2 K(t',t),\nonumber \\
i\widehat L(t,t')&=&\frac{1}{2}\sigma^2C(t,t').\label{eq:sp_aux1_f}
\EE
However, the variation with respect to hatted order parameters now leads to
\BE
M(t)&=&\lim_{N\to\infty}N^{-1}\sum_i\avg{x(t)}_i, \nonumber \\
P(t)&=&\lim_{N\to\infty}N^{-1}\sum_i\avg{\widehat f(t)}_i, \nonumber \\
C(t,t')&=&\lim_{N\to\infty}N^{-1}\sum_i\avg{x(t)x(t')}_i,\nonumber \\
K(t,t')&=&\lim_{N\to\infty}N^{-1}\sum_i\avg{x(t)\widehat f(t')}_i, \nonumber \\
L(t,t')&=&\lim_{N\to\infty}N^{-1}\sum_i\avg{\widehat f(t)\widehat f(t')}_i, 
\label{eq:var_wrt_hats_f}
\EE
This is the same as in Eq.~(\ref{eq:var_wrt_hats}), but with the replacement $\widehat \ulx_i \to \widehat \ulf_i$.

Following the steps of Sec.~\ref{sec:id_op_LV} one shows that $\ulP=\ulnull$ and $\ullL=\ullnull$ at the saddle point, and in the limit $\ulbpsi\to\ulbnull$. Using this and the remaining saddle-point relations in Eq.~(\ref{eq:sp_aux1_f}) one finds (after setting $h_i(t)\equiv h(t)$, and replacing $\ullK \to i\ullG$),
\be\label{eq:av_g_f}
\avg{g[\ulx,\widehat\ulx, \ulf,\widehat\ulf]}_*= \frac{\int \D\ulx \D\widehat\ulx  \D\ulf \D\widehat\ulf\, g[\ulx,\widehat\ulx, \ulf,\widehat\ulf]\, {\cal M}[\ulx,\widehat\ulx, \ulf,\widehat\ulf]}{\int  \D\ulx \D\widehat\ulx  \D\ulf \D\widehat\ulf \,  {\cal M}[\ulx,\widehat\ulx, \ulf,\widehat\ulf]}
\ee
with the single effective-species measure
\BE
{\cal M}[\ulx,\widehat\ulx, \ulf,\widehat\ulf]&=&p_{0}[x(0)] \,  \exp\left(i\int dt ~\widehat x(t)  \left[\dot x(t) - x(t) f(t) \right]\right)\nonumber \\
&&\times\exp\left(i\int dt ~\widehat f(t)  \left(f(t)-[1-x(t)+h(t)]\right)\right)\nonumber \\
&&\times \exp\left(-\sigma^2\int dt ~ dt' \left[ \frac{1}{2}C(t,t')\widehat f(t)\widehat f(t')+i\Gamma G(t',t) x(t)\widehat f(t')\right]\right).\nonumber \\\label{eq:m_effective_f}
\EE
This describes the process
\be\label{eq:eff_f_1}
\dot x(t) = x(t) f(t),
\ee
where
\be\label{eq:eff_f_2}
f(t)=1-x(t)+\Gamma\sigma^2 \int dt' G(t,t') x(t')+\eta(t)+h(t),
\ee
and where $\eta(t)$ is Gaussian noise with mean zero and $\avg{\eta(t)\eta(t')}_\star=\sigma^2 C(t,t')$. Eqs.~(\ref{eq:eff_f_1}) and (\ref{eq:eff_f_2}) can be combined to give

\be
\dot x(t)= x(t)\left[1-x(t)+\Gamma\sigma^2 \int dt' G(t,t') x(t')+\eta(t)+h(t)\right],
\ee
which is the effective process in Eq.~(\ref{eq:effective_copy}).

\medskip

From Eqs.~(\ref{eq:var_wrt_hats_f}) we have the self-consistency relations $M(t)=\avg{x(t)}_\star$, and $C(t,t')=\avg{x(t)x(t')}_\star$. Noting that a derivative of averages $\avg{\cdots}_*$ with respect to $h(t')$ now `brings down' a factor $-i\widehat f(t')$ we also have $G(t,t')=-iK(t,t')=-i\avg{x(t)\widehat f(t)}_\star=\frac{\delta}{\delta h(t')} \avg{x(t)}_\star$. 

\section{Equivalence of averages over the effective dynamics and averages in the original Lotka--Volterra problem}\label{app:equivalence}

In this appendix we demonstrate the validity of the identity in Eq.~(\ref{eq:general_equivalence}), which we repeat here for convenience,
\BE
\lim_{N\to\infty}\frac{1}{N}\sum_i \olavg{f[\ulx_i,\widehat\ulx_i]}=\lim_{N\to\infty}\frac{1}{N}\sum_i\left.\avg{f[\ulx,\widehat \ulx]}_i\right|_{\mbox{SP},\ulpsi_i=0}.\label{eq:general_equivalence_app}
\EE
The object on the left corresponds to an observable in the original Lotka--Volterra problem. The object on the right is a quantity to be evaluated from the effective single-particle measure at the saddle-point of the generating functional calculation. 

It is perhaps not immediately clear in what sense the object on the right is an observable in the original problem. First we notice that $\avg{\cdots}$ is an average over realisations of the original dynamics for fixed disorder, and the overbar stands for the average over the disorder. So this is manifestly for the original Lotka--Volterra problem. The issue is the occurence of the variables $\widehat x_i$, which don't have a direct physical meaning. What we mean by the notation on the left is 
\BE\label{eq:app_olavg_f}
\olavg{f[\ulx_i,\widehat\ulx_i]}&=&\lim_{\ulbpsi\to\ulbnull} \Bigg\{\int \D\ulbx \D\widehat\ulbx \, p_0[\bx(0)]\,f[\ulx_i,\widehat\ulx_i] \exp\left(i\sum_i \int dt\, \psi_i(t)x_i(t)\right)\nonumber \\
&&\nonumber \\
&&\times\overline{\exp\Bigg(i\sum_i\int dt \Bigg[\widehat x_i(t)\Bigg(\frac{\dot x_i(t)}{x_i(t)}-[1-x_i(t)+\sum_{j\neq i}\alpha_{ij} x_j+h_i(t)]\Bigg)\Bigg]\Bigg)}\Bigg\}\nonumber \\
\EE
(careful not to confuse the overbar with a fraction).

Now, any reasonable function $f[\ulx_i,\widehat\ulx_i]$ can be expanded into a power series in terms of the $x_i(t)$ and $\widehat x_i(t)$, so we can assume that $f$ is a polynomial, with individual terms for example of the form $x_i(t_1)x_i(t_2)^2 \widehat x_i(t_3)^2 \widehat x_i(t_4)^3$. We can then make the replacement
\be
f[\ulx_i,\widehat\ulx_i] \to f\left[-i\frac{\delta}{\delta\ulpsi_i}, \frac{\delta}{-\delta \ulh_i}\right], 
\ee
i.e.,
\BE\label{eq:olavg_f_app}
\olavg{f[\ulx_i,\widehat\ulx_i]}&=&\lim_{\ulbpsi\to\ulbnull} \Bigg\{\int \D\ulbx \D\widehat\ulbx \, p_0[\bx(0)]\,f\left[-i\frac{\delta}{\delta\ulpsi_i}, \frac{\delta}{-\delta \ulh_i}\right] \exp\left(i\sum_i \int dt\, \psi_i(t)x_i(t)\right)\nonumber \\
&&\times\overline{\exp\Bigg(i\sum_i\int dt \Bigg[\widehat x_i(t)\Bigg(\frac{\dot x_i(t)}{x_i(t)}-[1-x_i(t)+\sum_{j\neq i}\alpha_{ij} x_j+h_i(t)]\Bigg)\Bigg]\Bigg)}\Bigg\}.\nonumber \\
\EE
For example $\olavg{x_i(t_1)x_i(t_2)^2 \widehat x_i(t_3)^3 \widehat x_i(t_4)^2}$ turns into $(-i)^5 \frac{\delta^5}{\delta h_i(t_3)^3 \delta h_i(t_4)^2}\olavg{x_i(t_1)x_i(t_2)^2}$, or alternatively into 
\BE
\olavg{x_i(t_1)x_i(t_2)^2 \widehat x_i(t_3)^3 \widehat x_i(t_4)^2}&=& \lim_{\ulbpsi\to\ulbnull} \Bigg\{\int \D\ulbx \D\widehat\ulbx \, p_0[\bx(0)]\frac{(-i)^8\delta^8}{\delta \psi_i(t_1)\delta\psi_i(t_2)^2\delta h_i(t_3)^3 \delta h_i(t_4)^2} \nonumber \\&& \times\exp\left(i\sum_i \int dt\, \psi_i(t)x_i(t)\right) \overline{\exp\left(\cdots\right)}\Bigg\} \nonumber \\
&=& \frac{(-i)^8\delta^8}{\delta \psi_i(t_1)\delta\psi_i(t_2)^2\delta h_i(t_3)^3 \delta h_i(t_4)^2}  \overline{Z}[\ulbpsi]\Big|_{\ulpsi=\ulbnull}.
 \EE

\medskip

The statement in Eq.~(\ref{eq:general_equivalence_app}) is limited to observables which can be written as a sum over species ($N^{-1}\sum_i\dots$) of quantities that relate to single species only. For example, the object $N^{-1}\sum_i \frac{\delta^2} {h_i(t_1) h_i(t_2)} \overline{\avg{x_i(t_3)^2}}$ qualifies, but $N^{-2} \sum_{ij} \overline{\avg{x_i(t_1)x_j(t_2)}}$ does not (or at least, it is not covered by our derivation).

\medskip

We therefore start from
\be
A=\lim_{N\to\infty} \left.\frac{1}{N}\sum_i {\cal A}_{\rm ind}\left(\frac{\delta}{\delta \ulpsi_i},\frac{\delta}{\delta \ulh_i}\right) \overline{Z}[\ulbpsi,\ulbh]\right|_{\ulpsi=\ulbnull},
\ee
assuming we are interested in the thermodynamic limit. The object ${\cal A}_{\rm ind}$ is to be understood as an `operator', composed of derivatives with respect to $\ulpsi_i$ and/or $\ulh_i$. The subscript `ind' indicates that the operation ${\cal A}_{\rm ind}$ acts on individual species only. We write ${\cal A}_i\equiv  {\cal A}_{\rm ind}\left(\frac{\delta}{\delta \ulpsi_i},\frac{\delta}{\delta \ulh_i}\right)$ in the following to keep the notation compact. We also introduce the operator
\be
{\cal A}=\lim_{N\to\infty} \frac{1}{N}\sum_i {\cal A}_i,
\ee
so that we have
\be
A={\cal A}\, \overline{Z}[\ulbpsi,\ulbh]\big|_{\ulpsi=\ulbnull}.
\ee

\medskip

From Eq.~(\ref{eq:sp}) we have
\be
\overline{Z}[\ulbpsi]=\int D[\ulM,\widehat \ulM] \D[\ulP,\widehat \ulP] \D[\ullC,\widehat\ullC]\D[\ullK,\widehat \ullK] \D[\ullL,\widehat \ullL] \,\exp\left(N\left[\Psi+\Phi+\Omega+{\cal O}(N^{-1})\right]\right),
\ee
and only $\Omega$ contains the $\{\psi_i(t)\}$ and $\{h_i(t)\}$.  From Eqs.~(\ref{eq:omega}) and (\ref{eq:z1_def}), we further have
\be
\Omega=\frac{1}{N}\sum_j \ln\,Z_1[\ulpsi_j,\ulh_j].\label{eq:omega_app}
\ee
Thus,
\be
\exp(N\Omega)=\prod_j Z_1[\ulpsi_j,\ulh_j].
\ee

Applying the operator ${\cal A}_i$ (for a given fixed $i$) we then have
\BE
{\cal A}_i \exp(N\Omega)&=& {\cal A}_i \left(\prod_j Z_1[\ulpsi_j,\ulh_j] \right)\nonumber \\
&=& \left(\prod_{j\neq i} Z_1[\ulpsi_j,\ulh_j]\right) {\cal A}_i Z_1[\ulpsi_i,\ulh_i] \nonumber \\
&=& \left(\prod_{j} Z_1[\ulpsi_j,\ulh_j]\right) \frac{{\cal A}_i Z_1[\ulpsi_i,\ulh_i]}{Z_1[\ulpsi_i,\ulh_i]} \nonumber \\
&=&  \exp(N\Omega) \frac{{\cal A}_i Z_1[\ulpsi_i,\ulh_i]}{Z_1[\ulpsi_i,\ulh_i]}.
\EE
Thus, \BE
{\cal A}\overline{Z}[\ulbpsi]&=&\lim_{N\to\infty}\frac{1}{N}\sum_i \int D[\ulM,\widehat \ulM] \D[\ulP,\widehat \ulP] \D[\ullC,\widehat\ullC]\D[\ullK,\widehat \ullK] \D[\ullL,\widehat \ullL] \,{\cal A}_i\exp\left(N\left[\Psi+\Phi+\Omega+{\cal O}(N^{-1})\right]\right) \nonumber \\
&=& \lim_{N\to\infty}\frac{1}{N}\sum_i \int D[\ulM,\widehat \ulM] \D[\ulP,\widehat \ulP] \D[\ullC,\widehat\ullC]\D[\ullK,\widehat \ullK] \D[\ullL,\widehat \ullL] \,\Bigg\{\frac{{\cal A}_i Z_1[\ulpsi_i,\ulh_i]}{Z_1[\ulpsi_i,\ulh_i]} \nonumber \\
&&\hspace{8em} \times\exp\left(N\left[\Psi+\Phi+\Omega+{\cal O}(N^{-1})\right]\right)\Bigg\}.
\EE
Using the rules of saddle-point integration (and the fact that the generating functional is normalised), we arrive at
\BE
{\cal A}\overline{Z}[\ulbpsi]&=&\lim_{N\to\infty}\frac{1}{N}\sum_i\left.\frac{{\cal A}_i Z_1[\ulpsi_i,\ulh_i]}{Z_1[\ulpsi_i,\ulh_i]}\right|_{\ulpsi=\ulbnull, \mbox{\footnotesize SP}}.
\EE
 
We next use the definition of $Z_1$ in Eq.~(\ref{eq:z1_def}) as well as that of the average $\avg{\cdots}_i$ in Eq.~(\ref{eq:avg_i}). The operator ${\cal A}_i$, applied to $Z_1[\ulpsi_i,\ulh_i]$, `brings down' the same powers of $\ulx$ and $\widehat \ulx$ as those of $\ulx_i$ and $\widehat \ulx_i$ in the original definition of $f[\ulx_i, \widehat\ulx_i]$ [see e.g. the text just after Eq.~(\ref{eq:olavg_f_app})]. We therefore have
\be
A=\lim_{N\to\infty}\frac{1}{N}\sum_i \avg{f[\ulx,\widehat\ulx]}_i,
\ee
which is what we wanted to show.
\end{appendix}

\end{document}